\documentclass[useAMS,usenatbib]{mn2e} 
\pdfoutput=1
\usepackage{graphicx,amsmath,amssymb,subfigure,rotating,epsfig}
\usepackage{longtable}
\usepackage{pdflscape}

\bibpunct{(}{)}{;}{a}{,}{,}

\def\simgt{\mathrel{\lower0.6ex\hbox{$\buildrel {\textstyle >} \over {\scriptstyle \sim}$}}}
\def\simlt{\mathrel{\lower0.6ex\hbox{$\buildrel {\textstyle <} \over {\scriptstyle \sim}$}}}

\begin{document}

\title[6.7GHz Methanol Maser Associated Outflows]{Methanol Maser Associated Outflows: Detection statistics and properties.} 

\author[H. M. de Villiers et al.~2014] 
{\parbox{\textwidth}{H. M. de Villiers,$^1$\thanks{Email: lientjiedv@gmail.com}
  A.\ Chrysostomou$^1$, M.\ A.\ Thompson$^1$, S.\ P.\ Ellingsen$^2$, J.\ S.\ Urquhart$^3$, 
  S.\ L.\ Breen$^4$, M.\ G.\ Burton$^5$,  T. Csengeri$^3$, D.\ Ward-Thompson$^6$ \\
\footnotesize 
$^1$Centre for Astrophysics Research, University of Hertfordshire, College Lane, Hatfield, Herts, AL10 9AB, United Kingdom\\
$^2$School of Physical Science, University of Tasmania, Private Bag 37, Hobart 7001, TAS, Australia\\
$^3$Max-Planck-Institut f\"ur Radioastronomie, Auf dem H\"ugel  69, D-53121 Bonn, Germany \\
$^4$CSIRO Astronomy and Space Science, Australia Telescope National Facility, PO Box 76, Epping, NSW 1710, Australia\\
$^5$School of Physics, University of New South Wales, Sydney, NSW 2052, Australia\\
$^6$Jeremiah Horrocks Institute, University of Central Lancashire, Preston, Lancashire, PR1 2HE, United Kingdom \\
}}

\maketitle
\begin{abstract}

We have selected the positions of 54 6.7GHz methanol masers from the Methanol Multibeam Survey catalogue, covering a range of longitudes between 20$^\circ$ and 34$^\circ$ of the Galactic Plane.  These positions were mapped in the $\rm{J=3-2}$ transition of both the $\rm{^{13}CO}$ and $\rm{C^{18}O}$ lines.  A total of 58 $\rm{^{13}CO}$ emission peaks are found in the vicinity of these maser positions.  We search for outflows around all $\rm{^{13}CO}$ peaks, and find evidence for high-velocity gas in all cases, spatially resolving the red and blue outflow lobes in 55 cases.  Of these sources, 44 have resolved kinematic distances, and are closely associated with the 6.7GHz masers, a sub-set referred to as Methanol Maser Associated Outflows (MMAOs).  We calculate the masses of the clumps associated with each peak using 870 $\rm{\mu m}$ continuum emission from the ATLASGAL survey.  A strong correlation is seen between the clump mass and both outflow mass and mechanical force, lending support to models in which accretion is strongly linked to outflow. We find that the scaling law between outflow activity and clump masses observed for low-mass objects, is also followed by the MMAOs in this study, indicating a commonality in the formation processes of low-mass and high-mass stars.

\end{abstract}


\section{Introduction}
\label{sec:introduction}

Massive stars ($\rm{>8 M_{\odot}}$) play a key role in the evolution of the Universe, as the principal sources of heavy elements and UV radiation.  Their winds, massive outflows, expanding H{\sc ii} regions and supernova explosions serve as an important source of enrichment, mixing and turbulence in the ISM of galaxies \citep{Zinnecker2007}.  Our understanding of the formation and evolution of young massive stars is made difficult by their rarity, large average distances that demands observations at higher angular resolution, deeply embedded formation within dense clusters resulting in confusing dynamics and obscuration, and rapid evolution with short-lived evolutionary phases \citep{Shepherd1996a, Zinnecker2007}.

The specific formation process of massive stars is not yet fully understood.  These stars reach the zero-age main sequence (ZAMS) while still accreting material from their parent molecular cloud.  Due to their high mass, they radiate strongly.  
This radiation pressure exceeds the gravitational pressure, and should the formation process be similar to low mass stars, the growing radiation pressure from the newborn stars will eventually become strong enough to stop the accretion, yielding an upper mass limit of $\sim 40 \rm{M_{\odot}}$ \citep{Wolfire1987,Stahler1993}.

Previously, two solutions were proposed to overcome this problem: (i) a formation process involving multiple lower mass stars, either via coalescence of low- to intermediate-mass protostars \citep[e.g.][]{Bonnell1998,Bally2005}, or competitive accretion in a clustered environment \citep[e.g.][]{Bonnell2004}, or (ii) a scaled-up version of the process found in low-mass star formation.  The latter solution can be sub-divided into the following main categories: (a) increased spherical accretion rates in turbulent cloud cores (order $10^{-4} - 10^{-3} \rm{M_{\odot} yr^{-1}}$), high enough to overcome the star's radiation pressure \citep[e.g.][]{McKee2003,Norberg2000} or (b) accretion via disks onto a single massive star \citep[e.g.][]{Jijina1996,Yorke2002}.  

A solution to overcome the radiation pressure barrier was proposed by \citet{Yorke2002}, that involved the generation of a strong anisotropic radiation field where an accretion disk reduces the effects of radiative pressure, by allowing photons to escape along the polar axis (the ``flashlight effect''). However, these simulations showed an early end of the disk accretion phase, with final masses limited to $\sim 42 \rm{M_{\odot}}$.  \citet{Krumholz2009} suggested that the early end of the accretion phase is because the disk loses its shielding property as it cannot be fed in an axially symmetric configuration.  Contrary to the stable radiation pressure-driven outflows in \citet{Yorke2002}, they proposed a three-dimensional radiation hydrodynamic simulation with a Rayleigh-Taylor instability in the outflow region, allowing further accretion onto the disk. 

\citet{Kuiper2010} took this further by introducing a dust sublimation front to their simulations. This preserves the shielding of the massive accretion disk and allows the protostar to grow to $\sim140 \rm{M_{\odot}}$.

The easiest way to verify the disk accretion models, would be with the detection of accretion disks around massive protostars, but this is difficult without specialized techniques \citep[e.g.][]{Pestalozzi2009}, because they are small (at most several hundred AU), short lived, and easily confused by envelopes \citep{Kim2006}. Few clear examples of such disks exist \citep[e.g.][]{Cesaroni2007, Zapata2010}.

However, we expect that if massive stars do form via accretion disks, they will generate massive and powerful outflows, similar to low-mass stars.  These outflows are necessary to transport angular momentum away from a forming star \citep{Shu1991,Shu2000,Konigl2000,Chrysostomou2008}. For massive stars, these outflows should be of much larger scale and easier to detect than the accretion disks \citep{Kim2006}. Studying outflows offers an alternative approach to probe the embedded core.

There have been many studies that collectively suggest outflows are ubiquitously associated with massive star formation \citep[e.g.][]{Molinari1998, Beuther2002, Shepherd1996b, Xu2006}.

\citet{Zhang2005} found outflow masses ($\sim 10$ to 100's $\rm{M_{\odot}}$), momenta (10-100 $\rm{M_{\odot} kms^{-1}}$) and energies ($\sim 10^{39}$ J) toward their sample of luminous IRAS point sources about a factor 10 higher than the values of low-mass outflows \citep{Bontemps1996}.  This suggests that outflows consist of accelerated gas that has been driven by a young accreting protostar, rather than swept-up ambient material \citep{Churchwell1999}.  It could also be material that originates from the accretion disk / young stellar object and is funnelled out of the central system \citep[e.g][]{Shepherd1996b}.  

To date, CO observations of molecular outflows have been made using mainly two methods: (1) single-point CO line surveys toward samples of massive young stellar objects (YSO's) in search of high-velocity molecular gas \citep[e.g.][]{Shepherd1996a, Sridharan2002} or (2) CO line mapping of carefully selected sources that exhibit high-velocity wings \citep[e.g.][]{Shepherd1996b, Beuther2002}.  Unless outflows are mapped, it is difficult to determine their physical properties.  Mapping outflows at sufficient sensitivity and high angular resolution is time-consuming, but the development of heterodyne focal plane arrays (e.g. HARP on JCMT or HERA on IRAM) has made it possible to map statistically significant samples of massive star-forming regions to search for outflows \citep[e.g.][]{Gottschalk2012,Lopez2009}.

Outflows are one of the earliest observable signatures of star formation, and are believed to develop from the central objects during the infrared bright stage called the ``hot core'' phase \citep{Cesaroni1992,Kurtz2000}, just before the UCH{\sc ii} phase \citep{Shepherd1996b, Wu1999, Molinari2002, Beuther2002, Zhang2001}.  

Another important signpost of the ``hot core'' phase, is the turn-on of radiatively pumped 6.7GHz (class II) methanol masers, the second brightest masers in the Galaxy \citep{Sobolev1997,Minier2003,Menten1991}.  Observations indicate that these masers are rarely associated with H{\sc ii} regions, but most of them are found to be associated with massive millimeter and submillimeter sources \citep[e.g.][]{Walsh2003,Beuther2002,Urquhart2013}.  It appears as if these masers occupy a brief phase in the pre-UCH{\sc ii} region, even as short as $\sim 10^4$ years, and disappear as the UCH{\sc ii} region evolves \citep{Hatchell1998,Codella2000,Codella2004,vanderWalt2005,Wu2010}. They are also known to be mostly associated with massive star formation, making them important signposts of massive star formation \citep{Minier2005, Ellingsen2006, Caswell2013, Breen2013}. 

However, there are limited simultaneous studies of methanol masers and outflow activity.  \citet{Minier2001} found that ten out of thirteen absolute positions for class II methanol maser sites coincided with typical tracers of massive star formation (e.g. UCH{\sc ii} regions, outflows and hot cores), while seven out of these ten were within less than 2000 AU ($\sim 10^{-2}$ pc) from outflows.  Their results supported the expected association between the occurrence of class II methanol masers and molecular outflows.

The {\em Spitzer} GLIMPSE survey \citep{Churchwell2009} revealed a new signpost for outflows in high-mass star formation regions in the form of extended emission which is bright in the 4.5-$\rm{\mu m}$ band.  These objects are generally referred to either as extended green objects (EGOs) \citep{Cyganowski2008} or green fuzzies \citep{Chambers2009}.  The enhanced emission in this wavelength range is believed to be due to shock-excited H$_2$ and/or CO band-head emission \citep{DeBuizer2010}.  \citet{Cyganowski2008} found that many EGOs are associated with 6.7GHz methanol masers, while \citet{Chen2009} showed a high rate of association with shock-excited class I methanol masers at 44 and 95 GHz.  Sensitive, high resolution searches for class II methanol masers towards a small sample of EGOs achieved a detection rates of $64 \%$ (although this should be considered an upper limit since most targets had known 6.7GHz methanol masers in their vicinity), with approximately $90 \%$ of these sources also having associated 44GHz class I methanol maser emission \citep{Cyganowski2009}.  These results demonstrate a close association between methanol masers and young high-mass stars with active outflows.

Molecular outflows are more visible than the YSO or its disk, and because of the association of 6.7GHz methanol masers with massive star formation, searching for outflows toward these masers and studying their physical properties can reveal information regarding the obscured massive cores they are associated with.  Moreover, by selecting outflows that are only associated with methanol masers, deliberately biases the resulting sample towards a narrower, relatively well-defined evolutionary range which allows constraints to be placed on the ``switch-on'' of the outflows and the study of their temporal development. In this paper we focus on the study of the physical properties of the outflows and the relationship of these properties with those of their embedding clumps. In a following publication (de Villiers et al 2014b, in prep.) we will explore the effects of the maser selection bias in our sample and the resulting behaviour in the dynamical ages of our maser selected sample.

We present a survey of $\rm{^{13}CO(J=3-2)}$ outflows toward a sample of 6.7GHz Methanol Multibeam (MMB) masers \citep[Breen et al. 2014 in prep.,][]{Green2009} using the HARP instrument on the JCMT.  Observations and data reduction are described in \S\ref{sec:observations}.  In \S\ref{sec:data_analysis} we describe the extraction and analysis of the spectra, as well as outflow mapping and outflow detection frequency.  The results are presented in \S\ref{sec:Results}, where we demonstrate the calculation of the outflows' physical properties and associated clump masses.  The relation between the outflow and associated clump properties are examined, and compared with some low-mass relations found in the literature.  We also inspect the correlation between outflow and 6.7GHz maser luminosities, as well as between maser luminosity and clump masses, as a probe of the relationship between the physical properties of the driving force, outflow and associated maser. The main results are summarized in \S\ref{sec:Conclusion}.  

Although the study of the properties of massive molecular outflows and their relation with associated clump masses is not novel \textit{per se}, the selection of the sources in this study is unique in terms of association with 6.7GHz masers.  This allows the selection of sources within a relatively well defined evolutionary phase, which potentially could limit the scatter in parameter space compared to previous work.  In this paper, we discuss and investigate the physical properties of the Methanol Maser Associated Outflows (MMAOs), and put them in context with other studies.  In a second forthcoming paper, we discuss the effect and implications of the 6.7GHz maser bias of our sample on our results.

\section{Observations and data reduction}
\label{sec:observations}

\begin{figure}
	\centering
	\includegraphics[width=0.85\columnwidth]{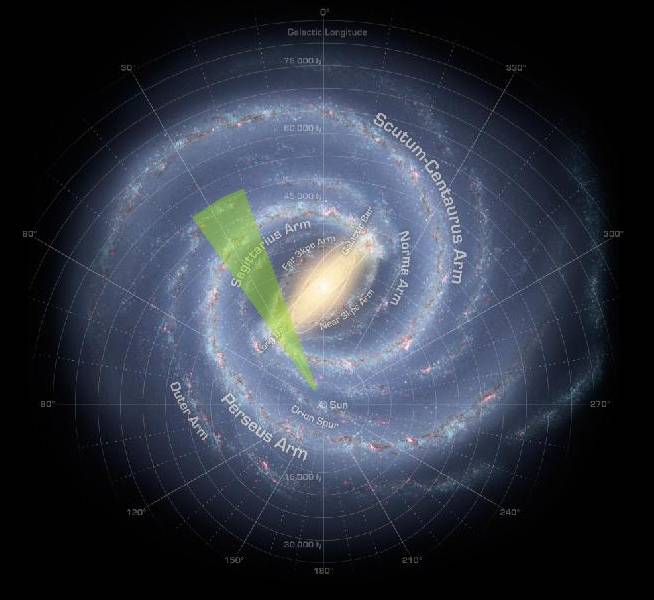}
	\caption{The shaded triangle indicates the approximate area from where the 6.7GHz methanol maser sample were selected for this study. The background sketch is by R. Hurt and R. Benjamin \citep{Churchwell2009}, and shows how the Galaxy is likely to appear face-on, based on radio, infrared and optical data.}
	\label{fig:methanolmaserregion}
	\end{figure}

A sample of 6.7GHz methanol masers were drawn from a preliminary catalogue of Northern Hemisphere masers from the Methanol Multibeam (MMB) Survey which has sub-arcsec positional accuracies \citep{Green2009}. The properties of these masers are described fully in  Breen et al.~(2014 in prep.).  The initial sample selection was chosen to have an even spread in maser luminosity, distance, association with UCH{\sc ii} regions and IR sources.  A sample of 70 sources were observed between $20^{\circ} < l < 34^{\circ}$.  

The targets were observed with the JCMT, on the summit of Mauna Kea, Hawaii on seven nights between 17 May 2007 and 22 July 2008. Targets were mapped in the $\rm{^{13}CO}$ and $\rm{C^{18}O (J = 3-2)}$ transitions (330.6 GHz and 329.3 GHz), using the 16-receptor Heterodyne Array Receiver Program (HARP). Only 14 of the 16 receptors were operational at the time of observation.  The receptors are laid out in a $4 \times 4$ grid separated by $30''$ and the beam size of the individual receptors at 345 GHz is $14''$.  All the data were corrected for a main-beam efficiency of $\eta\rm{_{mb} =0.66}$ \citep{Smith2008,Buckle2009}.  A HARP jiggle map \citep{Buckle2009} produces a fully sampled, 16-point rectangular map with a pixel scale of $6''$ and a spectral resolution of 0.06 $\rm{km s^{-1}}$.  The field-of-view is approximately $2' \times 2'$.  As the typical distance to the methanol maser target sources is $> 2$kpc, and with an estimated maser lifetime of $2.5-4 \times 10^4$ years \citep{vanderWalt2005}, the expected outflows from the maser-associated YSOs should be sampled in a single JCMT HARP jiggle-map. The pointing accuracy of the JCMT is of order $2''$ or better. Pointing checks were carried out regularly during observation runs to ensure and maintain accuracy.

The weather during the observations was mostly in JCMT-defined band 3, which implies a sky zenith opacity $\tau_{225}$ varying between 0.08 and 0.12 at 225 GHz as measured by the Caltech Submillimeter Observatory tipping radiometer \footnote{http://www.jach.hawaii.edu/weather/opacity/mk/}.

Out of the 70 observed maser coordinates, 16 observations did not meet one or more of the quality thresholds due to (a) too low signal-to-noise (less than $\sim 2$), (b) non-functioning  receptors (report unreliable temperatures), or (c) target positioning too close to the field-of-view border or a dead receptor.  The remaining 54 target coordinates are listed in Table \ref{tab:targetcoords} and occur in the shaded area in Figure \ref{fig:methanolmaserregion}.  

The $\rm{^{13}CO}$ and $\rm{C^{18}O}$ maps were simultaneously obtained using the multiple sub-band mode of the back-end Auto-Correlation Spectral Imaging System (ACSIS) \citep{Dent2000}.  The raw ACSIS data are in a HARP time series cube, giving the response of the receptors (x-axis) as a function of time (y-axis).  The third dimension is the velocity spectrum recorded at the time for that receptor.  Data were reduced with the Starlink ORAC-DR pipeline \citep{Cavanagh2008} using the $\rm{REDUCE\_SCIENCE\_NARROWLINE}$ recipe with minor modifications tailored for this dataset\footnote{http://www.oracdr.org}.  The pipeline reduction process automatically fits and subtracts polynomial baselines.  This was followed by truncation of the noisy spectral endpoints, removal of interference spikes and rebinning of the spectrum to a resolution of 0.5 $\rm{km s^{-1}}$.  Any receptors with high baseline variations compared to the bulk of the spectra, were flagged as bad in addition to those masked out by the pipeline.  Lastly the time series were then mapped onto a position-position-velocity (ppv) cube.  The reduced data antenna temperature ($T_A$) had an average RMS noise level of 0.24 K (per $6`` \times 6`` \times 0.5~\rm{km s^{-1}}$ pixel), or a main-beam efficiency corrected average RMS noise level of $T_{\rm{mb}}=0.36$ K.

\onecolumn
\begin{tiny}
\begin{table}
\centering
\caption{Complete list of 6.7GHz methanol maser coordinates used as pointing targets, including target names.  Suffixes ``A'' and ``B'' indicate separate clumps if more than one are detected.  The clump coordinates from where spectra were extracted are listed. Sources marked with * had their spectra extracted at the maser coordinate itself. The last column lists the noise RMS, integrated over the number of channels $n_{\rm{chan}}$ in each $\rm{C^{18}O}$ integrated map ($\phi = \sigma_{\rm{\small{RMS}}} \Delta v \sqrt{n_{\rm{chan}}}$, for a channel width $\Delta v$ of $0.5~\rm{kms^{-1}}$). When clumps were truncated at the edge of a map, or signal-to-noise was too low for significant $\rm{C^{18}O}$ detection, it is indicated.} \label{tab:targetcoords} 

\begin{tabular}{llrlrc} 
\hline
\multicolumn{1}{c}{Target name} & \multicolumn{2}{c}{Maser coord.} & \multicolumn{2}{c}{Clump coord.} & $\phi$ \\
{} & \multicolumn{1}{c}{$l (^{\circ})$} & \multicolumn{1}{c}{$b (^{\circ})$} & \multicolumn{1}{c}{$l (^{\circ})$} & \multicolumn{1}{c}{$b (^{\circ})$} & \multicolumn{1}{c}{K $\rm{km s^{-1}}$} \\\hline 
\hline

G 20.081-0.135 & 20.081 & -0.135 & 20.081 & -0.135 & 1.1 \\
G 21.882+0.013 & 21.882 & 0.013 & 21.875 & 0.008 & 0.9 \\
G 22.038+0.222 & 22.038 & 0.222 & 22.040 & 0.223 & 1.7 \\
G 22.356+0.066 & 22.356 & 0.066 & 22.356 & 0.068 & 2.0 \\
G 22.435-0.169 & 22.435 & -0.169 & 22.435 & -0.169 & 1.3 \\
G 23.003+0.124 & 23.003 & 0.124 & 23.002 & 0.126 & 1.1 \\
G 23.010-0.411 & 23.010 & -0.411 & 23.008 & -0.410 & 2.0 \\
G 23.206-0.378 & 23.206 & -0.378 & 23.209 & -0.378 & 1.0 \\
G 23.365-0.291 & 23.365 & -0.291 & 23.364 & -0.291 & 1.1 \\
G 23.437-0.184 & 23.437 & -0.184 & 23.436 & -0.183 & 1.4 \\
G 23.484+0.097 & 23.484 & 0.097 & 23.483 & 0.098 & 0.9 \\
G 23.706-0.198 & 23.706 & -0.198 & 23.706 & -0.197 & 1.3 \\
G 24.329+0.144 & 24.329 & 0.144 & 24.330 & 0.145 & 1.4 \\
G 24.493-0.039 & 24.493 & -0.039 & 24.493 & -0.039 & 1.4 \\
G 24.790+0.083A & 24.790 & 0.083 & 24.790 & 0.083 & 1.6 \\
G 24.790+0.083B & 24.790 & 0.083 & 24.799 & 0.097 & {cut off} \\
G 24.850+0.087 & 24.850 & 0.087 & 24.853 & 0.085 & 0.9 \\
G 25.650+1.050 & 25.650 & 1.050 & 25.649 & 1.051 & 1.2 \\
G 25.710+0.044 & 25.710 & 0.044 & 25.719 & 0.051 & 1.0 \\
G 25.826-0.178 & 25.826 & -0.178 & 25.824 & -0.179 & 1.2 \\
G 28.148-0.004 & 28.148 & -0.004 & 28.148 & -0.004 & 0.8 \\
G 28.201-0.049 & 28.201 & -0.049 & 28.201 & -0.049 & 1.0 \\
G 28.282-0.359 & 28.282 & -0.359 & 28.289 & -0.365 & 0.6 \\
G 28.305-0.387 & 28.305 & -0.387 & 28.307 & -0.387 & 0.8 \\
G 28.321-0.011 & 28.321 & -0.011 & 28.321 & -0.011 & 0.8 \\
G 28.608+0.018 & 28.608 & 0.018 & 28.608 & 0.018 & 0.7 \\
G 28.832-0.253 & 28.832 & -0.253 & 28.832 & -0.253 & 1.3 \\
G 29.603-0.625 & 29.603 & -0.625 & 29.600 & -0.618 & 1.1 \\
G 29.865-0.043 & 29.865 & -0.043 & 29.863 & -0.045 & 1.6 \\
G 29.956-0.016A & 29.956 & -0.016 & 29.956 & -0.017 & 1.6 \\
G 29.956-0.016B & 29.956 & -0.016 & 29.962 & -0.008 & 1.6 \\
G 29.979-0.047 & 29.979 & -0.047 & 29.979 & -0.048 & 1.7 \\
G 30.317+0.070* & 30.317 & 0.070 & 30.317 & 0.070 & 1.0 \\
G 30.370+0.482A & 30.370 & 0.482 & 30.370 & 0.484 & 0.6 \\
G 30.370+0.482B & 30.370 & 0.482 & 30.357 & 0.487 & {low S/N} \\
G 30.400-0.296 & 30.400 & -0.296 & 30.403 & -0.296 & 0.8 \\
G 30.419-0.232 & 30.419 & -0.232 & 30.420 & -0.233 & 1.1 \\
G 30.424+0.466 & 30.424 & 0.466 & 30.424 & 0.464 & 0.5 \\
G 30.704-0.068 & 30.704 & -0.068 & 30.701 & -0.067 & 1.2 \\
G 30.781+0.231 & 30.781 & 0.231 & 30.780 & 0.231 & 1.2 \\
G 30.788+0.204 & 30.788 & 0.204 & 30.789 & 0.205 & 1.4 \\
G 30.819+0.273 & 30.819 & 0.273 & 30.818 & 0.273 & 1.2 \\
G 30.851+0.123 & 30.851 & 0.123 & 30.865 & 0.114 & 1.2 \\
G 30.898+0.162 & 30.898 & 0.162 & 30.899 & 0.163 & 1.0 \\
G 30.973+0.562 & 30.973 & 0.562 & 30.972 & 0.561 & 1.2 \\
G 30.980+0.216 & 30.980 & 0.216 & 30.979 & 0.216 & 1.3 \\
G 31.061+0.094 & 31.061 & 0.094 & 31.060 & 0.092 & 0.9 \\
G 31.076+0.457 & 31.076 & 0.457 & 31.085 & 0.468 & 1.1 \\
G 31.122+0.063 & 31.122 & 0.063 & 31.124 & 0.063 & 1.0 \\
G 31.182-0.148A* & 31.182 & -0.148 & 31.182 & -0.148 & 1.2 \\
G 31.182-0.148B & 31.182 & -0.148 & 31.173 & -0.146 & {cut off} \\
G 31.282+0.062 & 31.282 & 0.062 & 31.281 & 0.063 & 0.9 \\
G 31.412+0.307 & 31.412 & 0.307 & 31.412 & 0.306 & 1.0 \\
G 31.594-0.192 & 31.594 & -0.192 & 31.593 & -0.193 & 1.2 \\
G 32.744-0.075 & 32.744 & -0.075 & 32.746 & -0.076 & 1.1 \\
G 33.317-0.360* & 33.317 & -0.360 & 33.317 & -0.360 & {low S/N} \\
G 33.486+0.040* & 33.486 & 0.040 & 33.486 & 0.040 & {low S/N} \\
G 33.634-0.021 & 33.634 & -0.021 & 33.649 & -0.024 & 1.4 \\

\hline
\end{tabular}
\end{table}
\end{tiny}
\twocolumn

\section{Data analysis}
\label{sec:data_analysis}

\subsection{Finding the peak emission}
\label{sec:clumpfind}

$\rm{^{13}CO}$ was used as an outflow tracer in this study.  It is a useful probe of the cloud structure and kinematics, because it traces the higher velocity gas, but has a lower abundance than $\rm{^{12}CO}$, and hence is less contaminated by other high velocity structures within the star forming complex \citep{Arce2010}.  Emission from the $\rm{(J=3-2)}$ transition was observed ($T\rm{_{trans}}=31.8$ K,\citet{Curtis2010}), which traces the warm, dense gas, close to the embedded YSO and also serves as a clearer tracer of warm outflow emission than lower J transitions.  Targets were simultaneously observed in the optically thin $\rm{C^{18}O}$ transition, which serves as a useful tracer of the column density.  The $\rm{C^{18}O}$ emission peak is most likely to coincide with the YSO core's position.  

Visual inspection indicated that the positions of peak emission in both $\rm{^{13}CO}$ and $\rm{C^{18}O}$ did not always coincide with the maser coordinate.  These offsets were larger than a beam size ($14''$) for seven maser coordinates, with a maximum offset of $1'$.  Although it is known that 6.7GHz methanol masers are mostly associated with massive YSOs, two competing and unresolved formation hypotheses state that either methanol masers are embedded in circumstellar tori or accretion disks around the massive protostars \citep{Pestalozzi2009}, or that they generally trace outflows \citep{DeBuizer2012}.  It thus seems possible that although the masers are in the close vicinity of the YSO, some could be offset, as was found in this study. 

Since the peak $\rm{^{13}CO}$ emission did not always coincide with the maser coordinates, and the exact coordinate of the peak emission was needed as the position from where the one dimensional spectrum would be extracted as part of the outflow detection method, an alternative method was needed to pin-point this position.  \textit{ClumpFind} \citep{Williams1994}, also used by \citet{Moore2007,Buckle2010,Parsons2012} was used to carry out a consistent search for the position of peak emission in this study. The search was undertaken on two dimensional images, intensity integrated over the emission peaks' velocity ranges.  

In a few cases \textit{ClumpFind} reported more than one clump coordinate per image, likely due to the irregular structure and crowded environment of massive star-forming regions.  The purpose of using \textit{ClumpFind} in this study was to find the position of the peak molecular emission in the vicinity of each methanol maser target. Multiple clumps were accepted if they were further than a beam width apart and not close to the edges of the image.  Multiple spectra per field-of-view were extracted at these positions.  In four cases (marked with asterisk in Table \ref{tab:targetcoords}) \textit{ClumpFind} did not detect any clumps, either due to low signal-to-noise or the physical area of the emission being too small to satisfy the \textit{ClumpFind} criteria (minimum seven pixels).  In these cases we did detect some emission at the maser coordinate, hence we used the maser coordinates as the location for spectrum extraction.

Where clumps were detected close to a dead receptor or the edge of the map, they were rejected from further analysis, as any extracted spectra and derived results will be incomplete.  This is the case for the maps of the targets associated with masers G 24.790+0.083 (clump 2), G 30.851+0.123 and G 31.182-0.148 (clump 2).   

Of the original 70 targets observed, reliable clump detections were obtained in 54 maps ($77\%$), and because more than one clump was found in some images, a total of 58 positions were analysed.  The positions of the observed clumps are summarized in Table \ref{tab:targetcoords}.  Intensity integrated spatial maps were created for these targets, and are shown online in Appendix A, where the integrated $\rm{C^{18}O}$ emission is contoured over the background of $\rm{^{13}CO}$ emission, the latter integrated over velocity ranges $v_{\rm{low}}$ to $v_{\rm{high}}$, listed in Table \ref{tab:targetvelocities}.  

Contour intervals are shown in steps of the integrated noise RMS, $\phi$, where $\phi = \sigma_{\rm{\small{RMS}}} \Delta v \sqrt{n_{\rm{chan}}}$, calculated for the same number of channels, $n_{\rm{chan}}$, over which the $\rm{C^{18}O}$ image is integrated, with $\sigma_{\rm{\small{RMS}}}$ the average RMS per channel and $\Delta v = 0.5~\rm{kms^{-1}}$ the velocity range per channel.  All values are listed in Table \ref{tab:targetcoords}.  Where contour intervals are larger than $2\phi$, this is indicated in the figure captions. The lowest level contour was manually selected for each image, because every image has a unique signal-to-noise and background. The lowest contour levels ranged between $3\phi$ and $14\phi$.  Both maser and clump coordinates are indicated on the maps, shown respectively as a star and circle symbol. 

Sometimes, $\rm{C^{18}O}$ noise levels were excessive due to poor atmospheric transmission, as this line is at the edge of the atmospheric window, which makes it susceptible to small changes in the water vapour column. Targets associated with masers G 30.370+0.482 (clump 2), G 31.182-0.148 (clump 1), G 33.317-0.360 and G 33.486+0.040 did not show sufficiently high signal-to-noise to isolate any clump emission above a $3\times \phi$ threshold. For these targets, the $\rm{C^{18}O}$ maps are not shown in Appendix A.

\subsection{Spectrum extraction and wing selection}
\label{sec:spectra}

\begin{figure}
		\begin{center}$
		\begin{array}{c}
		\includegraphics[width = 0.4\textwidth,clip]{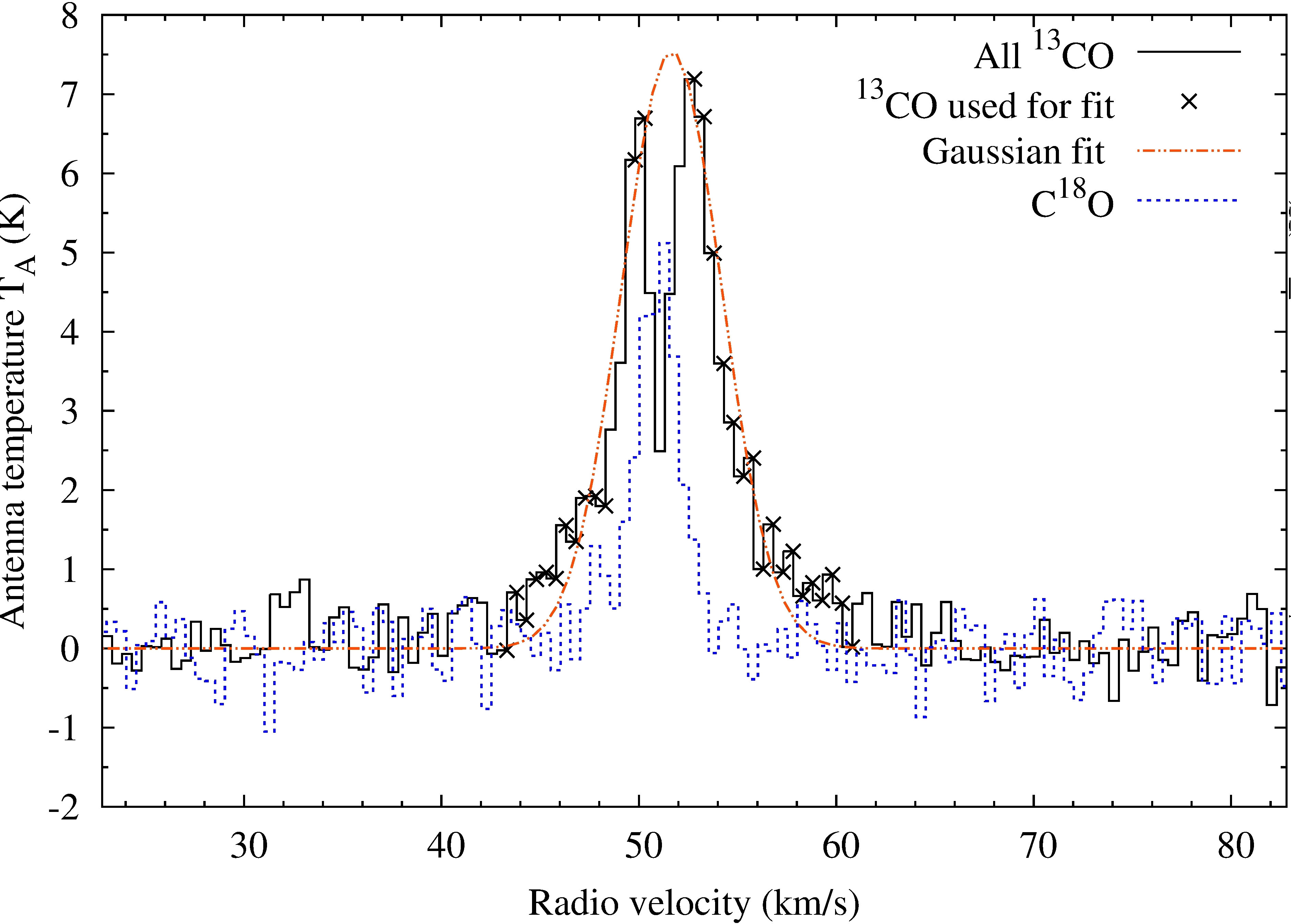} \\
		\includegraphics[width = 0.4\textwidth,clip]{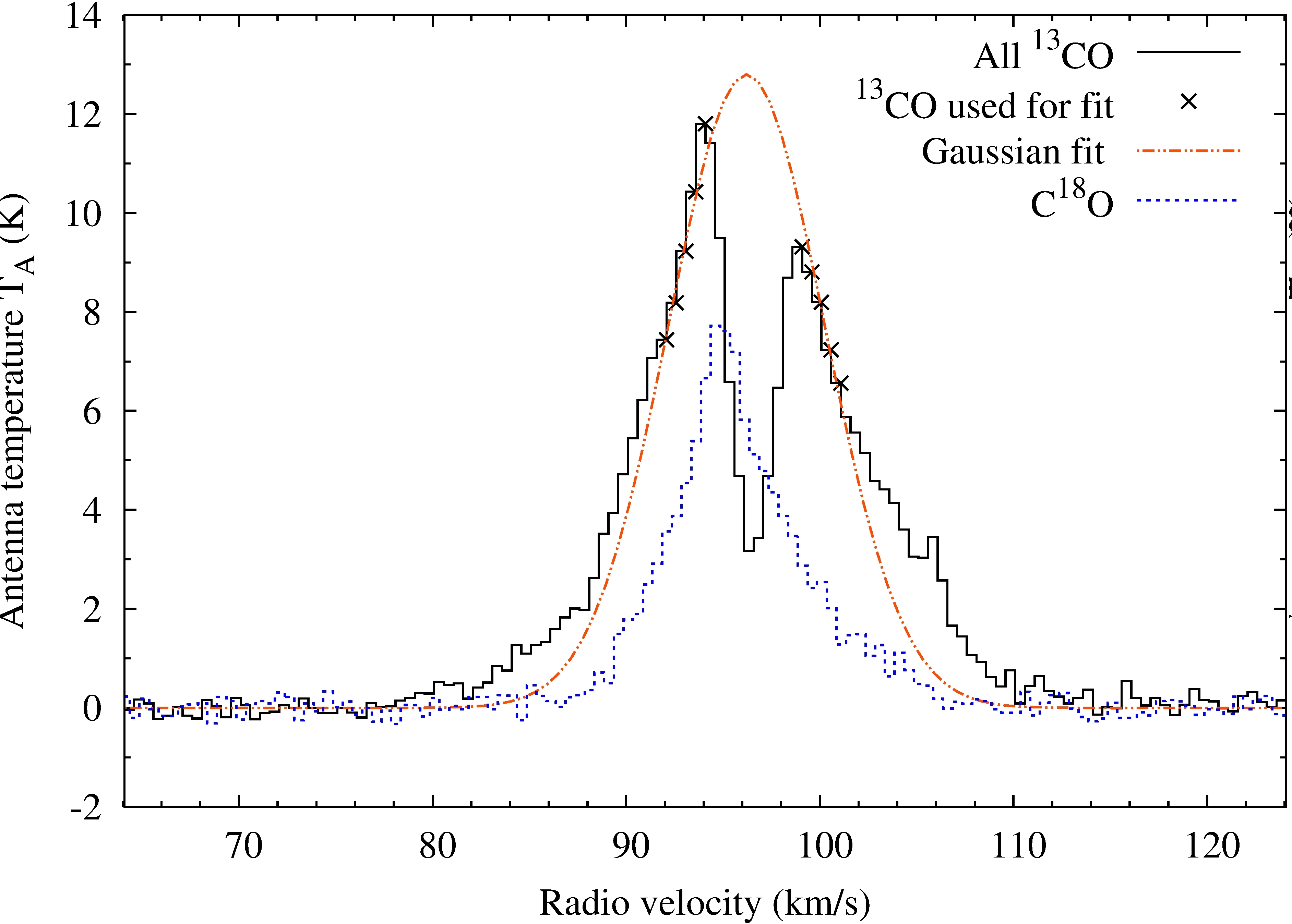} \\
		\end{array}$
		\end{center}
		\caption{ \small{Gaussian fits (dot-dot-dashed lines) to the shoulders of $\rm{^{13}CO}$ spectra (crosses) toward G 22.038+0.222 (top) and G 28.201-0.049 (bottom), whose profiles show clear evidence of self absorption. The Gaussian's peak is used as the estimated peak temperature. The $\rm{C^{18}O}$ spectra (short dashed lines) give an indication where the actual peak is expected.}}
		\label{fig:Gauss13CO}
\end{figure}

After locating the emission peak, the spectrum was extracted at this position from both the $\rm{^{13}CO}$ and $\rm{C^{18}O}$ cubes. Table \ref{tab:targetvelocities} lists the maser median velocity taken from the literature, or its associated IRDC velocity if the former was not available \citep{Simon2006} and literature references of these values, for each target.  It also gives the measured peak main-beam efficiency corrected temperatures $(T_{\rm{mb}})$ and corresponding velocities for both $\rm{^{13}CO}$ and $\rm{C^{18}O}$.  Sources marked with an asterisk exhibit self absorption dips in their $\rm{^{13}CO}$ spectra.  For these spectra, a Gaussian was fitted to the shoulders of the absorbed spectrum and the peak of this resultant profile was used as the estimate of peak temperature.  The peak from the Gaussian fit showed on average a $\sim 13\%$ increase with respect to the peak $T_{\rm{mb}}$ of the original, absorbed spectrum, with three extreme cases of a $30\%-40\%$ increase.  Two examples of these sources and their Gaussian fits are shown in Figure \ref{fig:Gauss13CO}. Plotted $\rm{C^{18}O}$ spectra give an indication where the optically thin peak is expected. In the case of the double-peaked target G 23.010-0.411, the values marked with an asterisk in Table \ref{tab:targetvelocities} represent peak values of fits to the individual peaks. The values of $\Delta v_{\rm{b}}$ and $\Delta v_{\rm{r}}$ in columns 9 and 10 are the blue and red velocities relative to the peak $\rm{C^{18}O}$ velocity, measured respectively from each wing extreme, to be discussed in \S\ref{sec:calculations}.  The use of $Int\rm{_{b}}$ and $Int\rm{_{r}}$ in columns 11 and 12 will be discussed in \S\ref{sec:contours}.

\setlength{\tabcolsep}{2pt}
\onecolumn
\begin{scriptsize}

\textbf{Table 2.} Literature $v_{\rm{lsr}}$ velocities (median velocity for 6.7GHz maser or associated IRDC or molecular cloud if no maser velocity is available) associated with each target. Observed peak $\rm{C^{18}O}$ and $\rm{^{13}CO}$ $v_{\rm{lsr}}$ velocities with corresponding temperatures as derived from the each target spectrum's peak antenna temperature at the clump coordinate.  These antenna temperatures are corrected for the main-beam efficiency ($\eta = 0.66$).  Temperatures marked with * are the peaks of Gaussians fitted to spectrum profiles excluding velocity ranges showing strong self-absorption in $\rm{^{13}CO}$, while for double-peaked G 23.010-0.411, they represent fits to the individual peaks (peak 1 indicated by ``(pk.1)'' and peak 2 by ``(pk.2)'').  The velocities over which the $\rm{^{13}CO}$ profile is integrated to obtain the background emission shown in Appendixes A and B, are chosen to include all emission and are given by $v_{\rm{low}}$ and $v_{\rm{high}}$. Where maximum integrated intensities $Int_{\rm{b}}$ and $Int_{\rm{r}}$ are available for respectively blue and red $\rm{^{13}CO}$ integrated maps (corrected for main-beam efficiency), they are listed.  For monopolar outflows, only one value is given.  These intensities are used to determine contour intervals in Appendix B, published online.  $\Delta v_{\rm{b}}$ and $\Delta v_{\rm{r}}$ are used in \S\ref{sec:calculations}, equations \ref{eq:momentum}, \ref{eq:Energy} and \ref{eq:timescale}.  These are the velocity extents measured from the peak velocity (as defined by $\rm{C^{18}O}$) to the maximum velocity along the blue or red $\rm{^{13}CO}$ line wing (as defined in the text). \\
\end{scriptsize}
\begin{tiny}
\begin{center}

\begin{longtable}{lrcrrrccrrrr} 
\caption*{} \label{tab:targetvelocities} \\

\hline \\
\multicolumn{1}{c}{Target} & \multicolumn{1}{c}{Maser $v$} & \multicolumn{1}{c}{Vel. Ref.} & \multicolumn{1}{c}{$\rm{C^{18}O}$ $v_p$} & \multicolumn{1}{c}{$\rm{C^{18}O}$ $T_{\rm{mb}}$} & \multicolumn{1}{c}{$\rm{^{13}CO}$ $v_p$} & \multicolumn{1}{c}{$\rm{^{13}CO}$ $T_{\rm{mb}}$} & ($v_{\rm{low}} \rightarrow v_{\rm{high}}$) & \multicolumn{1}{c}{$\Delta v_{\rm{b}}$} & \multicolumn{1}{c}{$\Delta v_{\rm{r}}$} & \multicolumn{1}{c}{$Int_{\rm{b}}$} & \multicolumn{1}{c}{$Int_{\rm{r}}$}\\

{} & \multicolumn{1}{c}{($\rm{km s^{-1}}$)} & {} & \multicolumn{1}{c}{($\rm{km s^{-1}}$)} & (K) & \multicolumn{1}{c}{($\rm{km s^{-1}}$)} & \multicolumn{1}{c}{(K)} & \multicolumn{1}{c}{($\rm{km s^{-1}}$)} & \multicolumn{1}{c}{($\rm{km s^{-1}}$)} & \multicolumn{1}{c}{($\rm{km s^{-1}}$)} & \multicolumn{1}{c}{($\rm{Kkm s^{-1}}$)} & \multicolumn{1}{c}{($\rm{Kkm s^{-1}}$)} \\\hline \\ 
\endhead

\hline\\

\multicolumn{12}{l}{First column supercripts: $b =$ blue lobe only, $r =$ red lobe only.  Second column superscripts: $m =$ mid-line velocity, $p =$ peak-velocity, $c =$ cloud velocity.} \\
\multicolumn{12}{l}{1. \citet{Green2011}, 2. \citet{Roman2009}, 3. \citet{Szymczak2012}, 4. \citet{Simon2006}} \\ 
\hline
\endlastfoot

G 20.081-0.135 & $ 43.8^m$ & 1 & 41.6 & 9.1 & 42.3 & 14.4* & ( 20 $\rightarrow$ 60 ) & 11.1 & 10.9 & 35.0 & 48.4 \\
G 21.882+0.013 & $ 20.7^m$ & 1 & 20.2 & 4.0 & 19.8 & 11.3 & ( 10 $\rightarrow$ 35 ) & 4.5 & 6.5 & 20.3 & 9.7 \\
G 22.038+0.222 & $ 50.4^m$ & 1 & 51.5 & 7.8 & 51.7 & 11.5* & ( 40 $\rightarrow$ 70 ) & 9.3 & 8.7 & 20.8 & 35.8 \\
G 22.356+0.066 & $ 82.4^m$ & 1 & 84.2 & 5.8 & 84.4 & 11.5 & ( 75 $\rightarrow$ 95 ) & 5.0 & 3.5 & 13.9 & 4.3 \\
G 22.435-0.169 & $ 31.2^m$ & 1 & 27.9 & 3.3 & 27.8 & 7.1 & ( 20 $\rightarrow$ 40 ) & 2.0 & 4.5 & 10.6 & 5.2 \\
G 23.003+0.124$^r$ &  - & - & 107.4 & 3.5 & 108.5 & 4.7 & ( 102 $\rightarrow$ 112 ) & - & 4.0 & - & 7.7 \\
G 23.010-0.411$\rm{_{pk.1}}$ & $ 77.7^m$ & 1 & 76.4 & 4.0 & 75.6 & 9.2* & ( 60 $\rightarrow$ 90 ) & 11.5 & 11.0 & 37.7 & 27.0 \\
G 23.010-0.411$\rm{_{pk.2}}$ &  - & - & 78.4 & 3.8 & 79.6 & 9.4* & ( 60 $\rightarrow$ 90 ) & - & - & - & {} \\
G 23.206-0.378 & $ 80.3^m$ & 1 & 77.8 & 4.0 & 77.6 & 5.3* & ( 65 $\rightarrow$ 95 ) & 12.0 & 11.5 & 18.0 & 18.9 \\
G 23.365-0.291 & $ 77.3^c$ & 4 & 78.3 & 4.1 & 77.8 & 5.9* & ( 72 $\rightarrow$ 85 ) & 3.9 & 4.6 & 8.2 & 11.7 \\
G 23.437-0.184 & $ 101.5^m$ & 1 & 100.6 & 8.0 & 101.2 & 12.7 & ( 90 $\rightarrow$ 115 ) & 14.5 & 9.0 & 31.7 & 46.1 \\
G 23.484+0.097 & $ 87.2^m$ & 1 & 84.2 & 5.7 & 84.2 & 6.7* & ( 75 $\rightarrow$ 95 ) & 4.1 & 7.4 & 11.1 & 10.0 \\
G 23.706-0.198 & $ 76.8^m$ & 1 & 69.1 & 5.4 & 68.3 & 8.3 & ( 60 $\rightarrow$ 80 ) & 3.5 & 5.5 & 17.2 & 13.7 \\
G 24.329+0.144 & $ 115.4^m$ & 1 & 112.7 & 4.0 & 112.8 & 7.9 & ( 105 $\rightarrow$ 130 ) & 10.0 & 6.5 & 15.7 & 8.0 \\
G 24.493-0.039 & $ 114.0^m$ & 1 & 111.8 & 6.1 & 109.8 & 14.2 & ( 100 $\rightarrow$ 120 ) & 6.5 & 7.5 & 33.7 & 17.9 \\
G 24.790+0.083A & $ 111.3^m$ & 1 & 110.5 & 9.3 & 111.6 & 15.8 & ( 100 $\rightarrow$ 125 ) & 7.0 & 6.5 & 15.1 & 30.9 \\
G 24.790+0.083B &  - & - & 110.5 & 6.4 & 111.1 & 11.5 & ( 100 $\rightarrow$ 120 ) & 9.5 & 9.5 & {} & {} \\
G 24.850+0.087 & $ 52.6^m$ & 1 & 108.9 & 7.3 & 109.0 & 14.6 & ( 105 $\rightarrow$ 115 ) & 4.0 & 4.0 & 19.2 & 14.2 \\
G 25.650+1.050 & $ 40.6^m$ & 1 & 42.3 & 8.7 & 43.1 & 18.0* & ( 35 $\rightarrow$ 55 ) & 12.5 & 10.5 & 32.5 & 45.3 \\
G 25.710+0.044 & $ 96.2^m$ & 1 & 101.2 & 4.2 & 101.3 & 14.7 & ( 95 $\rightarrow$ 110 ) & 9.0 & 2.5 & 32.7 & 25.6 \\
G 25.826-0.178 & $ 94.7^m$ & 1 & 93.2 & 6.5 & 91.8 & 10.9 & ( 80 $\rightarrow$ 105 ) & 8.0 & 10.0 & 22.2 & 13.8 \\
G 28.148-0.004 & $ 100.8^m$ & 1 & 98.7 & 6.4 & 99.0 & 8.0* & ( 90 $\rightarrow$ 115 ) & 7.7 & 5.8 & 12.7 & 12.5 \\
G 28.201-0.049 & $ 95.9^m$ & 1 & 94.9 & 11.7 & 96.2 & 19.4* & ( 78 $\rightarrow$ 115 ) & 15.6 & 15.4 & 62.3 & 87.1 \\
G 28.282-0.359 & $ 41.6^m$ & 1 & 47.4 & 10.6 & 49.1 & 17.9* & ( 40 $\rightarrow$ 55 ) & 9.3 & 4.7 & 29.9 & 35.8 \\
G 28.305-0.387 & $ 80.9^m$ & 1 & 85.6 & 8.0 & 85.9 & 27.5 & ( 78 $\rightarrow$ 95 ) & 5.5 & 3.0 & 37.1 & 50.3 \\
G 28.321-0.011 & $ 99.0^c$ & 4 & 99.6 & 5.2 & 99.8 & 12.4 & ( 85 $\rightarrow$ 110 ) & 8.0 & 4.5 & 22.7 & 17.0 \\
G 28.608+0.018 &  - & - & 103.1 & 9.0 & 103.8 & 22.3 & ( 90 $\rightarrow$ 115 ) & 11.5 & 7.0 & 38.1 & 25.8 \\
G 28.832-0.253 & $ 86.1^m$ & 1 & 87.2 & 6.1 & 88.4 & 10.6 & ( 72 $\rightarrow$ 110 ) & 8.0 & 11.0 & 20.2 & 35.1 \\
G 29.603-0.625$^r$ &  - & - & 77.2 & 4.0 & 76.9 & 7.2 & ( 70 $\rightarrow$ 90 ) & - & 4.0 & - & 11.0 \\
G 29.865-0.043 &  - & - & 101.8 & 9.0 & 101.1 & 19.6 & ( 90 $\rightarrow$ 110 ) & 6.5 & 6.0 & 67.9 & 22.5 \\
G 29.956-0.016A & $ 99.9^m$ & 1 & 97.8 & 17.6 & 97.6 & 31.7 & ( 90 $\rightarrow$ 110 ) & 10.5 & 9.0 & 63.2 & 38.8 \\
G 29.956-0.016B & $ 99.9^m$ & 1 & 97.8 & 5.8 & 97.6 & 20.4 & ( 90 $\rightarrow$ 108 ) & 5.5 & 9.0 & 30.8 & 38.8 \\
G 29.979-0.047 & $ 101.7^m$ & 1 & 101.8 & 4.3 & 101.7 & 8.2* & ( 85 $\rightarrow$ 110 ) & 11.1 & 5.4 & 40.1 & 18.2 \\
G 30.317+0.070 & $ 42.6^m$ & 1 & 44.6 & 2.9 & 44.1 & 6.0 & ( 30 $\rightarrow$ 50 ) & 5.5 & 4.5 & 9.7 & 5.9 \\
G 30.370+0.482A &  - & - & 17.4 & 1.4 & 17.4 & 6.0 & ( 10 $\rightarrow$ 28 ) & 2.5 & 5.5 & 4.4 & 7.7 \\
G 30.370+0.482B &  - & - & 17.9 & 1.5 & 17.4 & 4.6 & ( 10 $\rightarrow$ 22 ) & 2.0 & 2.0 & 2.5 & 3.6 \\
G 30.400-0.296 & $ 101.7^m$ & 3 & 103.0 & 3.9 & 102.7 & 12.8 & ( 90 $\rightarrow$ 115 ) & 10.0 & 5.0 & 23.8 & 16.4 \\
G 30.419-0.232 & $ 102.8^m$ & 1 & 104.5 & 7.7 & 104.3 & 17.2 & ( 95 $\rightarrow$ 110 ) & 5.0 & 10.0 & 30.2 & 41.4 \\
G 30.424+0.466 & $ 9.5^m$ & 3 & 15.5 & 3.6 & 15.6 & 6.8 & ( 10 $\rightarrow$ 25 ) & 3.5 & 6.0 & 7.4 & 9.1 \\
G 30.704-0.068$^b$ & $ 88.9^m$ & 1 & 90.1 & 9.5 & 88.9 & 32.1 & ( 80 $\rightarrow$ 102 ) & 9.0 & {} & 50.6 & - \\
G 30.781+0.231 & $ 49.5^m$ & 1 & 41.9 & 3.7 & 42.3 & 10.3 & ( 30 $\rightarrow$ 55 ) & 4.0 & 2.5 & 11.2 & 17.6 \\
G 30.788+0.204 & $ 82.8^m$ & 1 & 81.6 & 5.4 & 82.3 & 8.0* & ( 70 $\rightarrow$ 90 ) & 7.3 & 6.2 & 17.0 & 23.5 \\
G 30.819+0.273$^r$ & $ 104.9^m$ & 1 & 98.1 & 3.0 & 98.1 & 7.8 & ( 90 $\rightarrow$ 110 ) & - & 5.5 & - & 10.2 \\
G 30.851+0.123 &  - & - & 39.4 & 4.8 & 40.4 & 15.6 & ( 30 $\rightarrow$ 50 ) & 6.5 & 5.5 & - & - \\
G 30.898+0.162$^r$ & $ 104.5^m$ & 1 & 105.3 & 4.2 & 105.8 & 9.5 & ( 100 $\rightarrow$ 115 ) & - & 2.5 & - & 13.0 \\
G 30.973+0.562 &  - & - & 23.4 & 3.6 & 23.5 & 9.2 & ( 10 $\rightarrow$ 30 ) & 3.0 & 3.0 & 23.6 & 9.8 \\
G 30.980+0.216$^r$ &  - & - & 107.4 & 3.2 & 107.1 & 7.2 & ( 100 $\rightarrow$ 120 ) & - & 4.5 & - & 8.4 \\
G 31.061+0.094 & $ 16.2^m$ & 1 & 19.2 & 1.9 & 17.7 & 13.6 & ( 10 $\rightarrow$ 25 ) & 4.0 & 5.0 & 19.0 & 1.0 \\
G 31.076+0.457$^b$ &  - & - & 28.3 & 1.9 & 24.5 & 5.8* & ( 15 $\rightarrow$ 30 ) & 5.1 & {} & 8.8 & - \\
G 31.122+0.063 &  - & - & 41.5 & 3.6 & 41.5 & 10.4 & ( 30 $\rightarrow$ 50 ) & 6.5 & 5.5 & 10.6 & 16.7 \\
G 31.182-0.148A &  - & - & 42.6 & 1.3 & 42.6 & 3.5 & ( 35 $\rightarrow$ 50 ) & 3.5 & 2.5 & 7.2 & 2.4 \\
G 31.182-0.148B &  - & - & 43.6 & 1.6 & 43.1 & 5.2 & ( 35 $\rightarrow$ 50 ) & 3.0 & 2.0 & - & - \\
G 31.282+0.062 & $ 108.0^m$ & 1 & 109.0 & 7.4 & 109.0 & 14.1* & ( 100 $\rightarrow$ 120 ) & 7.0 & 5.0 & 23.4 & 19.3 \\
G 31.412+0.307 & $ 96.7^m$ & 1 & 96.4 & 8.9 & 97.3 & 18.5* & ( 90 $\rightarrow$ 108 ) & 4.3 & 6.2 & 14.4 & 25.1 \\
G 31.594-0.192 &  - & - & 43.1 & 2.3 & 43.1 & 7.5 & ( 35 $\rightarrow$ 50 ) & 3.5 & 2.5 & 6.7 & 8.3 \\
G 32.744-0.075 & $ 34.8^m$ & 1 & 37.5 & 5.4 & 37.0 & 10.8 & ( 25 $\rightarrow$ 50 ) & 7.0 & 9.0 & 25.0 & 22.7 \\
G 33.317-0.360$^r$ &  - & - & 34.8 & 2.4 & 35.8 & 3.8 & ( 25 $\rightarrow$ 45 ) & - & 4.0 & - & 8.4 \\
G 33.486+0.040 &  - & - & 112.0 & 1.4 & 112.2 & 3.2 & ( 106 $\rightarrow$ 118 ) & 3.5 & 2.0 & 2.1 & 5.4 \\
G 33.634-0.021 & $ 105.900^m$ & 2 & 103.8 & 5.9 & 103.5 & 13.6 & ( 95 $\rightarrow$ 115 ) & 2.0 & 7.5 & 20.3 & 8.7 \\

\end{longtable}
\end{center}
\end{tiny}
\twocolumn
\setlength{\tabcolsep}{6pt}

\begin{figure}
		\begin{center}$
		\begin{array}{cc}
		\includegraphics[width = 0.4\textwidth,clip]{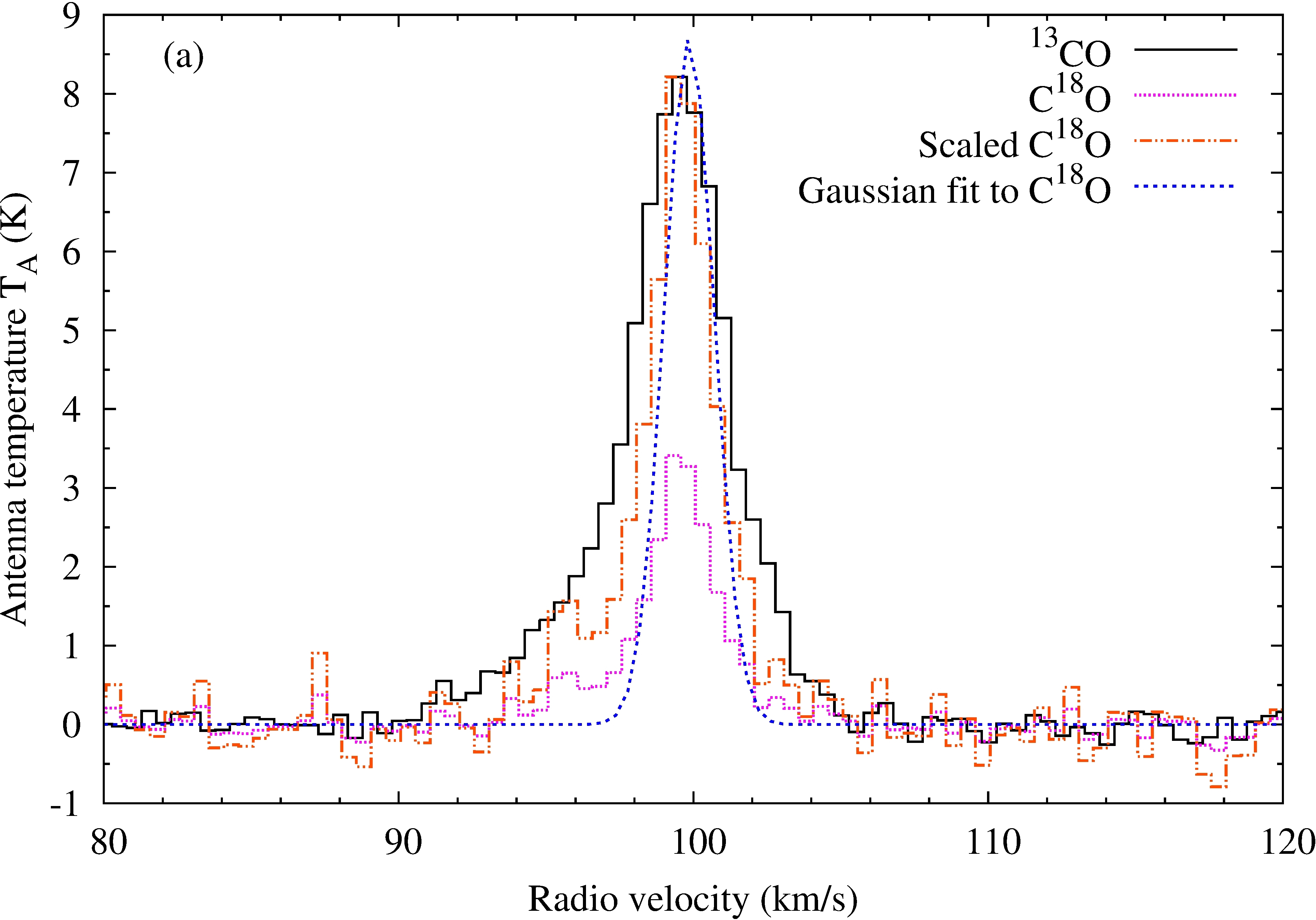} \\
		\includegraphics[width = 0.4\textwidth,clip]{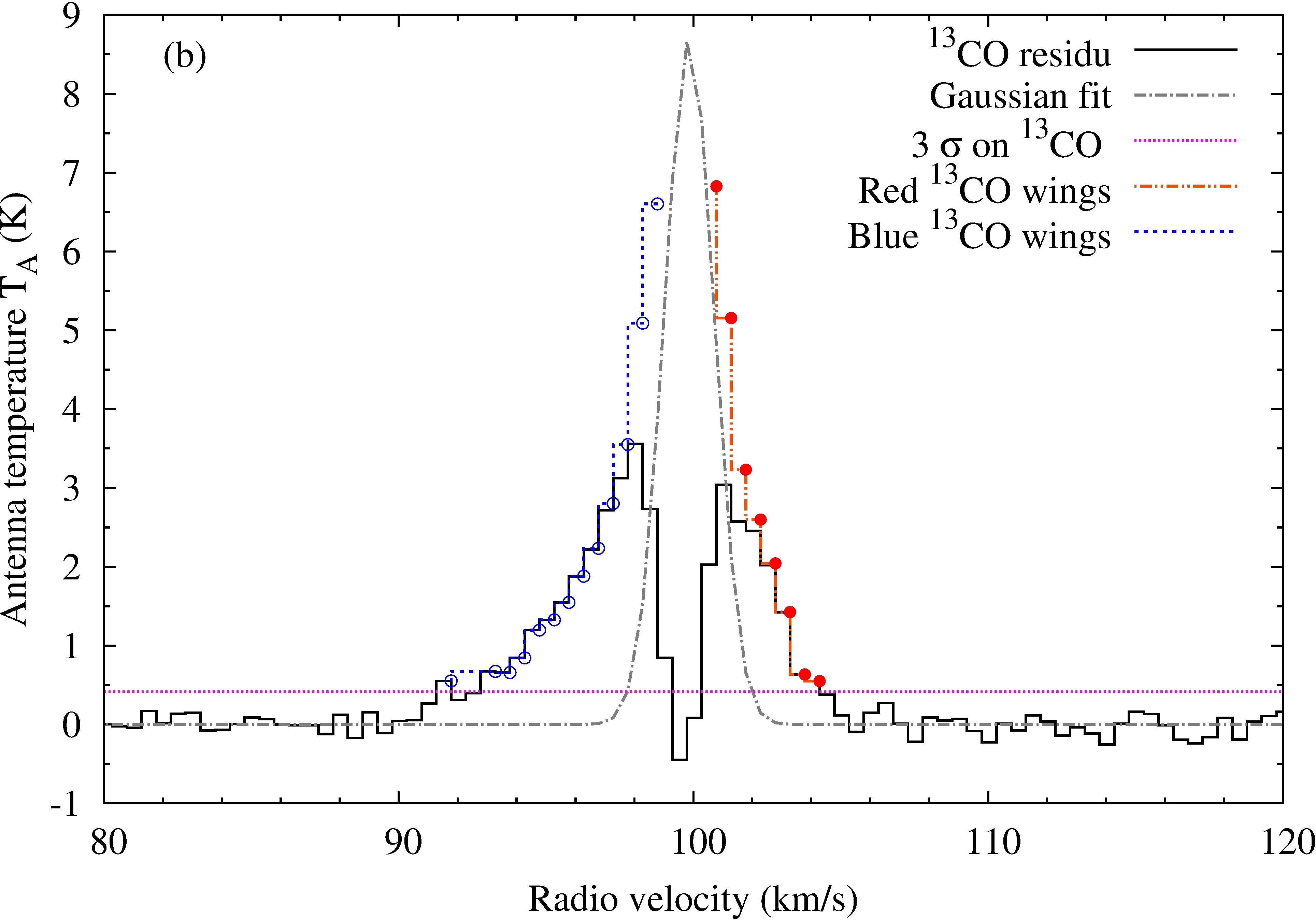} 
		\end{array}$
		\end{center}
		\caption{ \small{(a) Example of the $\rm{^{13}CO}$ spectrum (solid line) for the clump associated with maser G 28.321-0.011.  Its $\rm{C^{18}O}$ spectrum (dot-dashed line) is scaled to the $\rm{^{13}CO}$ peak (dot-dot-dashed line) and a Gaussian is fitted to the scaled spectrum (short dashed line). (b) The $\rm{^{13}CO}$ residuals following Gaussian subtraction is shown (solid line), along with the $3 \rm{\sigma}$ noise level (dotted line) and wing residuals satisfying the selection criteria. Blue wings are indicated by a short dashed line and empty circles, and red wings by a dot-dot-dashed line and solid circles.}}
		\label{fig:wingselection}
\end{figure}

Following \citet{Codella2004}, the optically thin $\rm{C^{18}O}$ profiles were used as tracers for the line cores of targets.  The $\rm{C^{18}O}$ spectra were scaled to the $\rm{^{13}CO}$ peak temperature.  To avoid subtracting any emission from higher velocity features that may be present in the $\rm{C^{18}O}$ if densities were sufficiently high, a Gaussian was fitted to the $\rm{C^{18}O}$ peak to approximate the line core-only emission.  This was done by gradually removing points from the outer (higher velocity) edges of the $\rm{C^{18}O}$ spectrum until the peak could be fitted, following the same approach as \citet{vanderWalt2007} (see Figure \ref{fig:wingselection} (a)).  The scaled Gaussian fit was then subtracted from the $\rm{^{13}CO}$ spectra to show the velocity ranges in the line wings where there is excess emission in $\rm{^{13}CO}$.

G 23.010-0.411 is a special case with a double peaked profile.  Assuming this is caused by two separate but closely associated clumps, we used two Gaussians, each fitted just to the highest velocity shoulder of each $\rm{C^{18}O}$ line peak.   Whenever absorption dips occur in the $\rm{^{13}CO}$ profiles, no natural profile peak existed.  Instead, the $\rm{C^{18}O}$ spectra were scaled to the the peak of the previous Gaussian fitted to the $\rm{^{13}CO}$.

This Gaussian was then subtracted from the $\rm{^{13}CO}$ profile.  The line wings are defined by the sections where the $\rm{^{13}CO}$ profile is broader than the scaled Gaussian representing the $\rm{C^{18}O}$ line core emission, \textit{provided} the $\rm{^{13}CO}$ corrected antenna temperature is higher than $\rm{3 \sigma}$ ($\rm{\sigma}$ is the noise per 0.5 $\rm{km s^{-1}}$ channel, averaged over a $\rm{30 km s^{-1}}$ section of the emission-free spectrum). An example of this wing selection process is shown in Figure \ref{fig:wingselection} (b), which shows the $\rm{^{13}CO}$ residual spectrum and discrete spectral points that satisfy the wing criteria (empty circles are blue, and solid circles are red).  

There is a risk that some blue and red emission might be missed by analysing a single spectrum at the location of the clump peak.  Therefore, when the position of peak intensity in both the blue and red integrated images was found (mapping of blue and red images is explained in \S \ref{sec:contours}), another two additional spectra, called the ``red-wing spectrum'' and ``blue-wing spectrum'', were extracted.  Once again blue and red residual spectra were calculated. If broader wing emission was found, the initial wing ranges were expanded to incorporate the ranges covered by the red-wing and blue-wing spectra.  The final velocity ranges for blue and red  wings are listed in Table \ref{tab:targetvelocities}.

\subsection{Mapping the outflows}
\label{sec:contours}

\begin{figure}
 	\begin{center}
		\includegraphics[width = 0.4\textwidth]{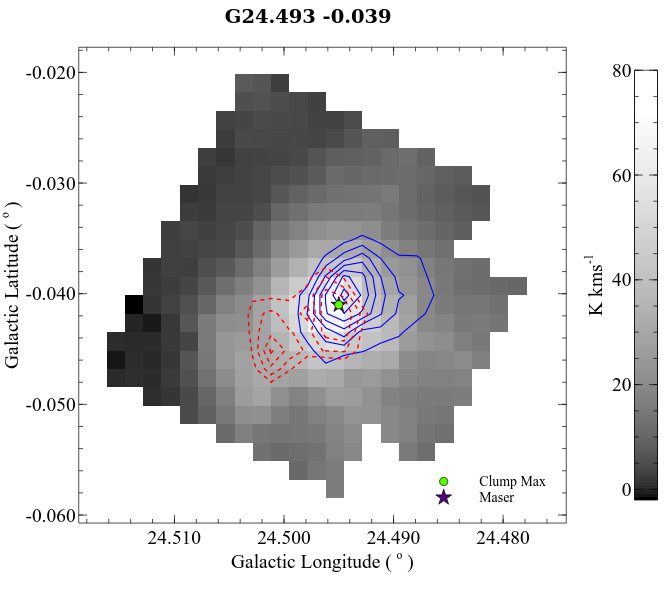}	\\
		\includegraphics[width = 0.4\textwidth]{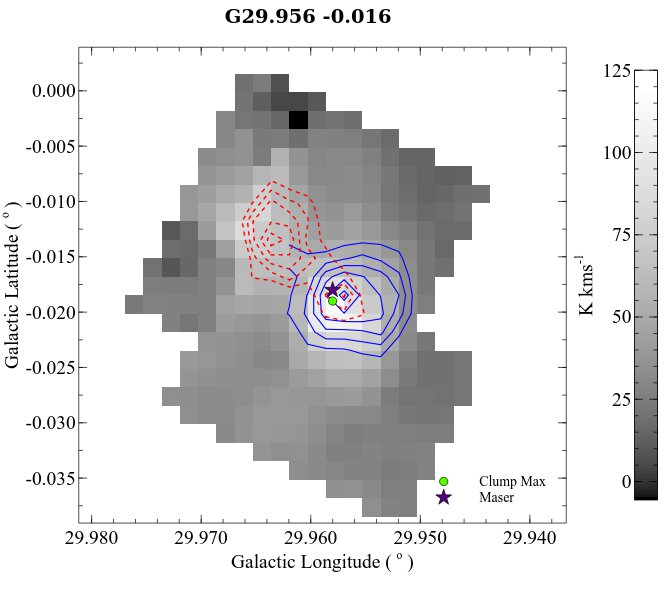}	
		\caption{Two examples of intensity integrated images of the blue and red wing, from top to bottom: G 24.493-0.039 and G 29.956-0.016 (clump 1).  Grey scale image shows $\rm{^{13}CO}$, integrated over the peak emission (velocity ranges listed in Table \ref{tab:targetcoords}), with blue and red contours representing blue and red wing integrated intensities respectively. Contour intervals are $10\%$ of the maximum intensity for each image, increasing up to $90\%$ of the maximum intensity. Lower contours are respectively at $60\%$ and $50\%$ for the two targets.}
		\label{fig:mapexamples}
 	\end{center}
 \end{figure}

The final blue- and redshifted velocity ranges are used to produce two dimensional $\rm{^{13}CO}$ intensity integrated images corresponding to each wing.  These are overlaid as solid blue- and dotted red contours onto the $\rm{^{13}CO}$ integrated intensity image, representing the outflow lobes. Two examples are shown in Figure \ref{fig:mapexamples}, showing target G 24.493-0.039 with the maser and clump coordinates overlapping, and target G 29.956-0.016A with an offset between the maser and clump coordinates. The remainder of the maps are shown online in Appendix B. Contours are plotted in $10\%$ intervals up to $90\%$ of the maximum intensity, $Int_{\rm{b}}$ or $Int_{\rm{r}}$, for each integrated image (values listed in columns 10 and 12 in Table  \ref{tab:targetvelocities}).  The lowest contour is never lower than $30\%$, but values differ for each image depending on the individual background brightness levels.  The lowest contour is selected by eye as the level which encompassed the outflow lobe clearly.   

As massive stars form in clusters, the observed targets often have contamination from similarly high velocity components as the outflow, but from different spatial structures in the field of view \citep{Shepherd1996b}.  This makes it difficult to isolate the outflow.  Therefore, if identified as belonging to such structures, these pixels were flagged to be bad in any further analysis.  Sometimes one or both of the outflows are partially cut-off where they are situated close to the edge of the field-of-view or to a dead receptor.  These sources are flagged as such in the second to last column in Table \ref{tab:properties} in \S\ref{sec:calculations} and their calculated properties only serve as a lower limit because a fraction of the emission is not included in the analysis.  

Three of the 58 analysed clumps have been too close to the edge of the field of view for any significant information to be derived and are excluded from further analysis.  Out of the remaining 55 maps, 47 outflows are clearly bipolar ($\rm{85 \%}$), with the eight exceptions marked with a superscript in Table \ref{tab:targetvelocities}. For a sample of high mass protostellar objects, \citet{Beuther2002} had a bipolar outflow detection frequency of $\rm{81 \%}$ in $\rm{^{12}CO}$, comparable with what we find.  

\section{Results}
\label{sec:Results}

\subsection{Detection Frequency}

All of the 58 spectra available for analysis (see \S{\ref{sec:clumpfind}}) were found to have high velocity outflow signatures, either in the spectra or in the contour maps, resulting in a 100\% detection rate.  Such a high detection rate of outflows toward massive YSOs is not uncommon.  \citet{Shepherd1996a} searched for $\rm{^{12}CO (J = 1-0)}$ high-velocity line wings toward 122 high-mass star forming regions and detected low-intensity line wings in 94 of them.  Of these 94, $\rm{90\%}$ were associated with high-velocity (HV) gas in the beam. The argument has already been made at that stage, that if the HV gas is due to bipolar outflows, molecular outflows are a common property of newly formed massive stars.  \citet{Sridharan2002} detected $\rm{84 \%}$ of sources with high-velocity gas from a $\rm{^{12}CO}$ $\rm{(J = 2-1)}$ survey of 69 protostellar candidates.  

Zhang et al. (2001,2005) observed a sample of 69 luminous \textit{IRAS} point sources in CO $\rm{(J = 2-1)}$ and detected 39 molecular outflows toward them ($57\%$). They found the search for outflows hampered for Galactic longitudes $< 50^{\circ}$ (due to confusion my multiple cloud components when observing in this transition). A total of 39 objects were outside of this region, toward which 35 outflows were detected, resulting in a $\rm{90\%}$ outflow detection rate.

\citet{Kim2006} observed 12 sources from the same \citet{Molinari1996} catalogue that \citet{Zhang2001} selected their sources from. They detected outflows in 10 sources and adding these sources to the detections from Zhang, results in a detection rate of $\rm{88 \%}$ ([$35+10-3=42$] out of [$39+12-3=48$]), taking into account that there are three sources in common between the two samples.   More recently, \citet{Lopez2009} searched for molecular outflows toward a sample of eleven very luminous massive YSOs.  They found high-velocity wings, indicative of outflow motions, in $\rm{100\%}$ of the sample. 

Three further studies have dealt specifically with class II methanol masers.  \citet{Codella2004} surveyed for molecular outflows towards 136 UCH{\sc ii} regions, out of which 56 positions showed either 6.7GHz methanol or 22.2GHz water maser emission.  Their overall outflow detection rate from $\rm{^{13}CO (J = 1-0)}$ and $\rm{(J = 2-1)}$ transition lines was $\rm{\sim 39 \%}$, but they found that in cases where observations were made toward 6.7GHz methanol or 22.2GHz water maser emission lines, the outflow detection rate increased to $\rm{50\%}$. As their observations were single pointings, they may have missed some outflows that were offset from the masers. \citet{Xu2006} studied molecular outflows using high-resolution CO $\rm{(J = 1-0)}$ mapping toward eight 6.7GHz methanol masers closer than 1.5 kpc. They found outflows in seven of them, an $\rm{88 \%}$ detection rate.  \citet{Wu2010} investigated the distinctions between low- and high-luminosity 6.7GHz methanol masers via multi-line  mapping observations of various molecular lines, including $\rm{^{12}CO (J = 1-0)}$, toward a sample of these masers.  They found outflows to be common among both sets of masers: of the low-luminosity masers, they found six outflows out of nine, and from the high-luminosity masers they found four outflows out of eight, an over-all detection rate of $\rm{59 \%}$. Note that the detection frequencies from both \citet{Xu2006} and \citet{Wu2010} are obtained from small number samples.

All these results suggest that the majority of massive YSO's have molecular outflows, and should 6.7GHz methanol masers be present, they are closely associated with the outflow phase.

\subsection{Maser distances}
\label{sec:Distance}

\citet{Green2011} published the kinematic distances for about $50 \%$ of the targets in this study, using the 6.7GHz maser mid-velocity as an estimate for the systemic velocity. They used respectively the presence / absence of self-absorption in H{\sc i} spectra in the proximity of the systemic velocity, to determine whether the source is at the near / far kinematic distance.  However, methanol maser emission often consists of a number of strong peaks spread over several $\rm{km s^{-1}}$.  As differences of only a few $\rm{kms^{-1}}$ in the velocity of the local standard of rest $v_{\rm{lsr}}$ can be enough to change the kinematic distance solution from near to far and vice versa in the H{\sc i} absorption feature method of resolving the former, using estimated $v_{\rm{lsr}}$ values form the maser emission could lead to an incorrect distance solution.  Therefore, molecular line observations provide more reliable measurements of the clump systemic velocity \citep{Urquhart2014b}.

For this reason, and to prevent the additional uncertainties introduced by adopting distances calculated using different techniques by different authors, we decided to recalculate all distances of the methanol masers using their associated $\rm{C^{18}O}$ peak velocities. This was done using the Galactic Rotation Curve (GRC) as fitted by \citet{Brand1993}, with the Sun's Galactocentric distance, $R_0$, assumed as 8.5 kpc and its circular rotation $\Theta_0$ as 220 $\rm{km s^{-1}}$.  Calculated values are listed in Table \ref{tab:distances}.  The average difference between distances calculated using the GRC, versus the maser distances listed by \citet{Green2011} for our targets, is $0.8 \pm 0.6$ kpc.

Alternative solutions to \citet{Brand1993} for calculating the kinematic distances, are \citet[e.g.][]{Reid2009} and \citet{Clemens1985}.  \citet{Urquhart2014b} found that for sources in the inner Galaxy, the distances given respectively by these rotation curves, all agree within a few tenths of a kpc, which are smaller than their associated uncertainty of the order $\pm 1$ kpc due to streaming motions \citep{Urquhart2011a, Urquhart2012}.  Consequently, the statistical results are robust against the choice of model.

The galactocentric distances obtained using the GRC from \citet{Brand1993} were geometrically converted to heliocentric distances, of which two solutions exist within the solar circle, called the kinematic distance ambiguity (KDA).  These distances are equally spaced on either side of the tangent position, and are generally referred to as the near and far distances. Sources with velocities within 10 $\rm{km s^{-1}}$ of the tangent velocity are placed at the tangent distance (indicated by TAN in the reference column), since the error in the distance is comparable to the difference between the near/far distance and the tangent distance.

As \citet{Green2011} could resolve the kinematic distance ambiguity for many masers using corresponding H{\sc i} self-absorption profiles, when available, we used their values to resolve the KDAs for our targets.  Where maser distances were not published in \citet{Green2011}, alternative sources were used where available, being other publications of 6.7GHzmasers \citep{Purcell2006,Caswell2011}, EGOs \citep{Cyganowski2009}, OH-masers \citep{Fish2003}, associated IRDCs \citep{Simon2006}, or molecular clouds \citep{Roman2009} (if the maser position fell within the cloud as well as within $\rm{\sim 5 ~ km s^{-1}}$ of the cloud's $v_{lsr}$) were used to resolve our KDAs.   The literature reference used to resolve the distance for each source, should it exist, is also listed in Table \ref{tab:distances}.
 
For four out of the 58 targets, no published distance could be found to resolve the KDA. In these cases both the near and far values are listed.  Three targets are rejected from this list in Table \ref{tab:distances}: G 24.790+0.083B, because of too high noise, and G 30.851+0.123 and G 31.182-0.148B, because the clumps are mostly cut off by the edge of the field-of-view.  The columns after the distance columns, list outflow lobe surface areas and lengths, values used for calculations discussed in \S \ref{sec:calculations}.

\onecolumn
\begin{scriptsize}
\begin{table}
\centering
\caption{Target $\rm{C^{18}O}$ velocities used to calculate their kinematic distances from the Galactic Rotation Curve \citep{Brand1993}.  Literature references used to resolve the far/near distance ambiguities are listed in the fourth column, being an assembly from published 6.7GHz masers, OHmasers, EGOs, IRCD's and molecular clouds.  Sources with velocities within 10 $\rm{km s^{-1}}$ of the tangent velocity are placed at the tangent distance, indicated by TAN.  Columns five to eight show information used in \S \ref{sec:calculations}: the surface areas $A$ for blue and red lobes as mapped in $\rm{^{13}CO}$ and lobe lengths $l$ as measured from the clump coordinate to each outflow's radial extreme.} \label{tab:distances} 

\begin{tabular}{l r r l c c c c}
\hline
\multicolumn{1}{c}{Target} & \multicolumn{1}{c}{$\rm{C^{18}O}$ $v$} & D & \multicolumn{1}{c}{Literature Reference} & $A_b$ & $A_r$ & $l_b$ & $l_r$ \\
{} & \multicolumn{1}{c}{($\rm{kms^{-1}}$)} & (kpc) & {} & ($\rm{pc^2}$) & ($\rm{pc^2}$) & (pc) & (pc) \\\hline \\

G 20.081-0.135 & 41.602 & 12.6 & \citet{Fish2003} & 1.48 & 1.75 & 1.28 & 0.92 \\
G 21.882+0.013 & 20.158 & 1.8 & \citet{Purcell2006} & 0.06 & 0.05 & 0.29 & 0.23 \\
G 22.038+0.222 & 51.533 & 3.8 & \citet{Cyganowski2009} & 0.29 & 0.29 & 0.60 & 0.44 \\
G 22.356+0.066 & 84.189 & 5.2 & \citet{Green2011} & 0.41 & 0.62 & 0.53 & 1.59 \\
G 22.435-0.169 & 27.895 & 13.4 & \citet{Roman2009} & 2.27 & 1.06 & 1.56 & 0.78 \\
G 23.003+0.124$^r$ & 107.445 & 6.2 & \citet{Roman2009} & - & 0.26 & - & 0.54 \\
G 23.010-0.411 & 76.380 & 4.8 & \citet{Green2011} & 1.03 & 0.68 & 1.48 & 0.99 \\
G 23.206-0.378 & 77.829 & 10.7 & \citet{Green2011} & 0.97 & 1.17 & 1.40 & 1.40 \\
G 23.365-0.291 & 78.292 & 4.9 & \citet{Roman2009} & 0.35 & 0.53 & 1.14 & 1.07 \\
G 23.437-0.184 & 100.630 & 5.9 & \citet{Green2011} & 0.52 & 0.90 & 0.77 & 0.85 \\
G 23.484+0.097 & 84.181 & 5.2 & \citet{Simon2006} & 0.50 & 0.25 & 0.68 & 0.53 \\
G 23.706-0.198 & 69.090 & 11.1 & \citet{Green2011} & 5.07 & 4.35 & 2.25 & 3.06 \\
G 24.329+0.144 & 112.743 & 7.7 & TAN & 0.86 & 0.81 & 1.01 & 0.68 \\
G 24.493-0.039 & 111.752 & 6.4 & \citet{Caswell2011} & 1.19 & 0.91 & 1.03 & 0.94 \\
G 24.790+0.083A & 110.548 & 9.1 & \citet{Green2011} & 0.49 & 0.56 & 0.92 & 0.79 \\
G 24.850+0.087 & 108.941 & 6.3 & \citet{Roman2009} & 0.40 & 1.48 & 0.82 & 1.19 \\
G 25.650+1.050 & 42.315 & 12.3 & \citet{Green2011} & 4.83 & 4.58 & 2.32 & 1.61 \\
G 25.710+0.044 & 101.214 & 9.4 & \citet{Green2011} & 1.86 & 2.83 & 2.46 & 1.77 \\
G 25.826-0.178 & 93.206 & 5.5 & \citet{Green2011} & 0.60 & 0.26 & 0.81 & 0.65 \\
G 28.148-0.004 & 98.665 & 5.9 & \citet{Green2011} & 0.68 & 0.59 & 0.86 & 0.95 \\
G 28.201-0.049 & 94.860 & 9.3 & \citet{Green2011} & 1.10 & 1.75 & 0.81 & 1.35 \\
G 28.282-0.359 & 47.354 & 3.2 & \citet{Green2011} & 0.24 & 0.44 & 0.61 & 0.76 \\
G 28.305-0.387 & 85.627 & 9.8 & \citet{Green2011} & 2.02 & 4.44 & 1.42 & 1.56 \\
G 28.321-0.011 & 99.570 & 6.0 & \citet{Roman2009} & 0.94 & 1.33 & 0.70 & 1.04 \\
G 28.608+0.018 & 103.075 & 7.5 & TAN & 0.80 & 1.23 & 1.09 & 1.09 \\
G 28.832-0.253 & 87.189 & 5.3 & \citet{Green2011} & 1.31 & 0.48 & 0.93 & 0.69 \\
G 29.603-0.625$^r$ & 77.185 & 4.8 & \citet{Roman2009} & - & 0.54 & - & 0.84 \\
G 29.865-0.043 & 101.849 & 7.4 & TAN & 1.79 & 1.98 & 2.36 & 1.61 \\
G 29.956-0.016A & 97.838 & 7.4 & TAN & 1.51 & 1.19 & 1.18 & 1.61 \\
G 29.956-0.016B & 97.838 & 7.4 & TAN & 0.55 & 0.92 & 0.64 & 1.18 \\
G 29.979-0.047 & 101.843 & 7.4 & TAN & 0.96 & 0.87 & 1.29 & 0.64 \\
G 30.317+0.070 & 44.645 & 11.6 & \citet{Green2011} & 1.83 & 1.49 & 1.52 & 1.52 \\
G 30.370+0.482A & 17.414 & 13.4 & \citet{Roman2009} & 3.03 & 1.36 & 1.95 & 1.36 \\
G 30.370+0.482B & 17.914 & 13.4 & \citet{Roman2009} & 2.58 & 1.67 & 2.92 & 1.17 \\
G 30.400-0.296 & 102.959 & 7.3 & TAN & 1.23 & 0.77 & 1.39 & 0.96 \\
G 30.419-0.232 & 104.549 & 7.3 & TAN & 0.73 & 1.59 & 2.00 & 2.03 \\
G 30.424+0.466 & 15.493 & 13.5 & \citet{Roman2009} & 4.64 & 4.95 & 3.74 & 2.36 \\
G 30.704-0.068$^b$ & 90.123 & 5.5 & \citet{Green2011} & 0.49 & - & 0.40 & - \\
G 30.781+0.231 & 41.851 & 2.9 & \citet{Green2011} & 0.13 & 0.09 & 0.29 & 0.25 \\
G 30.788+0.204 & 81.615 & 9.5 & \citet{Green2011} & 2.55 & 1.16 & 1.39 & 0.83 \\
G 30.819+0.273$^r$ & 98.128 & 6.1 & \citet{Green2011} & - & 0.58 & - & 0.80 \\
G 30.898+0.162$^r$ & 105.328 & 7.3 & TAN & - & 0.86 & - & 1.49 \\
G 30.973+0.562 & 23.396 & 12.89, 1.7 & - & 4.64, 0.08 & 2.25, 0.04 & 1.69, 0.22 & 1.12, 0.15 \\
G 30.980+0.216$^r$ & 107.365 & 7.3 & \citet{Roman2009} & - & 1.12 & - & 1.06 \\
G 31.061+0.094 & 19.227 & 13.2, 1.4 & - & 2.05, 0.02 & 0.29, 0.003 & 1.53, 0.16 & 0.96, 0.10 \\
G 31.076+0.457$^b$ & 28.310 & 12.5, 2.0 & - & 4.26, 0.11 & 3.59, 0.09 & 3.10, 0.50 &  \\
G 31.122+0.063 & 41.528 & 11.7 & \citet{Roman2009} & 5.11 & 3.25 & 3.58 & 3.58 \\
G 31.182-0.148A & 42.648 & 11.6 & \citet{Roman2009} & 2.30 & 1.26 & 1.02 & 0.85 \\
G 31.282+0.062 & 109.028 & 7.3 & TAN & 1.61 & 1.16 & 1.58 & 1.27 \\
G 31.412+0.307 & 96.428 & 7.3 & TAN & 0.58 & 1.42 & 0.63 & 0.84 \\
G 31.594-0.192 & 43.149 & 11.6 & \citet{Roman2009} & 3.17 & 5.09 & 1.18 & 3.70 \\
G 32.744-0.075 & 37.528 & 11.7 & \citet{Green2011} & 2.10 & 2.21 & 1.54 & 1.19 \\
G 33.317-0.360$^r$ & 34.814 & 11.8, 2.4 & - & - & 2.13, 0.09 & - & 1.38, 0.28 \\
G 33.486+0.040 & 111.986 & 7.1 & TAN & 0.17 & 0.64 & 0.52 & 1.13 \\
G 33.634-0.021 & 103.750 & 7.1 & TAN & 0.93 & 0.42 & 0.72 & 0.62 \\ 

\end{tabular}
\end{table}
\end{scriptsize}
\twocolumn

\subsection{Dealing with uncertainties}

Calculation of the physical properties of molecular outflows can provide useful information on the obscured driving source.  These calculations are subject to a number of uncertainties, most prominent of which is the outflow orientation. However, this is not easily  determined \citep[e.g.][]{Shepherd1996a,Curtis2010} and, as such, no correction is applied in this study. Although we will not apply any corrections, we discuss in the following paragraphs the corrections usually applied in the literature for outflow orientation. We also report the effects of such corrections on the calculated outflow properties, as well as additional contributors to uncertainties.

Our observations were some of the first to be carried out with the HARP instrument and some of the receptors exhibited poor performance and did not yield useful data.  Often more than two receptors had to be switched off. At times this resulted in some of the clump / outflow emission being missed.  Potentially, this could also result in outflow lobes not being detected at all.  Blue and red contour levels were determined by eye, since each image is uniquely characterised by the background noise, emission brightness and available receptors.  The $14''$ beam of the telescope places a limit on the size of outflows that can be resolved, especially for the more distant targets.

The most significant of the above uncertainties, is $\theta$, the angle of the outflow axis with respect to the line of sight.  As only a projection of the outflow is observed, any inclination with respect to the plane of sky will reduce the length of the outflow (not the width) by $\rm{sin}(\theta)$, and increase the observed Doppler broadening by $\rm{cos}(\theta)$.  \citet{Cabrit1990} give a detailed discussion of the effect of inclination angle. Due to the lack of a specific orientation for each outflow, many authors assume a mean inclination angle for their sample to correct the calculated outflow parameters.  The most commonly used angle is $57.3^o$, determined using the assumption that outflows are distributed uniformly and with random inclinations to the line of sight \citep{Bontemps1996,Beuther2002,Hatchell2007,Curtis2010}. Table \ref{tab:inclcorrections} summarises the corrections due to inclination for the outflow parameters calculated (see \S{\ref{sec:calculations}}).  Unknown inclinations mostly cause the outflow parameters to be under-estimated. Timescales $t_d$ are thus likely to represent a lower limit to the true age of the outflows \citep{Parker1991} and hence also to the time over which the embedded proto-stars responsible for the outflows have been accreting from their surroundings \citep{Beuther2002}. 

\begin{tiny}
\begin{table}
\centering
\caption{Inclination angle corrections on outflow parameters.  All values in column 3 calculated for $\theta = 57.3^{\circ}$.}
\label{tab:inclcorrections}
\begin{tabular}{cccc}
	\hline
	Flow parameters & Correction & Corr. Val. & Lit. Val.\\
	\hline
	$p$ & $1/\rm{cos}(\theta)$ & 1.9 & $2^{1,2,3}$ \\
	$E_m$ & $1/\rm{cos}^2(\theta)$ & 3.4 & $3^{1,2,3}$ \\
	$t_d$ & $\rm{cot}(\theta)$ & 0.6 & $*^4$ \\
	$\dot{M}_{\rm{out}}$ & $\tan(\theta)$ & 1.6 & \\
	$F_m$ & $\rm{sin}(\theta)/\rm{cos}^2(\theta)$ & 2.9 & $3^{5,6}$ \\
	$L_m$ & $\rm{sin}(\theta)/\rm{cos}^3(\theta)$ & 5.3 & \\
	\hline 
	\multicolumn{4}{l}{*for $20^o<\theta<70^o$, 0.4 to 2.7} \\	
	\multicolumn{4}{l}{1.\citet{Wu2004}, 2.\citet{Goldsmith1984},} \\
	\multicolumn{4}{l}{3.\citet{Curtis2010},4.\citet{Zhang2005},} \\
	\multicolumn{4}{l}{5.\citet{Henning2000}, 6.\citet{Beuther2002}} \\
	\hline
\end{tabular}
\end{table}
\end{tiny}

Other contributors to uncertainties are: possible difficulty separating the outflowing gas from the ambient gas; higher interstellar extinction toward the molecular ring in the inner Galaxy ($0^o < l < 50^o$) in addition to their internal extinction \citep{Zhang2005}; different $\rm{CO/H_2}$ abundance ratios used by different authors \citep[e.g.][]{Rodriguez1982,Cabrit1992,Herbst2009}. 

Some authors use the mean atomic weight of the mixture of hydrogen and helium gas \citep{Garden1991}, while others only consider pure hydrogen molecular gas \citep{Snell1984}, resulting in a difference of 0.36 amu in the mean atomic weight \citep{Wu2004}. 
The excitation temperature, $T\rm{_{ex}}$, is assumed to range from 30 to 50 K for high mass sources \citep{Shepherd1996a,Beuther2002,Wu2004}.  However, a constant temperature assumption will underestimate the kinetic energy for an outflow with high jet/ambient density contrast \citep{Downes2007}.  Contamination from additional unrelated velocity components within the telescope beam could make it difficult to isolate the outflow, unless the components have a different spatial distribution from the outflow gas \citep{Shepherd1996b}.  Finally, even though we used $\rm{^{13}CO}$ as a tracer, we note that for their $\rm{^{12}CO}$ observations, \citet{Cabrit1990} estimated typical errors in the outflow parameters that reflect uncertainties in  $\rm{^{12}CO/H_2}$, distance determinations, $T\rm{_{ex}}$, inclination angles, optical depth effects and possible low-level contamination of $\rm{^{12}CO}$ emission in the reference position.  These error values are a factor $\sim 3$ on outflow mass $M\rm{_{out}}$, a factor $\sim 10$ on mechanical force $F_m$,  and a factor $\sim 30$ on mechanical luminosity $L_m$. 

\subsection{Calculation of outflow physical properties}
\label{sec:calculations}

The physical properties of the outflows are calculated following \citet{Beuther2002}, with some adaptions given that $\rm{^{13}CO}$ was observed instead of $\rm{^{12}CO}$.  We refer to \citet{Curtis2010} for the derivation of $\rm{H_2}$ column density from $\rm{^{13}CO}$. It is assumed that $\rm{^{13}CO}$ line wings are optically thin.
The column density of $\rm{^{13}CO}$ is given by 
\begin{equation}
	N\rm{\left(^{13}CO\right)} = 5 \times 10^{12} T\rm{_{ex}} \exp \left( \frac{T\rm{_{trans}}}{T\rm{_{ex}}} \right) \int{T\rm{_{mb}} dv} \rm{cm^{-2}},
	\label{eq:N(CO)}
\end{equation}
with $T\rm{_{trans}}=31.8$ K, the upper level energy of the $\rm{J = 3-2}$ transition of $\rm{^{13}CO}$ \citep{Minchin1993}.  The excitation temperature of the outflow lobes, $T\rm{_{ex}}$, is taken as 35 K  \citep[e.g.][]{Shepherd1996b,Henning2000,Beuther2002}. $\rm{\int{T\rm{_{mb}}dv}}$ is the mean integrated emission (main-beam temperature) for the blue and red lobes. It is calculated by averaging the temperature of each lobe within an area defined by the lowest contour. 

The abundance ratio $\rm{[H_2]/[^{13}CO]}$ is used to convert to the $\rm{H_2}$ column density for each lobe, $N\rm{_{r/b}}$ (red or blue).  The isotopic ratio $\rm{[^{12}CO]/[^{13}CO]}$ is a function of the Galactocentric distance, $D_{\rm{gal}}$, of each source, given by \citet{Wilson1994} as $7.5D_{\rm{gal}}+7.6$, which is then converted to a $\rm{[H_2]/[^{13}CO]}$ ratio assuming $\rm{[CO]/[H_2]=10^{-4}}$ \citep{Frerking1982}.  These column densities are then used to calculate the mass of each lobe:
\begin{equation}
	M\rm{_{b/r}} = \left( N\rm{_{b/r}} \times A\rm{_{b/r}} \right)m\rm{_{H_2}}.
	\label{eq:mass}
\end{equation}
$A\rm{_{r/b}}$ is the surface area of each lobe and $m\rm{_{H_2}}$ is the mass of a hydrogen molecule.  This surface area (listed in Table \ref{tab:distances}) is calculated using the same threshold technique used to calculate $T\rm{_{mb}}$, followed by summing the total number of pixels in each lobe and converting to an area using the target's distance as given in Table \ref{tab:distances}.  Where a significant amount of emission was cut off due to a field-of-view edge or dead receptors, it is indicated in the second to last column of Table \ref{tab:properties}.  In these cases the estimated physical parameters should be regarded as lower limits.  Finally, the total mass $M\rm{_{out}}$ is obtained by adding the blue and red components: $M\rm{_{out}} = M_b + M_r$.  
Excluding outflows with distance ambiguities (and hence two possible values for $M\rm{_{out}}$), and multiplying monopolar outflow masses with two, to account for the missing lobe, outflow masses ranged from 4.0 $\rm{M_{\odot}}$ to 750 $\rm{M_{\odot}}$ with a median of 73 $\rm{M_{\odot}}$ and a mean of 120 $\rm{M_{\odot}}$.

Using the outflow masses and $\Delta v_{\rm{b}}$ and $\Delta v_{\rm{r}}$, which are the blue and red velocities relative to the peak $\rm{C^{18}O}$ velocity, measured respectively from each wing extreme (listed in Table \ref{tab:targetvelocities}), \citet{Beuther2002} calculated the outflow momentum $p$ and energy $E$ using:
\begin{equation}
	p = M_b \times \Delta v_{b} + M_r \times \Delta v_{r}
	\label{eq:momentum}
\end{equation}
\begin{equation}
	E = \frac{1}{2}M_b \times \Delta v^{2}_{b} + \frac{1}{2}M_r \times \Delta v^{2}_{r}.
	\label{eq:Energy}
\end{equation}
However, using the maximum wing velocities is likely to overestimate the momentum and energy of the outflows. Instead we make the more reasonable assumption that the material is moving at the observed velocity associated with it.  For each ``pixel'' in the defined outflow lobe area, we calculate the momentum/energy per velocity channel (width $\Delta v$), using the channel velocity relative to the systemic velocity ($v_i$), and the gas mass ($M_i$) corresponding to the emission in that channel.  This is followed by both summing over all velocity channels, and all pixels in the lobe area $A_{b/r}$.
\begin{equation}
	p = \sum_{A_b} \left[ \sum_{i=v_b} M_{b_i} v_i \right]\Delta v + \sum_{A_r} \left[ \sum_{i=v_r} M_{r_i} v_i \right]\Delta v
	\label{eq:momentum_me}
\end{equation}
A similar approach is followed for energy calculations.
\begin{equation}
	E = \frac{1}{2} \sum_{A_b} \left[ \sum_{i=v_b} M_{b_i} v^{2}_{i} \right]\Delta v  + \frac{1}{2} \sum_{A_r} \left[ \sum_{i=v_r} M_{r_i} v^{2}_{i} \right]\Delta v.
	\label{eq:Energy_me}
\end{equation}

$\rm{^{13}CO}$ is a less abundant molecule than $\rm{^{12}CO}$, thus exhibiting a narrower spectral profile.  A sample of 56 sources for which both $\rm{^{12}CO}$ and $\rm{^{13}CO}$ spectra were published, has been investigated and the average $\rm{^{12}CO/^{13}CO}$ full width zero intensity ratio is found to be $\sim 2$ with a standard deviation of 1.3 \citep{Cabrit1988, Shepherd1996b, Su2004, Bronfman2008, Narayanan2012, Ortega2012, Xu2013}.  All calculations containing wing velocities relative to the systemic velocity are scaled by this factor, implying a factor two increase in $p$ and factor four increase in $E$.

In order to calculate the dynamical timescale $t_d$, the length of each outflow lobe $l_b$ or $l_r$ is measured from the clump coordinate to the furthest radial distance.  Excluding sources with distance ambiguities, blue-red averaged lobe lengths varies between 0.3-3.6 pc with a mean of 1.2 pc (Table \ref{tab:distances}).  As the red/blue lobe lengths are often different, the maximum, $l_{max}$ is chosen and used to calculate $t_d$ as
\begin{equation}
	t_d = \frac{l_{max}}{\left( \Delta v_{b} + \Delta v_{r} \right) /2}.
	\label{eq:timescale}
\end{equation}
For monopolar outflow detections (e.g. red lobe only), the above formula is adapted to $t_d = l_r/v_r$.

The mass loss rate of the molecular outflow $\dot{M}\rm{_{out}}$, the mechanical force $F_m$ and the mechanical luminosity $L_m$ summed over both blue and red lobes for each target, are calculated using 
\begin{eqnarray}
	\dot{M}\rm{_{out}} & = & \frac{M\rm{_{out}}}{t} \\
	F_m & = & \frac{p}{t} \\
	L_m & = & \frac{E}{t},
	\label{eq:Mechanics}
\end{eqnarray}
where $\rm{^{12}CO/^{13}CO}$ scaling will again lead to a factor two increase in $\rm{F_m}$ and factor four increase in $\rm{L_m}$.  The results are summarised in Table \ref{tab:properties}. Peculiarities are indicated in the \textit{notes} column.  Monopolar target names are marked with a superscript $b$ or $r$ in Table \ref{tab:properties}, with the letter indicating which lobe (blue or red) is present.

For sources with unresolved distances, both values are shown and distinguished by the numbers next to the target names (1=far and 2=near).  Exclusions are then made in further analyses for targets which have (i) kinematic distance ambiguities, hence uncertainties in calculated physical parameters, or, (ii) offsets of more than 3 pixels ($18''$, of the order of a beam size) between the maser coordinate and peak CO emission.  These targets are marked as such in the notes of Table \ref{tab:properties}. Following the exclusions, we are left with 44 outflows in our sample that are positionally associated with methanol masers and for which we can calculate physical properties that are unaffected by distance ambiguities.  We refer to this sample as Methanol Maser Associated Outflows (MMAOs), indicated as such in Table \ref{tab:properties}, and base all further discussion on these outflows.

\onecolumn
\begin{scriptsize}
\textbf{Table 5.} Physical properties of all blue and red outflow lobes as detected in $\rm{^{13}CO}$.  Where multiple clumps exist, their target labels are distinguished by ``A'' and ``B''.  Both values are listed for sources with distance ambiguities, with (1) next to the target name indicating values for far distances, and (2) mark the values for near distances.  Application of the $\rm{^{12}CO/^{13}CO}$ scaling factor to wing velocity ranges will lead to a factor two increase in p and $F_m$ and factor four increase in E and $L_m$.  Column 11 lists any additional notes about the mapped lobes and column 12 indicates whether a target belongs to the Methanol Maser Associated Outflows (MMAOs) subset (as defined in \S \ref{sec:calculations}) or not. \\

\begin{center}
\setlength{\tabcolsep}{2pt}
\begin{longtable}{lrrrrrrrrrlc} 
\caption*{} \label{tab:properties} \\
\hline

\multicolumn{1}{c}{Target} & \multicolumn{1}{c}{$M_b$} & \multicolumn{1}{c}{$M_r$} & \multicolumn{1}{c}{$M_{out}$} & \multicolumn{1}{c}{$p$} & \multicolumn{1}{c}{$E$} & \multicolumn{1}{c}{$t$} & \multicolumn{1}{c}{$\dot{M}_{\rm{out}}$} & \multicolumn{1}{c}{$F_m$} & \multicolumn{1}{c}{$L_m$} & \multicolumn{1}{c}{Notes} & \multicolumn{1}{c}{MMAO?} \\

\multicolumn{1}{c}{} & \multicolumn{1}{c}{$\rm{M_{\odot}}$} & \multicolumn{1}{c}{$\rm{M_{\odot}}$} & \multicolumn{1}{c}{$\rm{M_{\odot}}$} & \multicolumn{1}{c}{$\rm{M_{\odot} km s^{-1}}$} & \multicolumn{1}{c}{J} & \multicolumn{1}{c}{yr} & \multicolumn{1}{c}{$\rm{10^{-4} M_{\odot} yr^{-1}}$} & \multicolumn{1}{c}{$\rm{M_{\odot}kms^{-1}yr}$} & \multicolumn{1}{c}{$\rm{L_{\odot}}$} & \multicolumn{1}{c}{} & \multicolumn{1}{c}{} \\\hline
\endfirsthead

\multicolumn{11}{c}{\tablename\ \thetable\ -- \textit{Continued}}\\
\hline

\multicolumn{1}{c}{Target} & \multicolumn{1}{c}{$M_b$} & \multicolumn{1}{c}{$M_r$} & \multicolumn{1}{c}{$M\rm{_{out}}$} & \multicolumn{1}{c}{$p$} & \multicolumn{1}{c}{$E$} & \multicolumn{1}{c}{$t$} & \multicolumn{1}{c}{$\dot{M}_{\rm{out}}$} & \multicolumn{1}{c}{$F_m$} & \multicolumn{1}{c}{$L_m$} & \multicolumn{1}{c}{Notes} & \multicolumn{1}{c}{MMAO?} \\

\multicolumn{1}{c}{} & \multicolumn{1}{c}{$\rm{M_{\odot}}$} & \multicolumn{1}{c}{$\rm{M_{\odot}}$} & \multicolumn{1}{c}{$\rm{M_{\odot}}$} & \multicolumn{1}{c}{$\rm{M_{\odot} km s^{-1}}$} & \multicolumn{1}{c}{J} & \multicolumn{1}{c}{yr} & \multicolumn{1}{c}{$\rm{10^{-4} M_{\odot} yr^{-1}}$} & \multicolumn{1}{c}{$\rm{M_{\odot}kms^{-1}yr}$} & \multicolumn{1}{c}{$\rm{L_{\odot}}$} & \multicolumn{1}{c}{} & \multicolumn{1}{c}{} \\\hline
\endhead

\hline \multicolumn{11}{l}{Notes key: R=red lobe, B=blue lobe; RR=red ridge; Offset=clump-maser coordinate offset; X=reject} \\\hline
\endfoot

\hline \multicolumn{11}{l}{Notes key: R=red lobe, B=blue lobe; RR=red ridge morphology; Offset=clump-maser coordinate offset; X=reject} \\\hline
\endlastfoot

G 20.081-0.135 & 110 & 180 & 280 & 2700 & 1.5E+41 & 5.7E+04 & 50 & 4.6E-02 & 210 & B/R partly  c.o. & Y \\
G 21.882+0.013 & 3 & 1 & 4 & 23 & 8.4E+38 & 2.6E+04 & 2 & 8.9E-04 & 3 & Big offset-X  & N \\
G 22.038+0.222 & 12 & 18 & 30 & 170 & 7.3E+39 & 3.3E+04 & 9 & 5.3E-03 & 18 &  & Y \\
G 22.356+0.066 & 10 & 4 & 14 & 70 & 2.0E+39 & 1.8E+05 & 1 & 3.8E-04 & 1 &  & Y \\
G 22.435-0.169 & 55 & 16 & 71 & 240 & 5.8E+39 & 2.3E+05 & 3 & 1.0E-03 & 2 &  & Y \\
G 23.003+0.124$^r$ & - & 3 & 3 & 13 & 3.6E+38 & 6.6E+04 & 1 & 1.9E-04 & 0 & No B & Y \\
G 23.010-0.411 & 74 & 31 & 100 & 1200 & 8.1E+40 & 6.4E+04 & 16 & 1.8E-02 & 100 & 2 peaks,1clump  & Y \\
G 23.206-0.378 & 38 & 38 & 76 & 670 & 4.0E+40 & 5.8E+04 & 13 & 1.2E-02 & 57 &  & Y \\
G 23.365-0.291 & 7 & 12 & 20 & 83 & 2.1E+39 & 1.3E+05 & 2 & 6.3E-04 & 1 &  & Y \\
G 23.437-0.184 & 30 & 81 & 110 & 1100 & 6.6E+40 & 3.6E+04 & 31 & 3.1E-02 & 150 &  & Y \\
G 23.484+0.097 & 10 & 6 & 16 & 82 & 2.7E+39 & 5.8E+04 & 3 & 1.4E-03 & 4 & B/R partly c.o. & Y \\
G 23.706-0.198 & 160 & 110 & 270 & 1100 & 2.8E+40 & 3.3E+05 & 8 & 3.4E-03 & 7 & R partly c.o. & Y \\
G 24.329+0.144 & 24 & 11 & 35 & 230 & 1.0E+40 & 6.0E+04 & 6 & 3.8E-03 & 14 &  & Y \\
G 24.493-0.039 & 50 & 31 & 81 & 700 & 3.4E+40 & 7.2E+04 & 11 & 9.8E-03 & 39 &  & Y \\
G 24.790+0.083A & 17 & 31 & 48 & 400 & 1.8E+40 & 6.7E+04 & 7 & 6.0E-03 & 23 &  & Y \\
G 24.790+0.083B & - & - & - & - & - & - & - & - & - & clump c.o.-X & N \\
G 24.850+0.087 & 12 & 30 & 42 & 170 & 3.6E+39 & 1.5E+05 & 3 & 1.1E-03 & 2 & R partly c.o. & Y \\
G 25.650+1.050 & 400 & 350 & 750 & 7600 & 4.6E+41 & 9.9E+04 & 76 & 7.8E-02 & 380 &  & Y \\
G 25.710+0.044 & 110 & 110 & 220 & 880 & 2.9E+40 & 2.1E+05 & 10 & 4.2E-03 & 12 & Big offset-X & N \\
G 25.826-0.178 & 18 & 7 & 25 & 190 & 9.2E+39 & 4.4E+04 & 6 & 4.3E-03 & 17 &  & Y \\
G 28.148-0.004 & 14 & 12 & 26 & 140 & 4.4E+39 & 6.9E+04 & 4 & 2.0E-03 & 5 &  & Y \\
G 28.201-0.049 & 120 & 250 & 370 & 4400 & 3.4E+41 & 4.3E+04 & 86 & 1.0E-01 & 660 &  & Y \\
G 28.282-0.359 & 16 & 37 & 53 & 340 & 1.3E+40 & 5.3E+04 & 10 & 6.5E-03 & 21 & Big offset-X, B/R partly c.o. & N \\
G 28.305-0.387 & 160 & 410 & 570 & 1700 & 3.5E+40 & 1.8E+05 & 32 & 9.3E-03 & 16 & R partly c.o. & Y \\
G 28.321-0.011 & 34 & 36 & 70 & 280 & 7.6E+39 & 8.2E+04 & 9 & 3.4E-03 & 8 & R partly c.o. & Y \\
G 28.608+0.018 & 63 & 51 & 110 & 920 & 4.7E+40 & 5.7E+04 & 20 & 1.6E-02 & 67 & B/R partly c.o. & Y \\
G 28.832-0.253 & 43 & 38 & 81 & 660 & 3.8E+40 & 4.8E+04 & 17 & 1.4E-02 & 66 &  & Y \\
G 29.603-0.625$^r$ & - & 13 & 13 & 41 & 8.1E+38 & 1.0E+05 & 3 & 4.0E-04 & 1 & Big offset-X, no B & N \\
G 29.865-0.043 & 240 & 61 & 300 & 1900 & 6.5E+40 & 1.8E+05 & 16 & 1.0E-02 & 29 & B partly c.o. & Y \\
G 29.956-0.016A & 160 & 90 & 250 & 1800 & 7.6E+40 & 8.1E+04 & 31 & 2.2E-02 & 78 &  & Y \\
G 29.956-0.016B & 8 & 13 & 21 & 180 & 8.4E+39 & 7.9E+04 & 3 & 2.3E-03 & 9 & Big offset-X & N \\
G 29.979-0.047 & 93 & 26 & 120 & 1100 & 6.6E+40 & 7.6E+04 & 16 & 1.4E-02 & 72 &  & Y \\
G 30.317+0.070 & 42 & 25 & 67 & 320 & 9.1E+39 & 1.5E+05 & 5 & 2.1E-03 & 5 & R partly c.o. & Y \\
G 30.370+0.482A & 43 & 29 & 73 & 310 & 8.2E+39 & 2.4E+05 & 3 & 1.3E-03 & 3 &  & Y \\
G 30.370+0.482B & 3 & 2 & 5 & 15 & 2.5E+38 & 7.1E+05 & 0 & 2.1E-05 & 0 & Big offset-X, B mostly c.o. & N \\
G 30.400-0.296 & 65 & 20 & 84 & 500 & 2.1E+40 & 9.0E+04 & 9 & 5.5E-03 & 20 &  & Y \\
G 30.419-0.232 & 53 & 140 & 190 & 810 & 2.4E+40 & 1.3E+05 & 15 & 6.1E-03 & 15 & B  mostly c.o. & Y \\
G 30.424+0.466 & 100 & 130 & 230 & 990 & 2.4E+40 & 3.8E+05 & 6 & 2.6E-03 & 5 & B/R partly c.o. & Y \\
G 30.704-0.068$^b$ & 67 & - & 67 & 200 & 9.1E+39 & 2.2E+04 & 61 & 9.1E-03 & 34 & RR-X red lobe & Y \\
G 30.781+0.231 & 4 & 4 & 8 & 24 & 4.6E+38 & 4.4E+04 & 2 & 5.5E-04 & 1 &  & Y \\
G 30.788+0.204 & 76 & 48 & 120 & 830 & 3.3E+40 & 1.0E+05 & 12 & 8.2E-03 & 27 &  & Y \\
G 30.819+0.273$^r$ & - & 11 & 11 & 59 & 1.9E+39 & 7.2E+04 & 3 & 8.2E-04 & 2 & no B & Y \\
G 30.851+0.123 & - & - & - & - & - & - & - & - & - & clump c.o., Big offset-X & N \\
G 30.898+0.162$^r$ & - & 26 & 26 & 170 & 3.8E+39 & 2.9E+05 & 2 & 5.7E-04 & 1 & no B, RR-adapted shape  & Y \\
G 30.973+0.562(1) & 250 & 58 & 310 & 590 & 9.0E+39 & 2.7E+05 & 11 & 2.1E-03 & 3 &  & N \\
G 30.973+0.562(2) & 4 & 1 & 5 & 10 & 1.5E+38 & 3.6E+04 & 2 & 2.8E-04 & 0 &  & N \\
G 30.980+0.216$^r$ & - & 19 & 19 & 850 & 4.7E+41 & 1.2E+05 & 3 & 7.3E-03 & 330 & B separated-X, partly c.o. & Y \\
G 31.061+0.094(1) & 110 & 1 & 110 & 780 & 2.9E+40 & 1.7E+05 & 7 & 4.7E-03 & 15 & Sub-resolution R & N \\
G 31.061+0.094(2) & 1 & 0 & 1 & 9 & 3.3E+38 & 1.8E+04 & 1 & 5.0E-04 & 2 & Sub-resolution R & N \\
G 31.076+0.457$^b$(1) & 110 & - & 110 & 1500 & 1.1E+41 & 3.0E+05 & 7 & 5.1E-03 & 30 & Big offset-X, RR-X red lobe & N \\
G 31.076+0.457$^b$(2) & 3 & - & 12 & 40 & 2.8E+39 & 3.8E+04 & 6 & 1.1E-03 & 6 & Big offset-X, RR-X red lobe & N \\
G 31.122+0.063 & 170 & 130 & 300 & 2000 & 7.8E+40 & 2.9E+05 & 10 & 6.9E-03 & 22 & B/R partly c.o. & Y \\
G 31.182-0.148A  & 43 & 7 & 50 & 190 & 4.1E+39 & 1.7E+05 & 3 & 1.2E-03 & 2 &  & Y \\
G 31.182-0.148B & - & - & - & - & - & - & - & - & - & clump c.o.-X & N \\
G 31.282+0.062 & 73 & 42 & 110 & 810 & 3.1E+40 & 1.3E+05 & 9 & 6.2E-03 & 20 &  & Y \\
G 31.412+0.307 & 15 & 58 & 73 & 680 & 3.3E+40 & 7.9E+04 & 9 & 8.6E-03 & 35 &  & Y \\
G 31.594-0.192 & 47 & 120 & 160 & 600 & 1.2E+40 & 6.0E+05 & 3 & 9.9E-04 & 2 & R partly c.o. & Y \\
G 32.744-0.075 & 110 & 110 & 220 & 1700 & 7.6E+40 & 9.4E+04 & 24 & 1.8E-02 & 67 &  & Y \\
G 33.317-0.360$^r$(1) & - & 46 & 46 & 240 & 6.7E+39 & 1.7E+05 & 6 & 1.4E-03 & 3 & no B  & N \\
G 33.317-0.360$^r$(2) & - & 2 & 2 & 10 & 2.7E+38 & 6.8E+04 & 1 & 1.4E-04 & 0 & no B & N \\
G 33.486+0.040 & 1 & 7 & 8 & 19 & 3.1E+38 & 2.0E+05 & 0 & 9.6E-05 & 0 & Sub-resolution B & Y \\
G 33.634-0.021 & 37 & 9 & 46 & 110 & 2.2E+39 & 7.4E+04 & 6 & 1.5E-03 & 3 & Big offset-X, B partly c.o. & N \\ 

\end{longtable}
\end{center}
\end{scriptsize}

\setlength{\tabcolsep}{6pt}
\twocolumn

\subsection{Clump masses}
\label{sec:coremass}

\setlength{\tabcolsep}{2pt}
\begin{center}
\begin{tiny}
\begin{table}
\caption[]{Coordinates and masses of the central clumps associated with the methanol masers, as derived from the $870 \rm{\mu m}$ dust flux measurements from ATLASGAL \citep{Csengeri2014}.  The last column lists the clump masses as calculated using $\rm{C^{18}O}$ maps.  (Suffixes ``A'' and ``B'' and numbers (1) and (2) next to some entries in column 1 have the same meaning as in Table \ref{tab:properties}).} 
\label{tab:dustcores} 
\begin{tabular}{lrrrrr} 
\hline
\multicolumn{1}{c}{Target} & \multicolumn{2}{c}{Clump coord.} & \multicolumn{1}{c}{Dust flux} & {$M\rm{_{870\mu m}}$} & {$M\rm{_{C^{18}O}}$}\\
{} & $l (^{\circ})$ & $b (^{\circ})$ & $S_{\nu}$ (Jy) & $\rm{M_{\odot}}$  & $\rm{M_{\odot}}$ \\\hline
 
G 20.081-0.135 & 20.081 & -0.135 & 10.5 & 9200 & 1800 \\
G 21.882+0.013 & 21.875 & 0.008 & 3.7 & 65 & 20 \\
G 22.038+0.222 & 22.040 & 0.223 & 5.5 & 430 & 41 \\
G 22.356+0.066 & 22.356 & 0.068 & 5.4 & 810 & 64 \\
G 22.435-0.169 & 22.435 & -0.169 & 2.3 & 2300 & 200 \\
G 23.003+0.124 & 23.002 & 0.126 & 0.9 & 200 & 18 \\
G 23.010-0.411 & 23.008 & -0.410 & 12.8 & 1700 & 110 \\
G 23.206-0.378 & 23.209 & -0.378 & 11.1 & 7100 & 370 \\
G 23.365-0.291 & 23.364 & -0.291 & 5.0 & 660 & 47 \\
G 23.437-0.184 & 23.436 & -0.183 & 11.9 & 2300 & 490 \\
G 23.484+0.097 & 23.483 & 0.098 & 4.2 & 620 & 120 \\
G 23.706-0.198 & 23.706 & -0.197 & 3.9 & 2600 & 340 \\
G 24.329+0.144 & 24.330 & 0.145 & 9.0 & 3000 & 91 \\
G 24.493-0.039 & 24.493 & -0.039 & 12.0 & 2700 & 300 \\
G 24.790+0.083A & 24.790 & 0.083 & 26.6 & 12000 & 670 \\
G 24.850+0.087 & 24.853 & 0.085 & 2.4 & 530 & 240 \\
G 25.650+1.050 & 25.649 & 1.051 & 16.6 & 14000 & 2200 \\
G 25.710+0.044 & 25.719 & 0.051 & 0.6 & 300 & 250 \\
G 25.826-0.178 & 25.824 & -0.179 & 12.1 & 2100 & 120 \\
G 28.148-0.004 & 28.148 & -0.004 & 3.6 & 700 & 170 \\
G 28.201-0.049 & 28.201 & -0.049 & 15.7 & 7500 & 1800 \\
G 28.282-0.359 & 28.289 & -0.365 & 8.8 & 510 & 250 \\
G 28.305-0.387 & 28.307 & -0.387 & 4.3 & 2300 & 1200 \\
G 28.321-0.011 & 28.321 & -0.011 & 3.4 & 670 & 150 \\
G 28.608+0.018 & 28.608 & 0.018 & 5.2 & 1600 & 680 \\
G 28.832-0.253 & 28.832 & -0.253 & 9.5 & 1500 & 130 \\
G 29.603-0.625 & 29.600 & -0.618 & 2.5 & 310 & 53 \\
G 29.865-0.043 & 29.863 & -0.045 & 4.2 & 1300 & 630 \\
G 29.956-0.016A & 29.956 & -0.017 & 17.5 & 5200 & 900 \\
G 29.956-0.016B & 29.962 & -0.008 & 3.5 & 1000 & 27 \\
G 29.979-0.047 & 29.979 & -0.048 & 6.5 & 1900 & 170 \\
G 30.317+0.070 & 30.317 & 0.070 & 1.2 & 930 & 160 \\
G 30.370+0.482A & 30.370 & 0.484 & 1.2 & 1200 & 140 \\
G 30.400-0.296 & 30.403 & -0.296 & 1.9 & 570 & 120 \\
G 30.419-0.232 & 30.420 & -0.233 & 7.2 & 2100 & 290 \\
G 30.424+0.466 & 30.424 & 0.464 & 1.9 & 1900 & 950 \\
G 30.704-0.068 & 30.701 & -0.067 & 22.0 & 3700 & 790 \\
G 30.781+0.231 & 30.780 & 0.231 & 0.7 & 30 & 10 \\
G 30.788+0.204 & 30.789 & 0.205 & 5.9 & 3000 & 320 \\
G 30.819+0.273 & 30.818 & 0.273 & 1.8 & 380 & 53 \\
G 30.898+0.162 & 30.899 & 0.163 & 3.7 & 1100 & 140 \\
G 30.973+0.562(1) & 30.972 & 0.561 & 0.7 & 660 & 150 \\
G 30.973+0.562(2) & - & - & - & 11 & 3 \\
G 30.980+0.216 & 30.979 & 0.216 & 2.7 & 780 & 150 \\
G 31.061+0.094(1) & 31.060 & 0.092 & 1.0 & 930 & 130 \\
G 31.061+0.094(2) & - & - & - & 10 & 1 \\
G 31.076+0.457(1) & 31.085 & 0.468 & 1.5 & 1300 & 230 \\
G 31.076+0.457(2) & - & - & - & 34 & 6 \\
G 31.122+0.063 & 31.124 & 0.063 & 0.9 & 700 & 290 \\
G 31.182-0.148A & 31.182 & -0.148 & 1.1 & 830 & too low S/N \\
G 31.282+0.062 & 31.281 & 0.063 & 13.1 & 3800 & 520 \\
G 31.412+0.307 & 31.412 & 0.306 & 29.8 & 8700 & 1000 \\
G 31.594-0.192(1) & 31.593 & -0.193 & 1.0 & 720 & 180 \\
G 32.744-0.075 & 32.746 & -0.076 & 7.8 & 6000 & 1100 \\
G 33.317-0.360(1) & 33.317 & -0.360 & 0.6 & 500 & too low S/N \\
G 33.317-0.360(2) & - & - & - & 20 & too low S/N \\
G 33.634-0.021 & 33.649 & -0.024 & 2.3 & 630 & 190 \\ 

\hline

\end{tabular}
\end{table}
\end{tiny}
\end{center}
\setlength{\tabcolsep}{6pt}

The evolutionary sequence of massive stars begins with prestellar clumps and cores, which are gravitationally bound overdensities inside a molecular cloud that show signs of inward motion before a protostar starts forming \citep{Zinnecker2007,Dunham2011}.  Most massive stars form in star clusters \citep[e.g.][]{Clarke2000,Lada2003}, which are part of a hierarchical structure, defined by \citet{Williams2000} and summarized by \citet{Bergin2007}.  In this approach, the largest structure is a molecular \textit{cloud}, with masses of the order $10^3$ - $10^4 \rm{M_{\odot}}$ and diameters ranging from 2 to 15 pc.  Clouds contain subunits of enhanced density gas and dust, called \textit{clumps}, wherein the earliest stages of massive star formation take place. Clumps will typically form stellar clusters \citep{Williams2000}.

Studies of massive star formation regions showed that clumps generally have sizes of the order $\sim 1$ pc, and masses ranging from order $10 \rm{M_{\odot}}$ to $\sim 10^3 - 10^4 \rm{M_{\odot}}$ \citep{Kurtz2000,Smith2009}. They are defined to be coherent in position-velocity space. Smith et al. showed that the gravitational potential of these clumps causes global collapse, which channels mass from large radii towards the center of the cluster, where protostars with the greatest gravitational radius accrete it, causing them to become massive.  Stars (or multiple systems such as binaries) eventually form from gravitationally bound sub-units in the clumps, called \textit{cores} \citep{Williams2000}.  Cores have sizes typically $\leq 0.1$ pc and masses ranging from $0.5 \rm{M_{\odot}}$ up to $\sim 10^2 - 10^3 \rm{M_{\odot}}$ \citep{Kurtz2000,Smith2009}.

We have corresponding $\rm{C^{18}O}$ maps for 51 out of the 55 $\rm{^{13}CO}$ maps.  The optically thin $\rm{C^{18}O}$ serves as a useful tracer of the central clump \citep[e.g.][]{Lopez2009}.  With a median source distance of 7.2 kpc and telescope beam of $14''$, our resolving power is of the order 0.5 pc, which, given the above definitions, implies the traced structures are more likely clumps than cores.  

The $\rm{C^{18}O}$ maps are used to calculate the $\rm{H_2}$ clump masses.  The $\rm{C^{18}O}$ column density is calculated for each clump, again using equation \ref{eq:N(CO)} with $T\rm{_{trans} = 31.6}$ K, $T\rm{_{ex}}$ unchanged, and $T\rm{_{mb}}$ the mean main-beam temperature for each clump's area, as derived from the intensity integrated image of each clump.  The $\rm{C^{18}O}$ column density of each clump is then converted to an $\rm{H_2}$ column density using the Galactocentric distance dependant isotopic abundance ratio given by \citet{Wilson1994} as $\rm{[C^{16}O]/[C^{18}O]} = 58.8D_{\rm{gal}}+37.1$, with the $\rm{[CO]/[H_2]}$ ratio the same as described before.  Finally the clump mass is calculated as 
\begin{equation}
	M\rm{_{clump}} = \left( N\rm{_{H_2}} \times A\rm{_{clump}} \right)m\rm{_{H_2}},
	\label{eq:c18omass}
\end{equation}
where $A\rm{_{clump}}$ is the surface area of each clump.  These clump masses are listed in the last column of Table \ref{tab:dustcores}, and excluding sources with distance ambiguities, they have values ranging between 10 - 2200 $\rm{M_{\odot}}$ with a mean of $\sim 420 \rm{M_{\odot}}$ and median of $\sim 190 \rm{M_{\odot}}$.  The clump masses associated with the MMAO sub-set, have a mean of $\rm{\sim 460 M_{\odot}}$.

However, \citet{Lopez2009} noted that their dust clump masses are a factor $\sim 5$ larger than the corresponding $\rm{C^{18}O}$ masses, and stated that this difference might be explained by the fact that $\rm{C^{18}O}$ and the sub-mm continuum are tracing different parts of the clump.  They found the angular FWHM measured in the sub-mm continuum surveys to be a factor $\sim 2.5$ larger than what is mapped by $\rm{C^{18}O}$ (J = 2-1), and speculated that this played a main role in the difference between the two mass estimates.  \citet{Hofner2000} also concluded from their survey that masses derived from sub-mm dust emission, tend to be systematically higher than masses derived from $\rm{C^{18}O}$ by a factor $\sim 2$. They pointed out that contributing sources of uncertainty to this discrepancy could be $\rm{C^{18}O}$ abundance, optical depth estimates, and the dust grain emissivity adopted.

Therefore, we also use continuum measurements to calculate the clump masses associated with MMAOs, and use the latter in all further discussions.  The 870 $\rm{\mu m}$ flux measurements  were obtained from the ATLASGAL survey \citep{Schuller2009,Contreras2013}, using offsets within a FWHM beam (beam size $19''$) as matching criteria.  Using the matching fluxes from \citet{Csengeri2014} for the targets in this study, clump masses were calculated following \citet{Urquhart2013}, with a gas-to-dust mass ratio assumed to be 100, dust absorption coefficient $\kappa_{\nu}$ of 1.85 $\rm{cm^2 g^{-1}}$ and dust temperature of 20 K.  All values are listed in Table \ref{tab:dustcores}. Two clump masses are listed for sources with distance ambiguities, marked with (1) next to the name for the far distance, and (2) for the near distance.  Excluding all targets with distance ambiguities, these clump masses range from $\sim 30 \rm{M_{\odot}}$ to $1.4 \times 10^4 \rm{M_{\odot}}$, have a mean value of $\sim (2.5 \pm 0.5) \times 10^3 \rm{M_{\odot}}$ and median of $\sim 1.3 \times 10^3 \rm{M_{\odot}}$.  The clump masses associated with the MMAO sub-set, have a mean of $\rm{\sim 2.8 \times 10^3 {\odot}}$. For the targets with resolved distances, $96\%$ have masses $>10^2 \rm{M_{\odot}}$ and $49\%$ have masses of the order $10^3 - 10^4 \rm{M_{\odot}}$.  This confirms that the majority of these targets are likely classified as clumps, as per definitions given above.

We find that the clump masses derived from dust measurements for our targets, are on average a factor 8 higher than masses derived using their $\rm{C^{18}O}$ emission, in agreement with \citet{Lopez2009} and \citet{Hofner2000}.

\section{Discussion}
\label{sec:Discussion}

\subsection{Clump and outflow mass relations}

\begin{figure*}
		\begin{center}$
		\begin{array}{ccc}
		\includegraphics[width = 0.445\textwidth,clip]{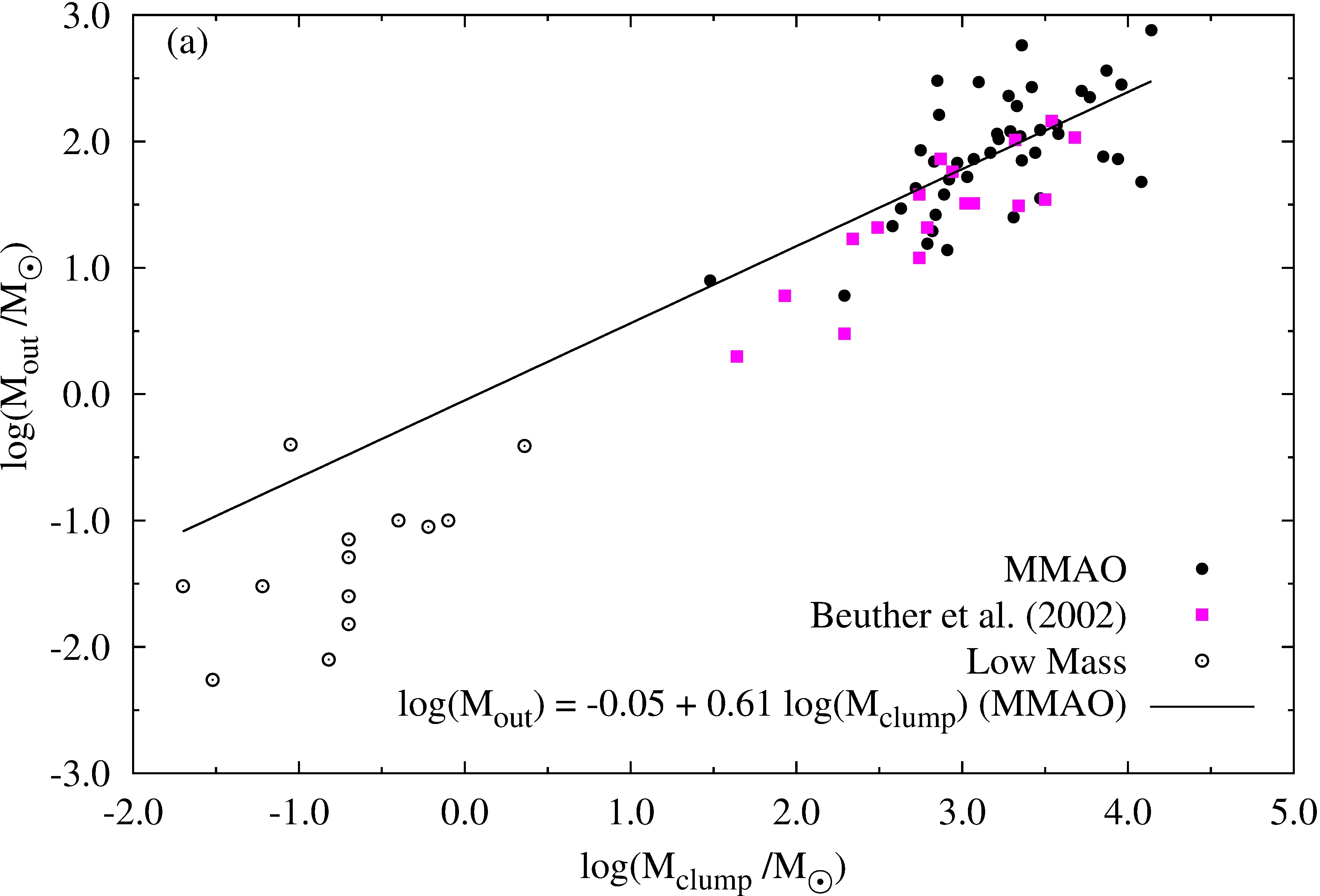} & {  } &
		\includegraphics[width = 0.455\textwidth,clip]{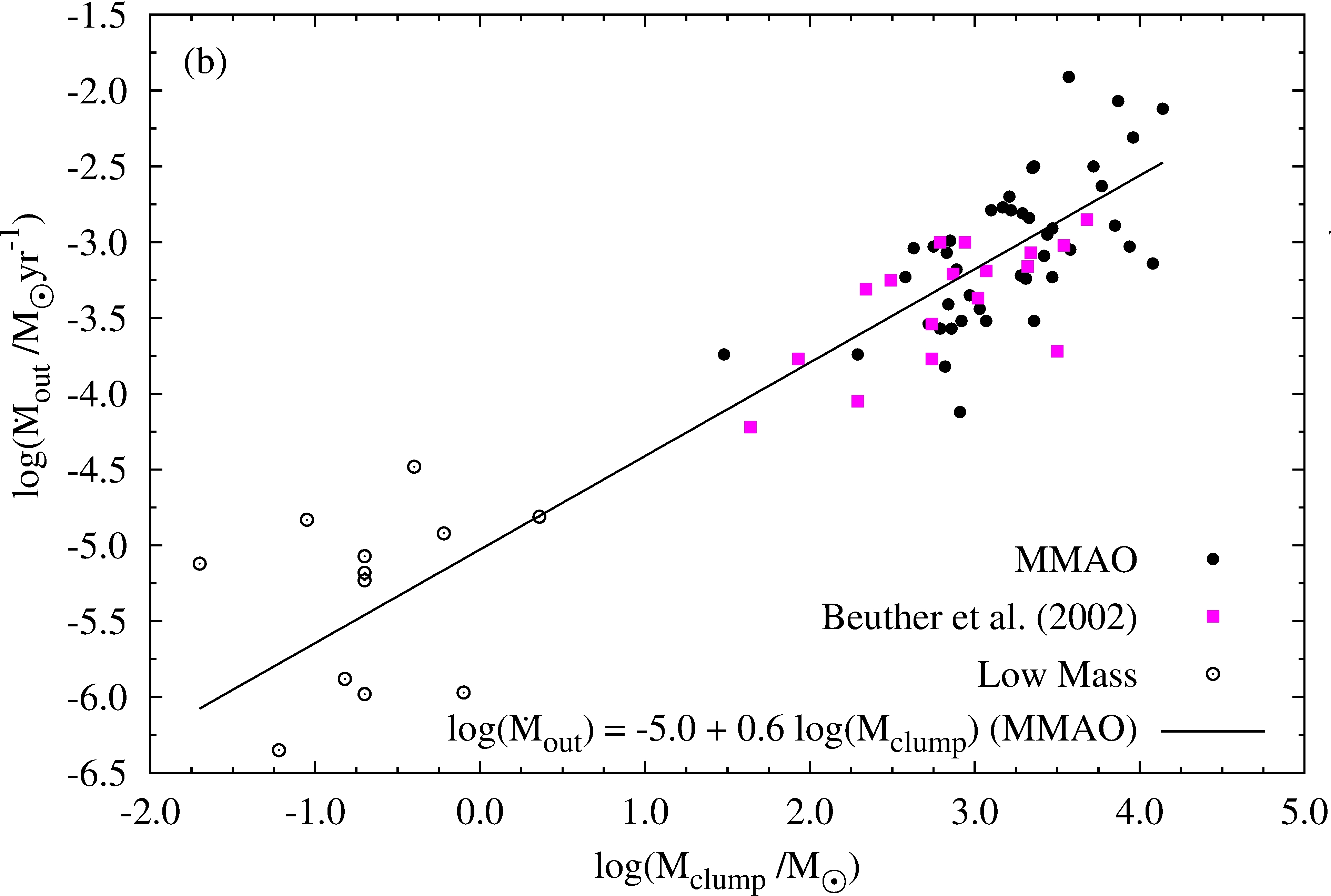} \\
		\end{array}$
		\end{center}
		\caption{ \small{Relation between (a) outflow and clump masses and (b) outflow mass loss rate and clump masses, for MMAOs (solid circles), with the best-fit power laws shown.  Empty circles indicate values for low mass YSOs, with core envelope masses from \citet{Bontemps1996} and associated outflow mass and mass loss rates from \citet{Wu2004}, \citet{Narayanan2012} and \citet{Davis2010}. Pink squares show the data from \citet{Beuther2002}.}} 
		\label{fig:MoutvsMcore}
\end{figure*}

Here we investigate the relationships between outflow properties and the mass of the clumps that they are associated with. While it is not possible to resolve the contribution from individual stars or protostellar cores in our data we can at least infer a relationship between clump mass and the most massive star present in the clump \citep[e.g.][]{Urquhart2013b}. 

\citet{McKee2003} derived a relation for the accretion rate of a free-falling envelope as a function of time, wherein it is proportional to the clump surface density $\Sigma^{0.75}$.  This implies higher mass accretion in the most massive and dense clumps, compared to those forming in lower mass clumps.  The former will arrive at the main-sequence sooner, and hence form an H{\sc ii} region much more quickly than less massive stars forming in lower-mass clumps.  \citet{Urquhart2014b} found their results to confirm this hypothesis, being consistent with the decreasing timescale with increasing massive YSO luminosity found by \citet{Mottram2011}.

In addition to this, Urquhart et al. also found that the most massive stars form predominantly towards the centres of their spherical, centrally condensed host clumps, where the highest densities exists and the gravitational potential is the deepest.  These stars evolve much quicker than those outside the central region, and reach the main sequence in $\sim 10^5$ yr, ahead of the lower-mass stars which can take $\sim 10$ times longer. Thus, while the most massive star, traced by the 6.7GHz methanol maser in this study, is near to joining the main sequence, it is likely that the lower-mass members of the proto-cluster still need to evolve to a stage where they make a significant contribution to the observed luminosity.

\citet{Urquhart2013b} found that for the H{\sc ii} regions they investigated, the bolometric luminosities (which are effectively a measure of the whole proto-cluster luminosity), were very similar to values estimated from the radio continuum flux (which only trace the most massive stars).  This suggests that it is likely that the bolometric luminosity is actually dominated by the most massive stars.  \citet{Urquhart2014b} also found that there is a strong correlation between the clump masses and the bolometric luminosities of the most massive stars, but that the total clump luminosities are much lower than would be expected from the fully formed cluster. These findings agree with the stated hypothesis, that the most massive stars have very rapid evolution times, and are consequently likely to dominate the observed clump properties.  Thus, if the luminosity of a clump is dominated by the most massive stars, then it is reasonable to assume that so too, is the luminosity, and consequently the energetics, of the outflow.

In the following, we compensate for the loss of one lobe's mass in monopolar targets by doubling the values of the detected lobe for $\rm{M_{out}}$, $p$, $E$, $\dot{M}\rm{_{out}}$, $F_m$ and $L_m$, as they all depend on the outflow mass. Figure \ref{fig:MoutvsMcore} (a) and (b) shows the relation between respectively outflow masses and mass loss rates, and clump masses for the MMAO sample (solid circles). In order to compare these relations with low-mass YSO's, we obtained associated outflow masses from \citet{Wu2004}, \citet{Narayanan2012} and \citet{Davis2010} for 13 of the core envelope masses listed by \citet{Bontemps1996} (empty circles).  The pink squares represent the clump masses from \citet{Beuther2002a}, derived from 1.2 mm dust continuum data, together with their associated outflow masses and mass loss rates, as calculated in \citet{Beuther2002}.  These values are corrected according to the erratum that was later published by \citet{Beuther2005E}, where the authors explain that their grain emissivity approximation should be a factor two higher than initially calculated, which would cause their derived clump masses to be a factor two lower than reported.  
 
In Figure \ref{fig:MoutvsMcore} (a), a best-fit power-law to the MMAO sample (circles) is given by $\rm{log}(M\rm{_{out}} /\rm{M_{\odot}}) = (-0.8 \pm 0.3)+(0.8 \pm 0.1)\rm{log}(M\rm{_{clump}} /\rm{M_{\odot}})$, holding over three orders of magnitude for massive outflows, and extending toward the low mass regime to cover six orders of magnitude in total. 

\citet{Lopez2009} converted their clump masses derived from $\rm{C^{18}O}$ to dust masses, and found a tight correlation between $M\rm{_{clump}}$ and $M\rm{_{out}}$, described by $M\rm{_{out}} = 0.3 M^{0.8}\rm{_{clump}}$.  \citet{SanchezMonge2013} did a similar fit, but they calculated their outflow masses from SiO observations and clump masses from SED fits to Hi-GAL data.  They found the same relation as \citet{Lopez2009}.  The power-law found for MMAOs agrees with these authors within uncertainties.  It is interesting to note that the outflow masses and mass loss rates for MMAOs are generally higher than those estimated by \citet{Beuther2002}.  We will discuss this property more detail in an upcoming second paper (de Villiers et al. 2014b in prep.).

Figure \ref{fig:MoutvsMcore} (b) shows the best-fit linear relation between the logarithmic values for the mass loss rates and clump masses for MMAOs, given by $\rm{log}(\dot{M}_{\rm{out}} /\rm{M_{\odot} yr^{-1}}) = (-5.0 \pm 0.4)+(0.6 \pm 0.1)\rm{log}(M\rm{_{clump}} /\rm{M_{\odot}})$.  It again holds over three orders of magnitude for massive outflows, and extend to the low mass regime to cover six orders of magnitude in total. 

As both $M\rm{_{out}}$ and $\dot{M}_{\rm{out}}$ depend on mass and distance, a nonparametric measure of the statistical dependence between these parameters and $M\rm{_{clump}}$ is needed.  The Spearman-rank test is used, similarly to \citet{Ridge2001}, where a perfect positive or negative correlation exists between the ranks when $r_s$ is $\pm 1$.  No correlation exists when $r_s=0$. The Spearman parameter $r_s$ is related to the two-tailed Student's $t$-test via, $t_s = r_s \sqrt{(n-2)/(1-r^{2}_{s})}$ for a sample size $n$.  For no significant correlation, $|t_s| < |t_{s\rm{_{crit}}|}$.  For a sample size of 43 (MMAO sample of 44 had 43 matches from ATLAGAL for clump masses), $t_{s\rm{_{crit}}}$ is $\rm{\pm 2.02}$.  The relation between $M\rm{_{out}}$ and $M\rm{_{clump}}$ for MMAOs is statistically significant with $r_s=0.59$ ($t_s=4.66$).  The same is true for the relation between $\dot{M}_{\rm{out}}$ and $M\rm{_{clump}}$, where $r_s=0.64$ ($t_s=5.28$).  

The result that both relations in Figure \ref{fig:MoutvsMcore} are found to hold over six orders of magnitude when extrapolated to the low-mass regime, suggests that a similar process causes outflows in both low and high mass star formation.
 
\begin{figure*}
	\begin{center}$
		\begin{array}{cc}
		\includegraphics[width = 0.44\textwidth,clip]{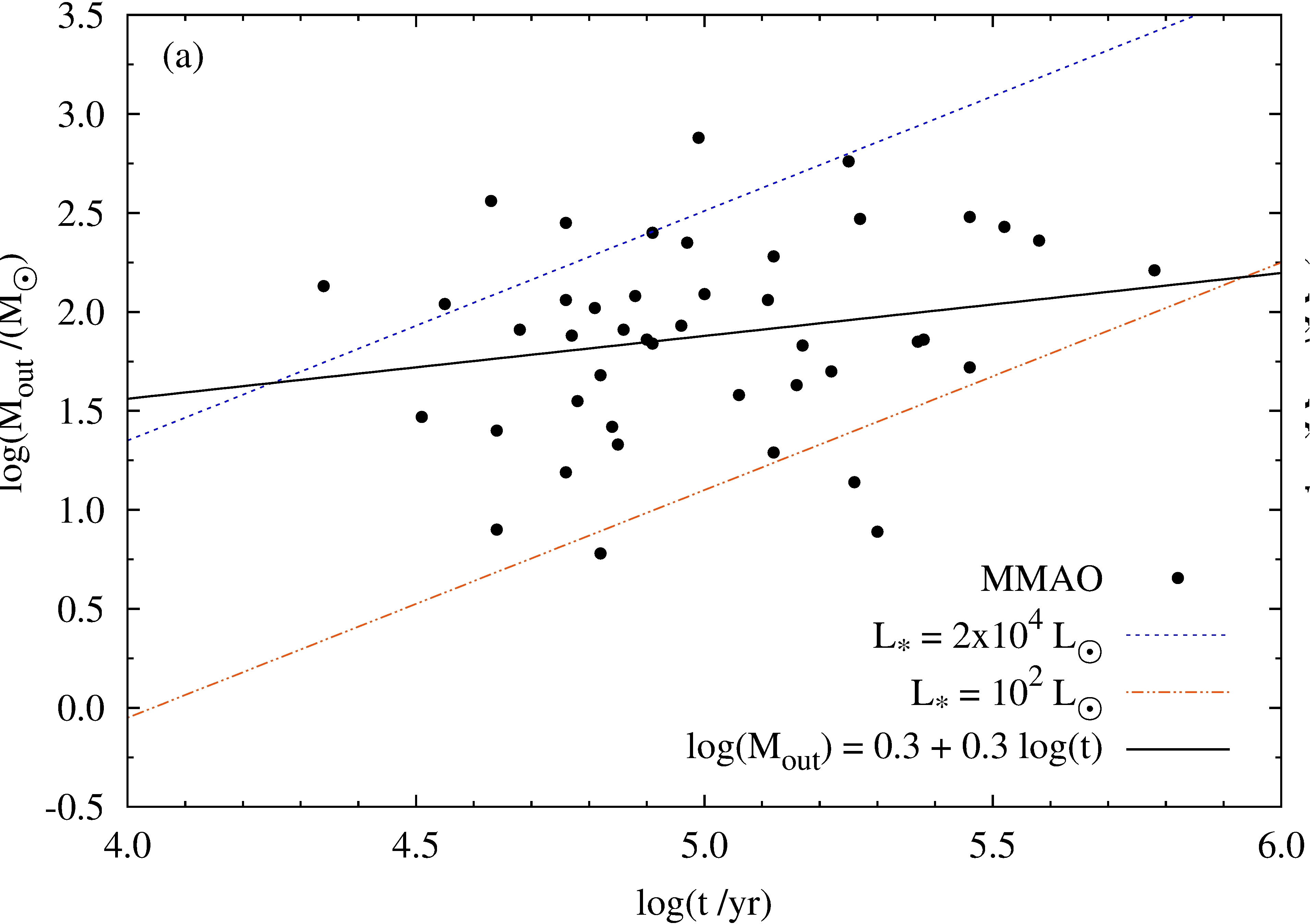} &
		\includegraphics[width = 0.46\textwidth,clip]{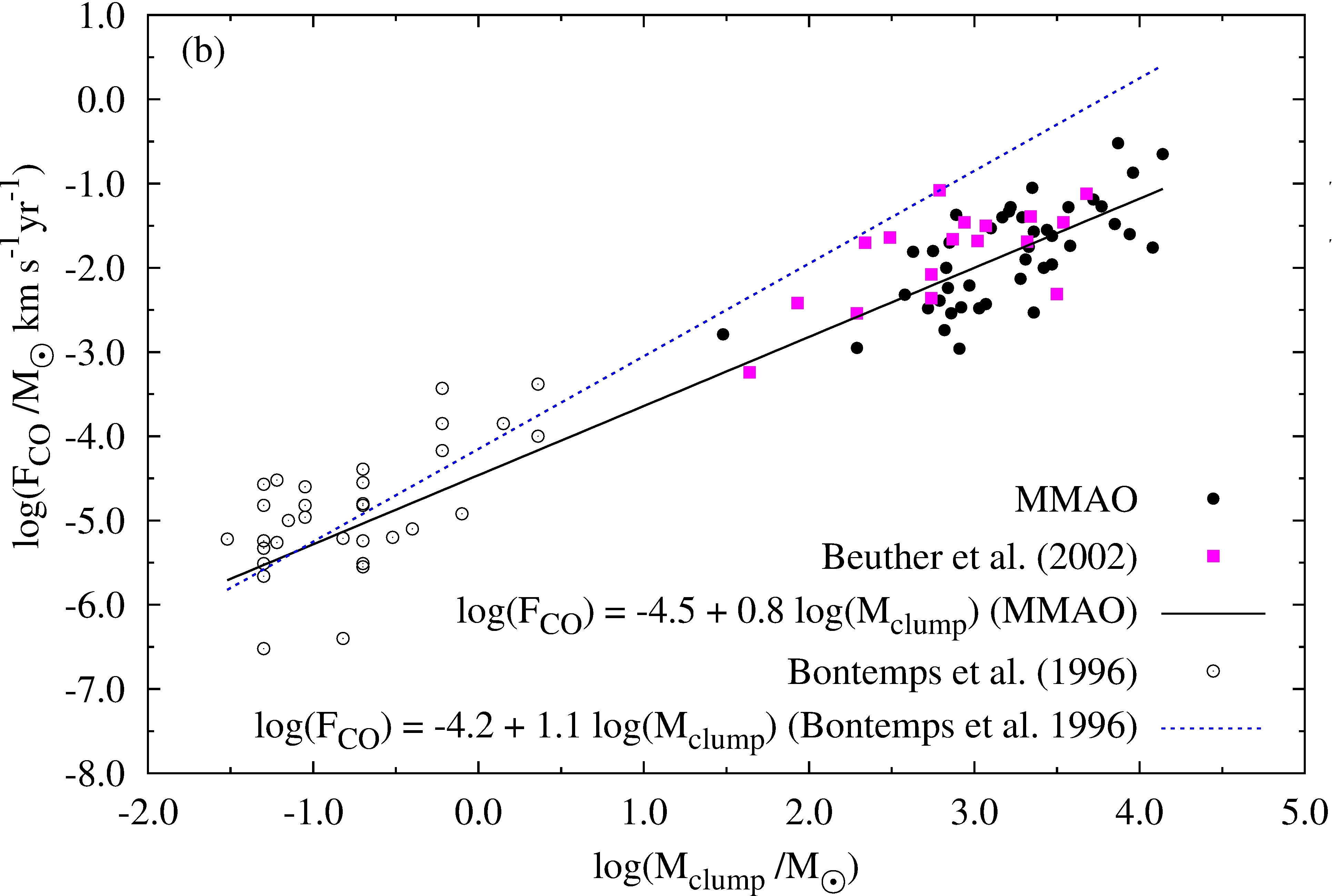} \\
	\end{array}$
		\end{center}
		\caption{ \small{(a) Outflow masses versus dynamical time as derived from $\rm{^{13}CO}$ for MMAOs, with the best-fit power law indicated by the solid line. Dot-dashed and dotted lines show respectively the lower and upper luminosity boundaries as defined by \citet{Shepherd1996b}. (b) Outflow mechanical force derived from $\rm{^{13}CO}$ versus associated clump masses for MMAOs (solid circles).  The solid line is a power-law fit to MMAOs and the dotted line the fit from \citet{Bontemps1996}. Empty circles represent this relation for low mass YSOs \citep{Bontemps1996}, and pink squares show the data from \citet{Beuther2002}.}}
		\label{fig:Mrelations2}
\end{figure*}

Figure \ref{fig:Mrelations2} (a) shows the relation between the outflow mass and dynamical timescales for MMAOs. A power law fitted to the data resulted in a poor correlation given by $\rm{log}(M\rm{_{out}} /\rm{M_{\odot}}) = (0.3 \pm 1.2)+(0.3 \pm 0.3)\rm{log}(t /\rm{yr^{-1}})$, with $r_s = 0.18$ ($t_s=1.18$, for a complete MMAO sample of 44) implying no significant correlation, as also found by \citet{Wu2004}. Our range of dynamical times is possibly too small for any significant evolutionary trends, such as the increase in outflow mass, to be observed.

\citet{Shepherd1996b} plotted the same relation for their ten massive star forming regions.  They divided their plot into three regions based on the bolometric luminosities of the outflow sources in \citet{Cabrit1992}. We show the same luminosity boundaries in Figure \ref{fig:Mrelations2} (a). Using data from \citet{Cabrit1992}, \citet{Shepherd1996b} found that sources with $L_* < 10^2 L_{\odot}$ were located below the bottom line and sources with $L_* > 2 \times 10^4 L_{\odot}$ above the top.  This sectioning contributes a useful estimate of the expected bolometric luminosity of the YSO associated with each outflow studied - a property that could not be derived directly from the available data.  Consistent with what is expected for massive YSO's, the majority ($95\%$) of the sources occur above the $L_* = 10^2 L_{\odot}$ boundary, with $\sim 79\%$ of them between $L_* = 10^2 L_{\odot}$ and $L_* = 2 \times 10^4 L_{\odot}$, and $\sim 16\%$ above $L_* = 2 \times 10^4 L_{\odot}$.  Also, we find that higher luminosity (and hence more massive) YSO's will drive the higher mass outflows we study in this paper.  A more massive accreting YSO is likely to have higher angular momentum, hence power larger outflows.  In addition to this, star formation regions that form O and B stars tend to have larger reservoirs of material available to be entrained into a molecular outflow.

Figure \ref{fig:Mrelations2} (b) shows the relation between the CO momentum flux / mechanical force $F\rm{_{CO}}$ and the clump masses for MMAOs (circles), together with the values from \citet{Beuther2002} (pink squares), as well as with 33 low mass sources from \citet{Bontemps1996} (triangles).  $F\rm{_{CO}}$ is the inclination corrected mechanical force, derived from $F_m$ by applying a correction factor of 2.9, corresponding to a mean inclination angle of $57.3^o$ (i.e. assuming random outflow orientations, \citet{Beuther2002} and Table \ref{tab:inclcorrections}).  Although we do not apply inclination corrections to the data in this study, in this specific case we correct $F_m$ values in order to compare the CO momentum flux values from MMAOs with those from \citet{Beuther2002} and \citet{Bontemps1996}.  Once again, as in Figure \ref{fig:MoutvsMcore}, MMAOs' momentum flux values are generally higher than those estimated by \citet{Beuther2002} (also to be discussed by de Villiers et al., 2014b in prep.).

The outflow's mechanical force is a very important parameter in studying the early phases of star formation.  It is a measure of the rate at which momentum is injected from the underlying driving agent, via interactions with the molecular gas in the core, into the envelope.  In other words, it is a measure of the outflow's strength and used to understand the driving mechanism of outflows \citep{Bachiller1999,Downes2007}. \citet{Hatchell2007} suggests that, although it is difficult to separate this effect from contamination due to the initial conditions, the correlation between $F\rm{_{CO}}$ and $M\rm{_{clump}}$ suggests that the outflow activity declines during the later stages of the accretion phase (when the clump mass decreases).

Many authors have studied this relation before, both for low mass outflows \citep[e.g.][]{Bontemps1996,Hatchell2007,Takahashi2008,Curtis2010,vanderMarel2013} and high mass outflows \citep[e.g.][]{Henning2000,Beuther2002}.
For our massive MMAOs, we find the relation to be $\rm{log} \left( F\rm{_{CO}} /\rm{M_{\odot} kms^{-1} yr^{-1}} \right) = (-4.5 \pm 0.4) + (0.8 \pm 0.1)\rm{log}\left( M\rm{_{clump}} /\rm{M_{\odot}} \right)$ (black solid line) with a Spearman-rank coefficient of $r_s=0.66 (t_s=5.62)$.  This relation found for MMAOs corresponds within uncertainties to what \citet{Bontemps1996} found for low mass sources, $\rm{log} \left( F\rm{_{CO}} /\rm{M_{\odot} kms^{-1} yr^{-1}} \right) = -(4.2 \pm 0.1) + (1.1 \pm 0.2) \rm{log} \left( M\rm{_{clump} /{M_{\odot}}} \right)$ (blue dotted line). 

If, as stated in the beginning of this section, the outflow energetics are dominated by the most massive cores in a clump harbouring massive star formation, the presence of the same relationship between the outflow parameters and the clump masses for both high and low masses, is consistent with the theory that massive star formation is an scaled-up version of the low-mass scenario.  

\subsection{Mass loss rate and accretion rates}
\label{sec:pardiscussion}

If massive outflows are produced by the same mechanism as low-mass outflows, it implies that the outflows are momentum driven by
the jet coming from the central YSO which entrains the surrounding molecular gas and forms the outflow.  One of the proposed solutions enabling disk accretion to be a possible formation process for massive stars, is when accretion proceeds through a disk with a high enough mass accretion rate to overcome the radiation pressure of the central massive star \citep{Jijina1996,Yorke2002}.  Even more involved disk accretion models including asymmetric configurations and Rayleigh-Taylor instabilities \citep{Krumholz2007,Krumholz2009}, assume accretion rates of the order $\rm{10^{-4} ~ M_{\odot}yr^{-1}}$.  High accretion rates ($\rm{\sim 10^{-5} - 10^{-3} ~ M_{\odot}yr^{-1}}$) are also used in more recent disk accretion models that incorporate a dust sublimation front and consequently yielded the growth of the highest-mass stars ever formed in multi-dimensional radiation hydrodynamic simulations \citep{Kuiper2010}.  

If one assumes that the momenta of the observed outflow and the jet entraining the outflow is conserved; \citep[if there is efficient mixing at the jet/molecular gas interface and zero loss of momentum to the ISM][]{Richer2000}, \citet{Beuther2002} stated that one could equate the momenta of the outflow and jet by
\begin{equation}
	M\rm{_{out}}v\rm{_{out}} = M\rm{_{jet}}v\rm{_{jet}}.
	\label{eq:momconserv}	
\end{equation}
Based on previous studies, they assumed that $v\rm{_{jet}}/v\rm{_{out}} \sim 20$.  Using equation \ref{eq:momconserv}, one can write an expression for the jet mass loss rate as $\dot{M}\rm{_{jet}} = M\rm{_{out}}v\rm{_{out}}/v\rm{_{jet}}t\rm{_{dyn}}$.  Together with Beuther's assumption for jet and outflow velocity ratios, this mass loss rate from jets is described by $\dot{M}\rm{_{jet}} = M\rm{_{out}}v\rm{_{out}}/20~v\rm{_{out}}t_{\rm{dyn}} = 0.05\dot{M}\rm{_{out}}$.  Furthermore, \citet{Beuther2002} also assumed that $\dot{M}\rm{_{jet}}/\dot{M}_{\rm{out}_{accr}} \sim 0.3$ based on \citet{Tomisaka1998, Shu1999}, which leads to the following expression for accretion rate in terms of outflow mass loss rate:
\begin{equation}
	\dot{M}_{\rm{out}_{accr}} \sim \frac{\dot{M}_{\rm{out}}}{6}.
\end{equation}

The mean $\dot{M}\rm{_{out}}$ for MMAOs is $\rm{\sim 1.7 \times 10^{-3} ~M_{\odot} / yr}$, which would lead to accretion rates of $\rm{\sim 3 \times 10^{-4} ~M_{\odot} / yr}$, being of the same order of magnitude as the $\rm{\sim 10^{-4} ~M_{\odot} / yr}$ found by both \citet{Beuther2002} and \citet{Kim2006}.  Our approximate accretion rate for MMAOs is also much higher than typical accretion rates of $\rm{10^{-7} - 10^{-5} ~M_{\odot} / yr}$ expected for low mass YSOs \citep{Shu1977}, and agrees with the theoretical rates used in disk accretion models for massive stars \citep{Krumholz2007,Kuiper2010}. 

\subsection{Comparison with Methanol masers}

Integrated 6.7GHz flux densities ($S_i$) are obtained from the MMB catalogue (Breen et al. in prep., 2014) for 38 of the MMAO masers.  Using the distance of each target, $S_i$ is converted to an integrated spectral luminosity $L\rm{_{6.7GHz}}$.

\begin{figure}
		\begin{center}
		\includegraphics[width = 0.45\textwidth,clip]{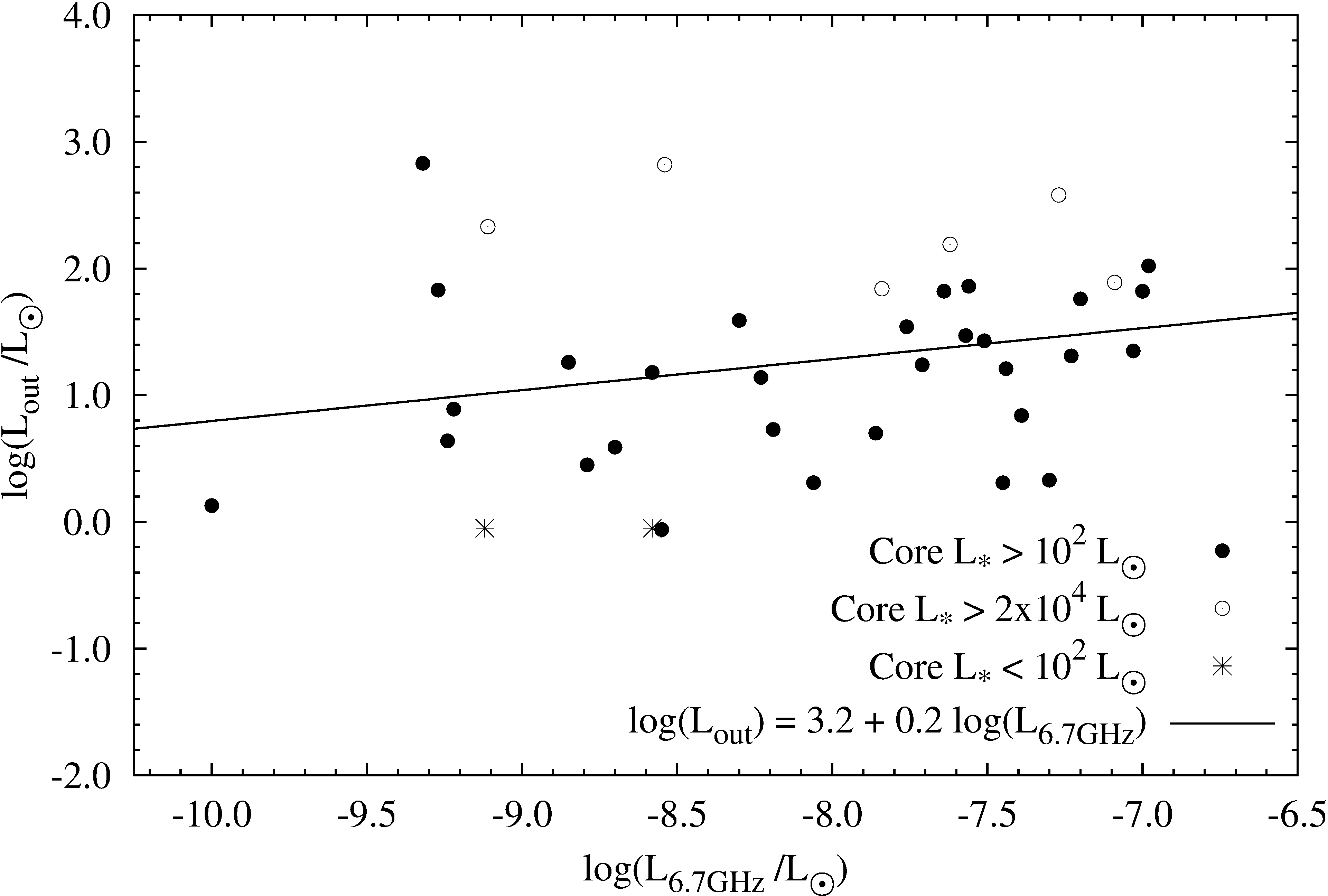}
		\end{center}
		\caption{ \small{Outflow mechanical luminosities $L_m$ derived from $\rm{^{13}CO}$ plotted versus 6.7GHz Maser luminosities.  The solid line indicates the best-fit power law to the data.  Empty circles indicate sources located above the $2 \times 10^4 \rm{L_{\odot}}$ bolometric luminosity boundary in Figure \ref{fig:Mrelations2} (a), solid circles are sources located between the $L_* = 10^2 L_{\odot}$ and $L_* = 2 \times 10^4 L_{\odot}$ boundaries, and star symbols indicate sources located below $L_* = 10^2 L_{\odot}$.}}
		\label{fig:Lum_OutvsMaser}
\end{figure}
Figure \ref{fig:Lum_OutvsMaser} shows the relation between the outflow mechanical luminosities (or energy supply rates) and maser luminosities for 38 masers from the MMAO sub-set, given as $\rm{log} \left( L_m /\rm{L_{\odot}} \right) = (3.2 \pm 1.3) + (0.2 \pm 0.2)\rm{log} \left(L\rm{_{6.7GHz} /\rm{L_{\odot}}} \right)$ and a Spearman-Rank coefficient of $r_s = 0.28 (t_s=1.78)$, which, although not quite, is at the margin of being statistically significant as the $5\%$ acceptance interval for t is $\pm 2.02$. 

Furthermore, following their positions in Figure \ref{fig:Mrelations2} (a), we divide all the outflows plotted in Figure \ref{fig:Lum_OutvsMaser} into three categories; outflows associated with clumps whose luminosities above $2 \times 10^4 \rm{L_{\odot}}$, between the $L_* = 10^2 \rm{L_{\odot}}$ and $L_* = 2 \times 10^4 \rm{L_{\odot}}$, and below $L_* = 10^2 \rm{L_{\odot}}$.  The data points in Figure \ref{fig:Lum_OutvsMaser} are marked accordingly. Although bright outflows are mostly associated with bright clumps, and low luminosity outflows associated with lower luminosity clumps, there is no such preference for maser brightness.

\begin{figure}
		\begin{center}
		\includegraphics[width = 0.45\textwidth,clip]{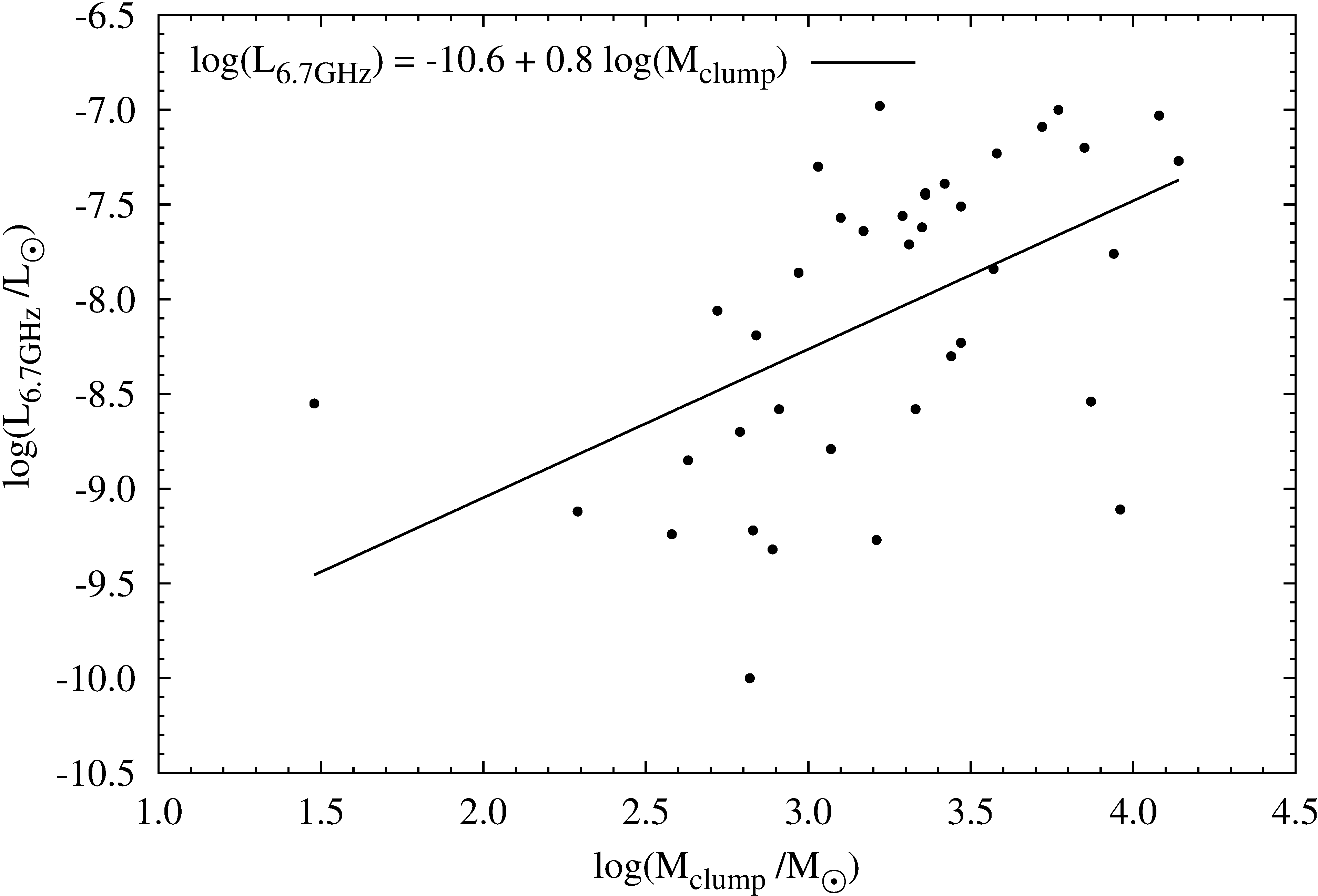}
		\end{center}
		\caption{ \small{6.7GHz Maser luminosities versus clump masses derived from 870 $\rm{\mu m}$ continuum emission.}}
		\label{fig:MaserL_vs_Mcore}
\end{figure}

Finally we plot the relation between the 6.7GHz maser luminosities against clump masses in Figure \ref{fig:MaserL_vs_Mcore} and found the best-fit linear relation between the logarithmic values of these parameters to be $\rm{log}(L\rm{_{6.7GHz}} /\rm{L_{\odot}}) = (-10.6 \pm 0.7) + (0.8 \pm 0.2)log(M\rm{_{clump}} /\rm{M_{\odot}})$. The associated value for $r_s$ is 0.59 ($t_s = 4.36$), which implies a statistically significant positive correlation.  

\citet{Breen2010} studied the relation between 101 1.2mm dust clumps and their associated 6.7GHz methanol masers. Their study showed that more luminous 6.7GHz methanol masers are likely to be associated with dust clump sources that have bigger radii, higher mass, and lower clump densities (the latter possibly due to the assumption of a constant dust temperature).  These results are represented graphically in their Figure A2.  \citet{Wu2010} also found molecular clumps (derived from N$\rm{H_3}$ lines) associated with high-luminosity 6.7GHz masers to be larger and $\sim 10$ times more massive than those associated with low-luminosity masers.  They also found that outflows associated with high-luminosity masers have wider line wings and larger sizes than those associated with low-luminosity masers. This lead to their interpretation that masers with higher luminosities are associated with YSO's with larger masses.

\citet{Urquhart2013} found a weak linear correlation between the maser luminosity and clump masses for 442 ATLASGAL-MMB associations. They speculated that this correlation may be related to the fact that the most massive clumps are likely to form more massive stars, and that a higher isotropic maser luminosity is somehow related to a higher stellar luminosity. 

The results from Figures \ref{fig:MaserL_vs_Mcore} and \ref{fig:Lum_OutvsMaser} agrees with above authors and suggest that there is some correlation between the mass (hence also brightness) of a massive YSO and both the luminosity of the outflow it generates as well as the total 6.7GHz maser luminosity it pumps.  Note that in this case, we know that a higher maser luminosity could not be due to a larger number of masers in more massive regions, as all our MMB-sample selected masers have supporting interferometric observations that reveal they are principally single maser spots.

\section{Summary and conclusions}
\label{sec:Conclusion}

We analysed the $\rm{^{13}CO}$ and $\rm{C^{18}O}$ spectra extracted at 58 $\rm{^{13}CO}$ clump coordinates toward 6.7GHz methanol maser coordinates between $20^o < l < 34^o$. All spectra showed high velocity outflow signatures. Of these, the high velocity emission was mapped for 55 with 47 showing bipolar structures.  A sub-set containing 44 targets is referred to as MMAOs (Methanol Maser Associated Outflows), with the criteria that all members have to have resolved kinematic distances and be closely associated with a methanol maser.  Only MMAOs were used for further analysis.   

The wing spectra and spatial maps were used to calculate the physical properties of all 55 mapped outflows, generally following \citet{Beuther2002}, with a few adjustments in their approach for this study.  The associated clump masses for MMAOs were calculated from the 870 $\rm{\mu m}$ flux measurements towards these targets, of the ATLASGAL survey \citep{Csengeri2014}.

The main results for this study can be summarized as follows: \newline
(i)  A statistically significant relation over three orders of magnitude was found between the outflow and clump masses for the MMAOs, given by $\rm{log}(M\rm{_{out}} /\rm{M_{\odot}}) = (-0.8 \pm 0.3)+(0.8 \pm 0.1)\rm{log}(M\rm{_{clump}} /\rm{M_{\odot}})$. This relation agreed, within uncertainties, with \citet{Beuther2002}, \citet{Lopez2009} and \citet{SanchezMonge2013}.  Low-mass sources \citep{Bontemps1996,Wu2004,Narayanan2012,Davis2010} were found to follow a similar trend as the massive sources, expanding the power-law relationship to over six orders of magnitude.  

(ii) The relation between the outflow mass loss rate and clump masses was described by $\rm{log}(\dot{M}_{\rm{out}} /\rm{M_{\odot} yr^{-1}}) = (-5.0 \pm 0.4)+(0.6 \pm 0.1)\rm{log}(M\rm{_{clump}} /\rm{M_{\odot}})$, which held over six magnitudes down to the low-mass regime.  

(iii) When plotting outflow masses against dynamical timescales, $95\%$ of the MMAOs occurred above the $L_* = 10^2 \rm{L_{\odot}}$ boundary for YSO luminosities, with $\sim 79\%$ of them between $L_* = 10^2 \rm{L_{\odot}}$ and $L_* = 2 \times 10^4 \rm{L_{\odot}}$, and $\sim 16\%$ above $L_* = 2 \times 10^4 \rm{L_{\odot}}$.  This suggests that higher luminosity (and hence more massive) YSOs will generate higher mass outflows.

(iv) The relationship between the mechanical force (CO momentum flux) $F\rm{_{CO}}$ and the clump masses was described by $\rm{log} \left( F\rm{_{CO}} /\rm{M_{\odot} kms^{-1} yr^{-1}} \right) = (-4.5 \pm 0.4) + (0.8 \pm 0.1)\rm{log}\left( M\rm{_{clump}} /\rm{M_{\odot}} \right)$, a trend which held over six orders of magnitude if we included low mass outflows. 

(v)  We derived an approximate accretion rate of $\rm{\sim 3 \times 10^{-4} ~M_{\odot} / yr}$ from our mean mass loss rate.  This is of the same order of magnitude than both the mean accretion rates found by \citet{Beuther2002} and \citet{Kim2006} and agrees with the theoretical rates used in disk accretion models for massive stars \citep{Krumholz2007,Kuiper2010}.

(vi) An investigation of the relation between the outflow mechanical luminosities and maser luminosities for 38 MMAO targets showed a weak positive correlation wherein bright outflows were associated with brighter masers and clumps.  However, low luminosity outflows and clumps showed no preference for maser brightness.  The relation between the 6.7GHz maser luminosities and clump masses was given by $\rm{log}(L\rm{_{6.7GHz}} /\rm{L_{\odot}}) = (-10.6 \pm 0.7) + (0.8 \pm 0.2)log(M\rm{_{clump}} /\rm{M_{\odot}})$. Although weakly correlated, these relations suggest that there is a correlation between the mass (hence also brightness) of a massive YSO and both the luminosity of the outflow it generates as well as the 6.7GHz maser it pumps. It agrees with the speculation from \citet{Urquhart2013} that such a correlation may be related to the fact that the most massive clumps are likely to form more massive stars, and that a higher isotropic maser luminosity is somehow related to a higher stellar luminosity.  

Results (i), (ii) and (vi) indicate that both the outflow mass relations and energetics follow a common relation for low- and high mass YSOs.  This lends evidence to the hypothesis that a similar process is responsible for outflows in both mass regimes, i.e. high mass star formation is a scaled-up version of the low mass scenario.  In addition to this, result (v) shows that the approximate mass accretion rate for MMAOs is sufficiently high to overcome the radiation pressure of a massive central star.  

Although this evidence suggests that massive stars form in a similar fashion to low mass stars, to determine whether each massive YSO indeed drives it's own outflow, high resolution imaging is required.

In a following paper (de Villiers et al. 2014b in prep.), the outflow property distributions are compared to those from other surveys, focusing on the evolutionary sequence of 6.7GHz methanol masers and molecular outflows during the hot core phase.

\section{Acknowledgements}

We want to thank James Caswell for useful comments and contributions to the paper.  Timea Csengeri acknowledges financial support for the ERC Advanced Grant GLOSTAR under contract no. 247078.  The JCMT is operated by the Joint Astronomy Centre on behalf of the STFC of the United Kingdom, the National Research Council of Canada, and (until 31 March 2013) the Netherlands Organisation for Scientific Research. The program ID for the observations is M07AU20. We also thank the anonymous referee for constructive comments on the paper.

\label{sec:acknowledgements}

\bibliographystyle{mn2e}
\bibliography{Sources}

\begin{thebibliography}{131}
\expandafter\ifx\csname natexlab\endcsname\relax\def\natexlab#1{#1}\fi

\bibitem[{{Arce} {et~al.}(2010){Arce}, {Borkin}, {Goodman}, {Pineda}, \&
  {Halle}}]{Arce2010}
{Arce}, H.~G., {Borkin}, M.~A., {Goodman}, A.~A., {Pineda}, J.~E., \& {Halle},
  M.~W. 2010, ApJ, 715, 1170

\bibitem[{{Bachiller} \& {Tafalla}(1999)}]{Bachiller1999}
{Bachiller}, R. \& {Tafalla}, M. 1999, in NATO ASIC Proc. 540: The Origin of
  Stars and Planetary Systems, ed. {C.~J.~Lada \& N.~D.~Kylafis}, 227--+

\bibitem[{{Bally} \& {Zinnecker}(2005)}]{Bally2005}
{Bally}, J. \& {Zinnecker}, H. 2005, AJ, 129, 2281

\bibitem[{{Bergin} \& {Tafalla}(2007)}]{Bergin2007}
{Bergin}, E.~A. \& {Tafalla}, M. 2007, ARA\&A, 45, 339

\bibitem[{{Beuther} {et~al.}(2002{\natexlab{a}}){Beuther}, {Schilke}, {Menten},
  {Motte}, {Sridharan}, \& {Wyrowski}}]{Beuther2002a}
{Beuther}, H., {Schilke}, P., {Menten}, K.~M., {Motte}, F., {Sridharan}, T.~K.,
  \& {Wyrowski}, F. 2002{\natexlab{a}}, ApJ, 566, 945

\bibitem[{{Beuther} {et~al.}(2005){Beuther}, {Schilke}, {Menten}, {Motte},
  {Sridharan}, \& {Wyrowski}}]{Beuther2005E}
{Beuther}, H., {Schilke}, P., {Menten}, K.~M., {Motte}, F., {Sridharan}, T.~K.,
  \& {Wyrowski}, F. 2005, ApJ, 633, 535

\bibitem[{{Beuther} {et~al.}(2002{\natexlab{b}}){Beuther}, {Schilke},
  {Sridharan}, {Menten}, {Walmsley}, \& {Wyrowski}}]{Beuther2002}
{Beuther}, H., {Schilke}, P., {Sridharan}, T.~K., {Menten}, K.~M., {Walmsley},
  C.~M., \& {Wyrowski}, F. 2002{\natexlab{b}}, A\&AS, 383, 892

\bibitem[{{Bonnell} {et~al.}(1998){Bonnell}, {Bate}, \&
  {Zinnecker}}]{Bonnell1998}
{Bonnell}, I.~A., {Bate}, M.~R., \& {Zinnecker}, H. 1998, MNRAS, 298, 93

\bibitem[{{Bonnell} {et~al.}(2004){Bonnell}, {Vine}, \& {Bate}}]{Bonnell2004}
{Bonnell}, I.~A., {Vine}, S.~G., \& {Bate}, M.~R. 2004, MNRAS, 349, 735

\bibitem[{{Bontemps} {et~al.}(1996){Bontemps}, {Andre}, {Terebey}, \&
  {Cabrit}}]{Bontemps1996}
{Bontemps}, S., {Andre}, P., {Terebey}, S., \& {Cabrit}, S. 1996, A\&AS, 311,
  858

\bibitem[{{Brand} \& {Blitz}(1993)}]{Brand1993}
{Brand}, J. \& {Blitz}, L. 1993, A\&AS, 275, 67

\bibitem[{{Breen} {et~al.}(2010){Breen}, {Ellingsen}, {Caswell}, \&
  {Lewis}}]{Breen2010}
{Breen}, S.~L., {Ellingsen}, S.~P., {Caswell}, J.~L., \& {Lewis}, B.~E. 2010,
  MNRAS, 401, 2219

\bibitem[{{Breen} {et~al.}(2013){Breen}, {Ellingsen}, {Contreras}, {Green},
  {Caswell}, {Stevens}, {Dawson}, \& {Voronkov}}]{Breen2013}
{Breen}, S.~L., {Ellingsen}, S.~P., {Contreras}, Y., {Green}, J.~A., {Caswell},
  J.~L., {Stevens}, J.~B., {Dawson}, J.~R., \& {Voronkov}, M.~A. 2013, MNRAS,
  435, 524

\bibitem[{{Bronfman} {et~al.}(2008){Bronfman}, {Garay}, {Merello}, {Mardones},
  {May}, {Brooks}, {Nyman}, \& {G{\"u}sten}}]{Bronfman2008}
{Bronfman}, L., {Garay}, G., {Merello}, M., {Mardones}, D., {May}, J.,
  {Brooks}, K.~J., {Nyman}, L.-{\AA}., \& {G{\"u}sten}, R. 2008, ApJ, 672, 391

\bibitem[{{Buckle} {et~al.}(2010){Buckle}, {Curtis}, {Roberts}, {White},
  {Hatchell}, {Brunt}, {Butner}, {Cavanagh}, {Chrysostomou}, {Davis},
  {Duarte-Cabral}, {Etxaluze}, {di Francesco}, {Friberg}, {Friesen}, {Fuller},
  {Graves}, {Greaves}, {Hogerheijde}, {Johnstone}, {Matthews}, {Matthews},
  {Nutter}, {Rawlings}, {Richer}, {Sadavoy}, {Simpson}, {Tothill}, {Tsamis},
  {Viti}, {Ward-Thompson}, {Wouterloot}, \& {Yates}}]{Buckle2010}
{Buckle}, J.~V., {et~al.} 2010, MNRAS, 401, 204

\bibitem[{{Buckle} {et~al.}(2009){Buckle}, {Hills}, {Smith}, {Dent}, {Bell},
  {Curtis}, {Dace}, {Gibson}, {Graves}, {Leech}, {Richer}, {Williamson},
  {Withington}, {Yassin}, {Bennett}, {Hastings}, {Laidlaw}, {Lightfoot},
  {Burgess}, {Dewdney}, {Hovey}, {Willis}, {Redman}, {Wooff}, {Berry},
  {Cavanagh}, {Davis}, {Dempsey}, {Friberg}, {Jenness}, {Kackley}, {Rees},
  {Tilanus}, {Walther}, {Zwart}, {Klapwijk}, {Kroug}, \&
  {Zijlstra}}]{Buckle2009}
{Buckle}, J.~V., {et~al.} 2009, MNRAS, 399, 1026

\bibitem[{{Cabrit} \& {Bertout}(1990)}]{Cabrit1990}
{Cabrit}, S. \& {Bertout}, C. 1990, ApJ, 348, 530

\bibitem[{{Cabrit} \& {Bertout}(1992)}]{Cabrit1992}
{Cabrit}, S. \& {Bertout}, C. 1992, A\&AS, 261, 274

\bibitem[{{Cabrit} {et~al.}(1988){Cabrit}, {Goldsmith}, \&
  {Snell}}]{Cabrit1988}
{Cabrit}, S., {Goldsmith}, P.~F., \& {Snell}, R.~L. 1988, ApJ, 334, 196

\bibitem[{{Caswell}(2013)}]{Caswell2013}
{Caswell}, J.~L. 2013, in IAU Symposium, Vol. 292, IAU Symposium, ed. T.~{Wong}
  \& J.~{Ott}, 79--82

\bibitem[{{Caswell} \& {Green}(2011)}]{Caswell2011}
{Caswell}, J.~L. \& {Green}, J.~A. 2011, MNRAS, 411, 2059

\bibitem[{{Cavanagh} {et~al.}(2008){Cavanagh}, {Jenness}, {Economou}, \&
  {Currie}}]{Cavanagh2008}
{Cavanagh}, B., {Jenness}, T., {Economou}, F., \& {Currie}, M.~J. 2008,
  Astronomische Nachrichten, 329, 295

\bibitem[{{Cesaroni} {et~al.}(2007){Cesaroni}, {Galli}, {Lodato}, {Walmsley},
  \& {Zhang}}]{Cesaroni2007}
{Cesaroni}, R., {Galli}, D., {Lodato}, G., {Walmsley}, C.~M., \& {Zhang}, Q.
  2007, Protostars and Planets V, 197

\bibitem[{{Cesaroni} {et~al.}(1992){Cesaroni}, {Walmsley}, \&
  {Churchwell}}]{Cesaroni1992}
{Cesaroni}, R., {Walmsley}, C.~M., \& {Churchwell}, E. 1992, A\&AS, 256, 618

\bibitem[{{Chambers} {et~al.}(2009){Chambers}, {Jackson}, {Rathborne}, \&
  {Simon}}]{Chambers2009}
{Chambers}, E.~T., {Jackson}, J.~M., {Rathborne}, J.~M., \& {Simon}, R. 2009,
  ApJS, 181, 360

\bibitem[{{Chen} {et~al.}(2009){Chen}, {Ellingsen}, \& {Shen}}]{Chen2009}
{Chen}, X., {Ellingsen}, S.~P., \& {Shen}, Z.-Q. 2009, MNRAS, 396, 1603

\bibitem[{{Chrysostomou} {et~al.}(2008){Chrysostomou}, {Bacciotti}, {Nisini},
  {Ray}, {Eisl{\"o}ffel}, {Davis}, \& {Takami}}]{Chrysostomou2008}
{Chrysostomou}, A., {Bacciotti}, F., {Nisini}, B., {Ray}, T.~P.,
  {Eisl{\"o}ffel}, J., {Davis}, C.~J., \& {Takami}, M. 2008, A\&AS, 482, 575

\bibitem[{{Churchwell}(1999)}]{Churchwell1999}
{Churchwell}, E. 1999, in NATO ASIC Proc. 540: The Origin of Stars and
  Planetary Systems, ed. C.~J. {Lada} \& N.~D. {Kylafis}, 515

\bibitem[{{Churchwell} {et~al.}(2009){Churchwell}, {Babler}, {Meade},
  {Whitney}, {Benjamin}, {Indebetouw}, {Cyganowski}, {Robitaille}, {Povich},
  {Watson}, \& {Bracker}}]{Churchwell2009}
{Churchwell}, E., {et~al.} 2009, PASP, 121, 213

\bibitem[{{Clarke} {et~al.}(2000){Clarke}, {Bonnell}, \&
  {Hillenbrand}}]{Clarke2000}
{Clarke}, C.~J., {Bonnell}, I.~A., \& {Hillenbrand}, L.~A. 2000, Protostars and
  Planets IV, 151

\bibitem[{{Clemens}(1985)}]{Clemens1985}
{Clemens}, D.~P. 1985, ApJ, 295, 422

\bibitem[{{Codella} {et~al.}(2004){Codella}, {Lorenzani}, {Gallego},
  {Cesaroni}, \& {Moscadelli}}]{Codella2004}
{Codella}, C., {Lorenzani}, A., {Gallego}, A.~T., {Cesaroni}, R., \&
  {Moscadelli}, L. 2004, A\&AS, 417, 615

\bibitem[{{Codella} \& {Moscadelli}(2000)}]{Codella2000}
{Codella}, C. \& {Moscadelli}, L. 2000, A\&AS, 362, 723

\bibitem[{{Contreras} {et~al.}(2013){Contreras}, {Schuller}, {Urquhart},
  {Csengeri}, {Wyrowski}, {Beuther}, {Bontemps}, {Bronfman}, {Henning},
  {Menten}, {Schilke}, {Walmsley}, {Wienen}, {Tackenberg}, \&
  {Linz}}]{Contreras2013}
{Contreras}, Y., {et~al.} 2013, A\&AS, 549, A45

\bibitem[{{Csengeri} {et~al.}(2014){Csengeri}, {Urquhart}, {Schuller}, {Motte},
  {Bontemps}, {Wyrowski}, {Menten}, {Bronfman}, {Beuther}, {Henning}, {Testi},
  {Zavagno}, \& {Walmsley}}]{Csengeri2014}
{Csengeri}, T., {et~al.} 2014, A\&AS, 565, A75

\bibitem[{{Curtis} {et~al.}(2010){Curtis}, {Richer}, \& {Buckle}}]{Curtis2010}
{Curtis}, E.~I., {Richer}, J.~S., \& {Buckle}, J.~V. 2010, MNRAS, 401, 455

\bibitem[{{Cyganowski} {et~al.}(2009){Cyganowski}, {Brogan}, {Hunter}, \&
  {Churchwell}}]{Cyganowski2009}
{Cyganowski}, C.~J., {Brogan}, C.~L., {Hunter}, T.~R., \& {Churchwell}, E.
  2009, ApJ, 702, 1615

\bibitem[{{Cyganowski} {et~al.}(2008){Cyganowski}, {Whitney}, {Holden},
  {Braden}, {Brogan}, {Churchwell}, {Indebetouw}, {Watson}, {Babler},
  {Benjamin}, {Gomez}, {Meade}, {Povich}, {Robitaille}, \&
  {Watson}}]{Cyganowski2008}
{Cyganowski}, C.~J., {et~al.} 2008, AJ, 136, 2391

\bibitem[{{Davis} {et~al.}(2010){Davis}, {Chrysostomou}, {Hatchell},
  {Wouterloot}, {Buckle}, {Nutter}, {Fich}, {Brunt}, {Butner}, {Cavanagh},
  {Curtis}, {Duarte-Cabral}, {di Francesco}, {Etxaluze}, {Friberg}, {Friesen},
  {Fuller}, {Graves}, {Greaves}, {Hogerheijde}, {Johnstone}, {Matthews},
  {Matthews}, {Rawlings}, {Richer}, {Roberts}, {Sadavoy}, {Simpson}, {Tothill},
  {Tsamis}, {Viti}, {Ward-Thompson}, {White}, \& {Yates}}]{Davis2010}
{Davis}, C.~J., {et~al.} 2010, MNRAS, 405, 759

\bibitem[{{De Buizer} {et~al.}(2012){De Buizer}, {Bartkiewicz}, \&
  {Szymczak}}]{DeBuizer2012}
{De Buizer}, J.~M., {Bartkiewicz}, A., \& {Szymczak}, M. 2012, ApJ, 754, 149

\bibitem[{{De Buizer} \& {Vacca}(2010)}]{DeBuizer2010}
{De Buizer}, J.~M. \& {Vacca}, W.~D. 2010, AJ, 140, 196

\bibitem[{{Dent} {et~al.}(2000){Dent}, {Duncan}, {Ellis}, {Harris},
  {Lightfoot}, {Wall}, {Gibson}, {Hills}, {Richer}, {Smith}, {Withington},
  {Burgess}, {Casorso}, {Dewdney}, {Hovey}, {Redman}, {Yeung}, {Force}, \&
  {Pain}}]{Dent2000}
{Dent}, W., {et~al.} 2000, in Astronomical Society of the Pacific Conference
  Series, Vol. 217, Imaging at Radio through Submillimeter Wavelengths, ed.
  J.~G. {Mangum} \& S.~J.~E. {Radford}, 33

\bibitem[{{Downes} \& {Cabrit}(2007)}]{Downes2007}
{Downes}, T.~P. \& {Cabrit}, S. 2007, A\&AS, 471, 873

\bibitem[{{Dunham} {et~al.}(2011){Dunham}, {Robitaille}, {Evans}, {Schlingman},
  {Cyganowski}, \& {Urquhart}}]{Dunham2011}
{Dunham}, M.~K., {Robitaille}, T.~P., {Evans}, II, N.~J., {Schlingman}, W.~M.,
  {Cyganowski}, C.~J., \& {Urquhart}, J. 2011, ApJ, 731, 90

\bibitem[{{Ellingsen}(2006)}]{Ellingsen2006}
{Ellingsen}, S.~P. 2006, ApJ, 638, 241

\bibitem[{{Fish} {et~al.}(2003){Fish}, {Reid}, {Argon}, \& {Menten}}]{Fish2003}
{Fish}, V.~L., {Reid}, M.~J., {Argon}, A.~L., \& {Menten}, K.~M. 2003, ApJ,
  596, 328

\bibitem[{{Frerking} {et~al.}(1982){Frerking}, {Langer}, \&
  {Wilson}}]{Frerking1982}
{Frerking}, M.~A., {Langer}, W.~D., \& {Wilson}, R.~W. 1982, ApJ, 262, 590

\bibitem[{{Garden} {et~al.}(1991){Garden}, {Hayashi}, {Hasegawa}, {Gatley}, \&
  {Kaifu}}]{Garden1991}
{Garden}, R.~P., {Hayashi}, M., {Hasegawa}, T., {Gatley}, I., \& {Kaifu}, N.
  1991, ApJ, 374, 540

\bibitem[{{Goldsmith} {et~al.}(1984){Goldsmith}, {Snell}, {Hemeon-Heyer}, \&
  {Langer}}]{Goldsmith1984}
{Goldsmith}, P.~F., {Snell}, R.~L., {Hemeon-Heyer}, M., \& {Langer}, W.~D.
  1984, ApJ, 286, 599

\bibitem[{{Gottschalk} {et~al.}(2012){Gottschalk}, {Kothes}, {Matthews},
  {Landecker}, \& {Dent}}]{Gottschalk2012}
{Gottschalk}, M., {Kothes}, R., {Matthews}, H.~E., {Landecker}, T.~L., \&
  {Dent}, W.~R.~F. 2012, A\&AS, 541, A79

\bibitem[{{Green} {et~al.}(2009){Green}, {Caswell}, {Fuller}, {Avison},
  {Breen}, {Brooks}, {Burton}, {Chrysostomou}, {Cox}, {Diamond}, {Ellingsen},
  {Gray}, {Hoare}, {Masheder}, {McClure-Griffiths}, {Pestalozzi}, {Phillips},
  {Quinn}, {Thompson}, {Voronkov}, {Walsh}, {Ward-Thompson}, {Wong-McSweeney},
  {Yates}, \& {Cohen}}]{Green2009}
{Green}, J.~A., {et~al.} 2009, MNRAS, 392, 783

\bibitem[{{Green} \& {McClure-Griffiths}(2011)}]{Green2011}
{Green}, J.~A. \& {McClure-Griffiths}, N.~M. 2011, MNRAS, 417, 2500

\bibitem[{{Hatchell} {et~al.}(2007){Hatchell}, {Fuller}, \&
  {Richer}}]{Hatchell2007}
{Hatchell}, J., {Fuller}, G.~A., \& {Richer}, J.~S. 2007, A\&AS, 472, 187

\bibitem[{{Hatchell} {et~al.}(1998){Hatchell}, {Thompson}, {Millar}, \&
  {MacDonald}}]{Hatchell1998}
{Hatchell}, J., {Thompson}, M.~A., {Millar}, T.~J., \& {MacDonald}, G.~H. 1998,
  133, 29

\bibitem[{{Henning} {et~al.}(2000){Henning}, {Schreyer}, {Launhardt}, \&
  {Burkert}}]{Henning2000}
{Henning}, T., {Schreyer}, K., {Launhardt}, R., \& {Burkert}, A. 2000, A\&AS,
  353, 211

\bibitem[{{Herbst} \& {van Dishoeck}(2009)}]{Herbst2009}
{Herbst}, E. \& {van Dishoeck}, E.~F. 2009, ARA\&A, 47, 427

\bibitem[{{Hofner} {et~al.}(2000){Hofner}, {Wyrowski}, {Walmsley}, \&
  {Churchwell}}]{Hofner2000}
{Hofner}, P., {Wyrowski}, F., {Walmsley}, C.~M., \& {Churchwell}, E. 2000, ApJ,
  536, 393

\bibitem[{{Jijina} \& {Adams}(1996)}]{Jijina1996}
{Jijina}, J. \& {Adams}, F.~C. 1996, ApJ, 462, 874

\bibitem[{{Kim} \& {Kurtz}(2006)}]{Kim2006}
{Kim}, K.-T. \& {Kurtz}, S.~E. 2006, ApJ, 643, 978

\bibitem[{{Konigl} \& {Pudritz}(2000)}]{Konigl2000}
{Konigl}, A. \& {Pudritz}, R.~E. 2000, Protostars and Planets IV, 759

\bibitem[{{Krumholz} {et~al.}(2007){Krumholz}, {Klein}, \&
  {McKee}}]{Krumholz2007}
{Krumholz}, M.~R., {Klein}, R.~I., \& {McKee}, C.~F. 2007, ApJ, 656, 959

\bibitem[{{Krumholz} {et~al.}(2009){Krumholz}, {Klein}, {McKee}, {Offner}, \&
  {Cunningham}}]{Krumholz2009}
{Krumholz}, M.~R., {Klein}, R.~I., {McKee}, C.~F., {Offner}, S.~S.~R., \&
  {Cunningham}, A.~J. 2009, 323, 754

\bibitem[{{Kuiper} {et~al.}(2010){Kuiper}, {Klahr}, {Beuther}, \&
  {Henning}}]{Kuiper2010}
{Kuiper}, R., {Klahr}, H., {Beuther}, H., \& {Henning}, T. 2010, ApJ, 722, 1556

\bibitem[{{Kurtz} {et~al.}(2000){Kurtz}, {Cesaroni}, {Churchwell}, {Hofner}, \&
  {Walmsley}}]{Kurtz2000}
{Kurtz}, S., {Cesaroni}, R., {Churchwell}, E., {Hofner}, P., \& {Walmsley},
  C.~M. 2000, Protostars and Planets IV, 299

\bibitem[{{Lada} \& {Lada}(2003)}]{Lada2003}
{Lada}, C.~J. \& {Lada}, E.~A. 2003, ARA\&A, 41, 57

\bibitem[{{L{\'o}pez-Sepulcre} {et~al.}(2009){L{\'o}pez-Sepulcre}, {Codella},
  {Cesaroni}, {Marcelino}, \& {Walmsley}}]{Lopez2009}
{L{\'o}pez-Sepulcre}, A., {Codella}, C., {Cesaroni}, R., {Marcelino}, N., \&
  {Walmsley}, C.~M. 2009, A\&AS, 499, 811

\bibitem[{{McKee} \& {Tan}(2003)}]{McKee2003}
{McKee}, C.~F. \& {Tan}, J.~C. 2003, ApJ, 585, 850

\bibitem[{{Menten}(1991)}]{Menten1991}
{Menten}, K.~M. 1991, ApJ, 380, L75

\bibitem[{{Minchin} {et~al.}(1993){Minchin}, {White}, \&
  {Padman}}]{Minchin1993}
{Minchin}, N.~R., {White}, G.~J., \& {Padman}, R. 1993, A\&AS, 277, 595

\bibitem[{{Minier} {et~al.}(2005){Minier}, {Burton}, {Hill}, {Pestalozzi},
  {Purcell}, {Garay}, {Walsh}, \& {Longmore}}]{Minier2005}
{Minier}, V., {Burton}, M.~G., {Hill}, T., {Pestalozzi}, M.~R., {Purcell},
  C.~R., {Garay}, G., {Walsh}, A.~J., \& {Longmore}, S. 2005, A\&AS, 429, 945

\bibitem[{{Minier} {et~al.}(2001){Minier}, {Conway}, \& {Booth}}]{Minier2001}
{Minier}, V., {Conway}, J.~E., \& {Booth}, R.~S. 2001, A\&AS, 369, 278

\bibitem[{{Minier} {et~al.}(2003){Minier}, {Ellingsen}, {Norris}, \&
  {Booth}}]{Minier2003}
{Minier}, V., {Ellingsen}, S.~P., {Norris}, R.~P., \& {Booth}, R.~S. 2003,
  VizieR Online Data Catalog, 3403, 31095

\bibitem[{{Molinari} {et~al.}(1996){Molinari}, {Brand}, {Cesaroni}, \&
  {Palla}}]{Molinari1996}
{Molinari}, S., {Brand}, J., {Cesaroni}, R., \& {Palla}, F. 1996, A\&AS, 308,
  573

\bibitem[{{Molinari} {et~al.}(1998){Molinari}, {Brand}, {Cesaroni}, {Palla}, \&
  {Palumbo}}]{Molinari1998}
{Molinari}, S., {Brand}, J., {Cesaroni}, R., {Palla}, F., \& {Palumbo},
  G.~G.~C. 1998, A\&AS, 336, 339

\bibitem[{{Molinari} {et~al.}(2002){Molinari}, {Testi}, {Rodr{\'{\i}}guez}, \&
  {Zhang}}]{Molinari2002}
{Molinari}, S., {Testi}, L., {Rodr{\'{\i}}guez}, L.~F., \& {Zhang}, Q. 2002,
  ApJ, 570, 758

\bibitem[{{Moore} {et~al.}(2007){Moore}, {Bretherton}, {Fujiyoshi}, {Ridge},
  {Allsopp}, {Hoare}, {Lumsden}, \& {Richer}}]{Moore2007}
{Moore}, T.~J.~T., {Bretherton}, D.~E., {Fujiyoshi}, T., {Ridge}, N.~A.,
  {Allsopp}, J., {Hoare}, M.~G., {Lumsden}, S.~L., \& {Richer}, J.~S. 2007,
  MNRAS, 379, 663

\bibitem[{{Mottram} {et~al.}(2011){Mottram}, {Hoare}, {Davies}, {Lumsden},
  {Oudmaijer}, {Urquhart}, {Moore}, {Cooper}, \& {Stead}}]{Mottram2011}
{Mottram}, J.~C., {et~al.} 2011, ApJ, 730, L33

\bibitem[{{Narayanan} {et~al.}(2012){Narayanan}, {Snell}, \&
  {Bemis}}]{Narayanan2012}
{Narayanan}, G., {Snell}, R., \& {Bemis}, A. 2012, MNRAS, 425, 2641

\bibitem[{{Norberg} \& {Maeder}(2000)}]{Norberg2000}
{Norberg}, P. \& {Maeder}, A. 2000, A\&AS, 359, 1025

\bibitem[{{Ortega} {et~al.}(2012){Ortega}, {Paron}, {Cichowolski}, {Rubio}, \&
  {Dubner}}]{Ortega2012}
{Ortega}, M.~E., {Paron}, S., {Cichowolski}, S., {Rubio}, M., \& {Dubner}, G.
  2012, A\&AS, 546, A96

\bibitem[{{Parker} {et~al.}(1991){Parker}, {Padman}, \& {Scott}}]{Parker1991}
{Parker}, N.~D., {Padman}, R., \& {Scott}, P.~F. 1991, MNRAS, 252, 442

\bibitem[{{Parsons} {et~al.}(2012){Parsons}, {Thompson}, {Clark}, \&
  {Chrysostomou}}]{Parsons2012}
{Parsons}, H., {Thompson}, M.~A., {Clark}, J.~S., \& {Chrysostomou}, A. 2012,
  MNRAS, 424, 1658

\bibitem[{{Pestalozzi} {et~al.}(2009){Pestalozzi}, {Elitzur}, \&
  {Conway}}]{Pestalozzi2009}
{Pestalozzi}, M.~R., {Elitzur}, M., \& {Conway}, J.~E. 2009, A\&AS, 501, 999

\bibitem[{{Purcell} {et~al.}(2006){Purcell}, {Balasubramanyam}, {Burton},
  {Walsh}, {Minier}, {Hunt-Cunningham}, {Kedziora-Chudczer}, {Longmore},
  {Hill}, {Bains}, {Barnes}, {Busfield}, {Calisse}, {Crighton}, {Curran},
  {Davis}, {Dempsey}, {Derragopian}, {Fulton}, {Hidas}, {Hoare}, {Lee}, {Ladd},
  {Lumsden}, {Moore}, {Murphy}, {Oudmaijer}, {Pracy}, {Rathborne}, {Robertson},
  {Schultz}, {Shobbrook}, {Sparks}, {Storey}, \& {Travouillion}}]{Purcell2006}
{Purcell}, C.~R., {et~al.} 2006, MNRAS, 367, 553

\bibitem[{{Reid} {et~al.}(2009){Reid}, {Menten}, {Zheng}, {Brunthaler},
  {Moscadelli}, {Xu}, {Zhang}, {Sato}, {Honma}, {Hirota}, {Hachisuka}, {Choi},
  {Moellenbrock}, \& {Bartkiewicz}}]{Reid2009}
{Reid}, M.~J., {et~al.} 2009, ApJ, 700, 137

\bibitem[{{Richer} {et~al.}(2000){Richer}, {Shepherd}, {Cabrit}, {Bachiller},
  \& {Churchwell}}]{Richer2000}
{Richer}, J.~S., {Shepherd}, D.~S., {Cabrit}, S., {Bachiller}, R., \&
  {Churchwell}, E. 2000, Protostars and Planets IV, 867

\bibitem[{{Ridge} \& {Moore}(2001)}]{Ridge2001}
{Ridge}, N.~A. \& {Moore}, T.~J.~T. 2001, A\&AS, 378, 495

\bibitem[{{Rodriguez} {et~al.}(1982){Rodriguez}, {Carral}, {Ho}, \&
  {Moran}}]{Rodriguez1982}
{Rodriguez}, L.~F., {Carral}, P., {Ho}, P.~T.~P., \& {Moran}, J.~M. 1982, ApJ,
  260, 635

\bibitem[{{Roman-Duval} {et~al.}(2009){Roman-Duval}, {Jackson}, {Heyer},
  {Johnson}, {Rathborne}, {Shah}, \& {Simon}}]{Roman2009}
{Roman-Duval}, J., {Jackson}, J.~M., {Heyer}, M., {Johnson}, A., {Rathborne},
  J., {Shah}, R., \& {Simon}, R. 2009, ApJ, 699, 1153

\bibitem[{{S{\'a}nchez-Monge} {et~al.}(2013){S{\'a}nchez-Monge},
  {L{\'o}pez-Sepulcre}, {Cesaroni}, {Walmsley}, {Codella}, {Beltr{\'a}n},
  {Pestalozzi}, \& {Molinari}}]{SanchezMonge2013}
{S{\'a}nchez-Monge}, {\'A}., {L{\'o}pez-Sepulcre}, A., {Cesaroni}, R.,
  {Walmsley}, C.~M., {Codella}, C., {Beltr{\'a}n}, M.~T., {Pestalozzi}, M., \&
  {Molinari}, S. 2013, A\&AS, 557, A94

\bibitem[{{Schuller} {et~al.}(2009){Schuller}, {Menten}, {Contreras},
  {Wyrowski}, {Schilke}, {Bronfman}, {Henning}, {Walmsley}, {Beuther},
  {Bontemps}, {Cesaroni}, {Deharveng}, {Garay}, {Herpin}, {Lefloch}, {Linz},
  {Mardones}, {Minier}, {Molinari}, {Motte}, {Nyman}, {Reveret}, {Risacher},
  {Russeil}, {Schneider}, {Testi}, {Troost}, {Vasyunina}, {Wienen}, {Zavagno},
  {Kovacs}, {Kreysa}, {Siringo}, \& {Wei{\ss}}}]{Schuller2009}
{Schuller}, F., {et~al.} 2009, A\&AS, 504, 415

\bibitem[{{Shepherd} \& {Churchwell}(1996{\natexlab{a}})}]{Shepherd1996b}
{Shepherd}, D.~S. \& {Churchwell}, E. 1996{\natexlab{a}}, ApJ, 472, 225

\bibitem[{{Shepherd} \& {Churchwell}(1996{\natexlab{b}})}]{Shepherd1996a}
{Shepherd}, D.~S. \& {Churchwell}, E. 1996{\natexlab{b}}, ApJ, 457, 267

\bibitem[{{Shu}(1977)}]{Shu1977}
{Shu}, F.~H. 1977, ApJ, 214, 488

\bibitem[{{Shu} {et~al.}(1999){Shu}, {Allen}, {Shang}, {Ostriker}, \&
  {Li}}]{Shu1999}
{Shu}, F.~H., {Allen}, A., {Shang}, H., {Ostriker}, E.~C., \& {Li}, Z.-Y. 1999,
  in NATO ASIC Proc. 540: The Origin of Stars and Planetary Systems, ed. C.~J.
  {Lada} \& N.~D. {Kylafis}, 193

\bibitem[{{Shu} {et~al.}(2000){Shu}, {Najita}, {Shang}, \& {Li}}]{Shu2000}
{Shu}, F.~H., {Najita}, J.~R., {Shang}, H., \& {Li}, Z.-Y. 2000, Protostars and
  Planets IV, 789

\bibitem[{{Shu} {et~al.}(1991){Shu}, {Ruden}, {Lada}, \& {Lizano}}]{Shu1991}
{Shu}, F.~H., {Ruden}, S.~P., {Lada}, C.~J., \& {Lizano}, S. 1991, ApJ, 370,
  L31

\bibitem[{{Simon} {et~al.}(2006){Simon}, {Rathborne}, {Shah}, {Jackson}, \&
  {Chambers}}]{Simon2006}
{Simon}, R., {Rathborne}, J.~M., {Shah}, R.~Y., {Jackson}, J.~M., \&
  {Chambers}, E.~T. 2006, ApJ, 653, 1325

\bibitem[{{Smith} {et~al.}(2008){Smith}, {Buckle}, {Hills}, {Bell}, {Richer},
  {Curtis}, {Withington}, {Leech}, {Williamson}, {Dent}, {Hastings}, {Redman},
  {Wooff}, {Yeung}, {Friberg}, {Walther}, {Kackley}, {Jenness}, {Tilanus},
  {Dempsey}, {Kroug}, {Zijlstra}, \& {Klapwijk}}]{Smith2008}
{Smith}, H., {et~al.} 2008, in Society of Photo-Optical Instrumentation
  Engineers (SPIE) Conference Series, Vol. 7020, Society of Photo-Optical
  Instrumentation Engineers (SPIE) Conference Series

\bibitem[{{Smith} {et~al.}(2009){Smith}, {Longmore}, \& {Bonnell}}]{Smith2009}
{Smith}, R.~J., {Longmore}, S., \& {Bonnell}, I. 2009, MNRAS, 400, 1775

\bibitem[{{Snell} {et~al.}(1984){Snell}, {Scoville}, {Sanders}, \&
  {Erickson}}]{Snell1984}
{Snell}, R.~L., {Scoville}, N.~Z., {Sanders}, D.~B., \& {Erickson}, N.~R. 1984,
  ApJ, 284, 176

\bibitem[{{Sobolev} {et~al.}(1997){Sobolev}, {Cragg}, \&
  {Godfrey}}]{Sobolev1997}
{Sobolev}, A.~M., {Cragg}, D.~M., \& {Godfrey}, P.~D. 1997, A\&AS, 324, 211

\bibitem[{{Sridharan} {et~al.}(2002){Sridharan}, {Beuther}, {Schilke},
  {Menten}, \& {Wyrowski}}]{Sridharan2002}
{Sridharan}, T.~K., {Beuther}, H., {Schilke}, P., {Menten}, K.~M., \&
  {Wyrowski}, F. 2002, ApJ, 566, 931

\bibitem[{{Stahler} \& {Palla}(1993)}]{Stahler1993}
{Stahler}, S.~W. \& {Palla}, F. 1993, in Bulletin of the American Astronomical
  Society, Vol.~25, American Astronomical Society Meeting Abstracts \#182, 915

\bibitem[{{Su} {et~al.}(2004){Su}, {Zhang}, \& {Lim}}]{Su2004}
{Su}, Y.-N., {Zhang}, Q., \& {Lim}, J. 2004, ApJ, 604, 258

\bibitem[{{Szymczak} {et~al.}(2012){Szymczak}, {Wolak}, {Bartkiewicz}, \&
  {Borkowski}}]{Szymczak2012}
{Szymczak}, M., {Wolak}, P., {Bartkiewicz}, A., \& {Borkowski}, K.~M. 2012,
  Astronomische Nachrichten, 333, 634

\bibitem[{{Takahashi} {et~al.}(2008){Takahashi}, {Saito}, {Ohashi}, {Kusakabe},
  {Takakuwa}, {Shimajiri}, {Tamura}, \& {Kawabe}}]{Takahashi2008}
{Takahashi}, S., {Saito}, M., {Ohashi}, N., {Kusakabe}, N., {Takakuwa}, S.,
  {Shimajiri}, Y., {Tamura}, M., \& {Kawabe}, R. 2008, ApJ, 688, 344

\bibitem[{{Tomisaka}(1998)}]{Tomisaka1998}
{Tomisaka}, K. 1998, ApJ, 502, L163

\bibitem[{{Urquhart} {et~al.}(2012){Urquhart}, {Hoare}, {Lumsden}, {Oudmaijer},
  {Moore}, {Mottram}, {Cooper}, {Mottram}, \& {Rogers}}]{Urquhart2012}
{Urquhart}, J.~S., {et~al.} 2012, MNRAS, 420, 1656

\bibitem[{{Urquhart} {et~al.}(2014){Urquhart}, {Moore}, {Csengeri}, {Wyrowski},
  {Schuller}, {Hoare}, {Lumsden}, {Mottram}, {Thompson}, {Menten}, {Walmsley},
  {Bronfman}, {Pfalzner}, {K{\"o}nig}, \& {Wienen}}]{Urquhart2014b}
{Urquhart}, J.~S., {et~al.} 2014, ArXiv e-prints

\bibitem[{{Urquhart} {et~al.}(2011){Urquhart}, {Moore}, {Hoare}, {Lumsden},
  {Oudmaijer}, {Rathborne}, {Mottram}, {Davies}, \& {Stead}}]{Urquhart2011a}
{Urquhart}, J.~S., {et~al.} 2011, MNRAS, 410, 1237

\bibitem[{{Urquhart} {et~al.}(2013{\natexlab{a}}){Urquhart}, {Moore},
  {Schuller}, {Wyrowski}, {Menten}, {Thompson}, {Csengeri}, {Walmsley},
  {Bronfman}, \& {K{\"o}nig}}]{Urquhart2013}
{Urquhart}, J.~S., {et~al.} 2013{\natexlab{a}}, MNRAS, 431, 1752

\bibitem[{{Urquhart} {et~al.}(2013{\natexlab{b}}){Urquhart}, {Thompson},
  {Moore}, {Purcell}, {Hoare}, {Schuller}, {Wyrowski}, {Csengeri}, {Menten},
  {Lumsden}, {Kurtz}, {Walmsley}, {Bronfman}, {Morgan}, {Eden}, \&
  {Russeil}}]{Urquhart2013b}
{Urquhart}, J.~S., {et~al.} 2013{\natexlab{b}}, MNRAS, 435, 400

\bibitem[{{van der Marel} {et~al.}(2013){van der Marel}, {Kristensen},
  {Visser}, {Mottram}, {Y{\i}ld{\i}z}, \& {van Dishoeck}}]{vanderMarel2013}
{van der Marel}, N., {Kristensen}, L.~E., {Visser}, R., {Mottram}, J.~C.,
  {Y{\i}ld{\i}z}, U.~A., \& {van Dishoeck}, E.~F. 2013, A\&AS, 556, A76

\bibitem[{{van der Walt} {et~al.}(2007){van der Walt}, {Sobolev}, \&
  {Butner}}]{vanderWalt2007}
{van der Walt}, D.~J., {Sobolev}, A.~M., \& {Butner}, H. 2007, A\&AS, 464, 1015

\bibitem[{{van der Walt}(2005)}]{vanderWalt2005}
{van der Walt}, J. 2005, MNRAS, 360, 153

\bibitem[{{Walsh} {et~al.}(2003){Walsh}, {Macdonald}, {Alvey}, {Burton}, \&
  {Lee}}]{Walsh2003}
{Walsh}, A.~J., {Macdonald}, G.~H., {Alvey}, N.~D.~S., {Burton}, M.~G., \&
  {Lee}, J.-K. 2003, A\&AS, 410, 597

\bibitem[{{Williams} {et~al.}(2000){Williams}, {Blitz}, \&
  {McKee}}]{Williams2000}
{Williams}, J.~P., {Blitz}, L., \& {McKee}, C.~F. 2000, Protostars and Planets
  IV, 97

\bibitem[{{Williams} {et~al.}(1994){Williams}, {de Geus}, \&
  {Blitz}}]{Williams1994}
{Williams}, J.~P., {de Geus}, E.~J., \& {Blitz}, L. 1994, ApJ, 428, 693

\bibitem[{{Wilson} \& {Rood}(1994)}]{Wilson1994}
{Wilson}, T.~L. \& {Rood}, R. 1994, ARA\&A, 32, 191

\bibitem[{{Wolfire} \& {Cassinelli}(1987)}]{Wolfire1987}
{Wolfire}, M.~G. \& {Cassinelli}, J.~P. 1987, ApJ, 319, 850

\bibitem[{{Wu} {et~al.}(2004){Wu}, {Wei}, {Zhao}, {Shi}, {Yu}, {Qin}, \&
  {Huang}}]{Wu2004}
{Wu}, Y., {Wei}, Y., {Zhao}, M., {Shi}, Y., {Yu}, W., {Qin}, S., \& {Huang}, M.
  2004, A\&AS, 426, 503

\bibitem[{{Wu} {et~al.}(1999){Wu}, {Yang}, {Li}, {Lei}, {Sun}, {L{\"u}}, \&
  {Fu}}]{Wu1999}
{Wu}, Y., {Yang}, C., {Li}, Y., {Lei}, C., {Sun}, J., {L{\"u}}, J., \& {Fu}, H.
  1999, Science in China A: Mathematics, 42, 732

\bibitem[{{Wu} {et~al.}(2010){Wu}, {Xu}, {Pandian}, {Yang}, {Henkel}, {Menten},
  \& {Zhang}}]{Wu2010}
{Wu}, Y.~W., {Xu}, Y., {Pandian}, J.~D., {Yang}, J., {Henkel}, C., {Menten},
  K.~M., \& {Zhang}, S.~B. 2010, ApJ, 720, 392

\bibitem[{{Xu} \& {Wang}(2013)}]{Xu2013}
{Xu}, J.-L. \& {Wang}, J.-J. 2013, Research in Astronomy and Astrophysics, 13,
  39

\bibitem[{{Xu} {et~al.}(2006){Xu}, {Shen}, {Yang}, {Zheng}, {Miyazaki},
  {Sunada}, {Ma}, {Li}, {Sun}, \& {Pei}}]{Xu2006}
{Xu}, Y., {et~al.} 2006, AJ, 132, 20

\bibitem[{{Yorke} \& {Sonnhalter}(2002)}]{Yorke2002}
{Yorke}, H.~W. \& {Sonnhalter}, C. 2002, ApJ, 569, 846

\bibitem[{{Zapata} {et~al.}(2010){Zapata}, {Tang}, \& {Leurini}}]{Zapata2010}
{Zapata}, L.~A., {Tang}, Y.-W., \& {Leurini}, S. 2010, ApJ, 725, 1091

\bibitem[{{Zhang} {et~al.}(2005){Zhang}, {Hunter}, {Brand}, {Sridharan},
  {Cesaroni}, {Molinari}, {Wang}, \& {Kramer}}]{Zhang2005}
{Zhang}, Q., {Hunter}, T.~R., {Brand}, J., {Sridharan}, T.~K., {Cesaroni}, R.,
  {Molinari}, S., {Wang}, J., \& {Kramer}, M. 2005, ApJ, 625, 864

\bibitem[{{Zhang} {et~al.}(2001){Zhang}, {Hunter}, {Brand}, {Sridharan},
  {Molinari}, {Kramer}, \& {Cesaroni}}]{Zhang2001}
{Zhang}, Q., {Hunter}, T.~R., {Brand}, J., {Sridharan}, T.~K., {Molinari}, S.,
  {Kramer}, M.~A., \& {Cesaroni}, R. 2001, ApJ, 552, L167

\bibitem[{{Zinnecker} \& {Yorke}(2007)}]{Zinnecker2007}
{Zinnecker}, H. \& {Yorke}, H.~W. 2007, ARA\&A, 45, 481

\end{thebibliography}

\onecolumn
\include{AppendixA}
\include{AppendixB}

\end{document}